\documentclass[acmsmall]{acmart}

\AtBeginDocument{%
  \providecommand\BibTeX{{%
    \normalfont B\kern-0.5em{\scshape i\kern-0.25em b}\kern-0.8em\TeX}}}

\setcopyright{levonianArxiv} 
\copyrightyear{2020}
\acmYear{2020}
\acmDOI{XX.XXXX/XXXXXX.XXXXXXX}

\acmJournal{TOCHI}
\acmVolume{{CSCW}}
\acmNumber{1}
\acmArticle{1}
\acmMonth{10}

\usepackage{graphicx}
\usepackage{booktabs}
\usepackage{layouts}  
\usepackage{subcaption} 
\usepackage{makecell} 
\usepackage{multirow} 

\begin{document}

\title{Patterns of Patient and Caregiver Mutual Support Connections in an Online Health Community}

\author{Zachary Levonian}
\email{levon003@umn.edu}
\orcid{0000-0002-8932-1489}
\affiliation{%
  \institution{GroupLens Research}
  \streetaddress{University of Minnesota}
  \city{Minneapolis}
  \state{Minnesota}
  \postcode{55455}
}

\author{Marco Dow}
\email{dow00017@umn.edu}
\affiliation{%
  \institution{GroupLens Research}
  \streetaddress{University of Minnesota}
  \city{Minneapolis}
  \state{Minnesota}
  \postcode{55455}
}

\author{Drew Erikson}
\email{eriks074@umn.edu}
\affiliation{%
  \institution{GroupLens Research}
  \streetaddress{University of Minnesota}
  \city{Minneapolis}
  \state{Minnesota}
  \postcode{55455}
}

\author{Sourojit Ghosh}
\email{ghosh100@umn.edu}
\affiliation{%
  \institution{GroupLens Research}
  \streetaddress{University of Minnesota}
  \city{Minneapolis}
  \state{Minnesota}
  \postcode{55455}
}

\author{Hannah Miller Hillberg}
\email{hillbergh@uwosh.edu}
\affiliation{%
  \institution{University of Wisconsin - Oshkosh}
  \streetaddress{800 Algoma Boulevard}
  \city{Oshkosh}
  \state{Wisconsin}
  \postcode{54901}
}

\author{Saumik Narayanan}
\email{saumik.narayanan@outlook.com}
\affiliation{%
  \institution{GroupLens Research}
  \streetaddress{University of Minnesota}
  \city{Minneapolis}
  \state{Minnesota}
  \postcode{55455}
}

\author{Loren Terveen}
\email{terveen@umn.edu}
\affiliation{%
  \institution{GroupLens Research}
  \streetaddress{University of Minnesota}
  \city{Minneapolis}
  \state{Minnesota}
  \postcode{55455}
}

\author{Svetlana Yarosh}
\email{lana@umn.edu}
\affiliation{%
  \institution{GroupLens Research}
  \streetaddress{University of Minnesota}
  \city{Minneapolis}
  \state{Minnesota}
  \postcode{55455}
}

\renewcommand{\shortauthors}{Levonian et al.}

\begin{abstract}
Online health communities offer the promise of support benefits to users, in particular because these communities enable users to find peers with similar experiences.
Building mutually supportive connections between peers is a key motivation for using online health communities.  
However, a user's role in a community may influence the formation of peer connections.
In this work, we study patterns of peer connections between two structural health roles: \textit{patient} and \textit{non-professional caregiver}.
We examine user behavior in an online health community---CaringBridge.org---where finding peers is not explicitly supported.
This context lets us use social network analysis methods to explore the growth of such connections in the wild and identify users' peer communication preferences.
We investigated how connections between peers were initiated, finding that initiations are more likely between two authors who have the same role and who are close within the broader communication network.  Relationships---patterns of repeated interactions---are also more likely to form and be more interactive when authors have the same role.
Our results have implications for the design of systems supporting peer communication, e.g. peer-to-peer recommendation systems.
\end{abstract}



\maketitle


\section{Introduction} 

Online health communities (OHCs) give patients and caregivers the opportunity to mutually support one another.
To realize this opportunity, OHCs must be designed to facilitate supportive communication between peers.
Social support occurs within the context of an individual's relationships with others~\cite{albrecht_communicating_1987,berkman_social_2014}, so designing for new peer connections requires an understanding of the individual factors associated with the formation and growth of mutually-supportive relationships.
In this study, we explore these factors by analyzing an interaction network of peer connections in an OHC that lacks an elaborate technical infrastructure for finding and forming those peer connections.

CaringBridge\footnote{\url{https://www.caringbridge.org/}} is an OHC designed to enable patients and caregivers to communicate with their friends and family members about a health event such as an illness or injury \cite{smith_i_2020}.  
Patients and caregivers on CaringBridge author text posts for their extended support networks on individual blogs called ``sites''.
Interactions between authors---via comments on other authors' sites---constitute peer connections.
However, CaringBridge currently does not explicitly support authors in finding other authors to form connections with, providing only limited search functionality and no feed or recommendation system.
Despite this lack of affordance, there is substantial inter-author peer interaction on CaringBridge. 
Thus, studying CaringBridge represents a unique opportunity to observe users' preferences for peer connection when conventional social discovery mechanisms are absent: what connections are sufficiently desirable and accessible to users that they form without explicit design affordances?

We study this appropriative use of CaringBridge to learn about patient and caregiver preferences for peer connections and to understand the factors that lead peer authors to form connections ``in the wild''.
Identifying these factors opens pathways to designing digital interventions that are faithful to user preferences and that provision support more effectively~\cite{cross_knowing_2004}.
One key factor is the role adopted by an OHC user.
While there are many roles in OHCs, our study focuses on two common health roles: patient and caregiver.
Patients and caregivers have different motivations and needs, and the differences between patient and caregiver use of OHCs is understudied \cite{bloom_their_2019}.
Facilitating peer connections in OHCs is an area of active research \cite{eschler_im_2017,oleary_design_2017,yang_seekers_2019}, and understanding the impact of health role on peer connection formation and growth creates opportunities to facilitate these connections in a way that is responsive to patients' and caregivers' divergent needs.

We characterize connections between peer authors on CaringBridge with quantitative social network analysis.
To understand connection patterns, we analyze relationships between authors as they occur within the network formed by authors' interactions. 
The network context of an interaction can be as important as the content of that interaction: for example, research predicting cancer stage from users' OHC posts has found that network features were as predictive of cancer stage as the text of the posts themselves~\cite{jha_cancer_2010,wen_understanding_2012,tamersoy_characterizing_2015}.  Further, analyzing connections within an interaction network brings focus to the initiators of interactions and not just the support recipients.  In addition to forming the basis for supportive relationships, new connections provide benefits to both initiators and receivers~\cite{kim_process_2012,wills_downward_1981,riessman_helper_1965}.

To quantify the impact of authors' health role on connection formation, we use machine learning to identify patient and caregiver authors from their posts and identify differences in interaction behavior between patients and caregivers.
In addition to health role, we explore factors related to authors' level of activity, position within the interaction network, and health condition.
Our analysis explores these factors to address two broad research questions: \\
\textbf{RQ1 (Initiations):} What factors are associated with the initiation of a new connection by an author? \\
\textbf{RQ2 (Relationships):} What factors are associated with the reciprocation and growth of connections between peer authors?

To address RQ1 and identify factors associated with initiations, we first identify which authors engage in peer connection behavior. Second, we identify when authors' make their first peer connection relative to when they joined the site, as that first initiation is an explicit signal of peer finding behavior. 
Finally, we identify factors that make an author more likely to be the target of initiations.
Thus, we state three sub-questions for RQ1:
\begin{itemize}
\item \textbf{RQ1a:} Which authors initiate peer connections?
\item \textbf{RQ1b:} When do authors initiate peer connections?
\item \textbf{RQ1c:} With whom do authors initiate peer connections?
\end{itemize}

To address RQ2 and identify factors associated with reciprocation and the growth (or not) of dyadic relationships, we first identify which initiations are likely to be reciprocated. Secondly, we examine which reciprocated dyads are likely to be more interactive and more balanced. Thus, we state two sub-questions for RQ2:
\begin{itemize}
\item \textbf{RQ2a:} Which authors reciprocate? To which initiators?
\item \textbf{RQ2b:} What factors lead to more interactive relationships?
\end{itemize}

The contributions of this paper are the identification of factors associated with the formation and growth of peer connections and a comparison of connection behavior between two important health roles: patient and caregiver.
Specifically, we find that (1) patients are more likely than caregivers to initiate with other authors after receiving interactions, (2) patients who initiate do so earlier than caregivers, (3) authors are more likely to interact with others sharing the same author role, and (4) authors are more likely to reciprocate and form more interactive relationships with others sharing the same author role.
These differences in OHC use by author role have implications for the design of online health communities and other digital interventions that benefit both patients and caregivers~\cite{yang_seekers_2019}.  We discuss opportunities to integrate these results in peer recommendation systems to foster mutually supportive relationships.

\section{Related Work} 

In this paper, we aim to understand the communication preferences of health peers by studying interactions between CaringBridge authors.
To understand those preferences, we first discuss motivations for use of digital communications, specifically use of OHCs for peer support.  
Second, we discuss research on health roles, laying out the conceptual groundwork for a focus on patients and caregivers. 
Finally, we discuss connection dynamics on OHCs, with a focus on factors previously identified as important for the formation of new connections and their impact on the formation of supportive peer relationships.  


\subsection{Motivations for digital communication during health journeys}
Patients use the internet to find information and support \cite{hoybye_online_2005,chou_health-related_2011}.
For pursuing social connection specifically, patients use the internet to overcome isolation, identify others with similar experiences, reinforce existing relationships, and offset deficits in existing relationships \cite{rains_coping_2018}.
CaringBridge is designed primarily for reinforcing existing relationships~\cite{smith_i_2020}.
However, the existence of connections between authors indicates that authors also are using CaringBridge to address unmet needs~\cite{massimi_life_2014} and build supportive connections based on shared concerns~\cite{frost_social_2008}.
These support-seeking behaviors result in the formation of \textit{peer} connections, which we discuss next.



Connecting with experientially-similar others is a key motivation for patients to participate in online support communities \cite{rains_social_2016,thoits_mechanisms_2011,fox_social_2011}.
Experientially-similar others serve as important sources of support to both patients \cite{thoits_mechanisms_2011} and caregivers \cite{gage_social_2013,schorch_designing_2016}.
For CaringBridge authors, experiential similarity is entangled with the notion of authors as peers.
We adopt the four-point definition of peer suggested by Simoni et al.: (a) sharing personal circumstances i.e. some form of health challenge, (b) obtaining benefits from peer support that derive from their status as peers, (c) lacking professional training or medical credentials, and (d) ``intentionally setting out to interact with individuals they may or may not encounter in their everyday life''~\cite{simoni_peer_2011}.
Finding peers to communicate with online is complicated by many individual factors \cite{eschler_im_2017}, which we model quantitatively in this study.
Peer-finding behaviors have been implicated as an opportunity for technological innovation in the context of OHCs \cite{yang_seekers_2019,newman_its_2011} and more broadly \cite{cohen_social_2004,dunkel-schetter_social_1984}.
We study connections between peers---rather than other health relationships such as mentor/mentee and medical professional/patient---as an opportunity for exchanging support with experientially-similar others.

As CaringBridge most resembles individual health blogs, the motivations of CaringBridge authors may differ from users of other kinds of OHCs.
Blogging is fundamentally social~\cite{rains_social_2011}.
Health blogging fulfills both communicative and therapeutic roles~\cite{mccosker_living_2013}, with patients sharing their illness trajectories and processing their experiences through writing~\cite{huh_collaborative_2012}.
While blogging may provide benefits to patients due to the expressive self-disclosure involved in writing blog posts~\cite{ma_write_2017,hsiu-chia_can_2009}, having responsive and interactive readers provides additional benefits~\cite{rains_social_2011}. 
McCosker and Darcy argue that connectivity between blog authors has the potential to sustain health bloggers in their writings about their health journeys~\cite{mccosker_living_2013}. 
Our study focuses on communication between ``blog'' authors as a potent opportunity for understanding the dynamics that produce the benefits of these interactive connections.

\subsection{Patient \& caregiver: Structural health roles}\label{sec:related_lit_health_roles}

Users take varied roles in OHCs \cite{maloney-krichmar_multilevel_2005,yang_seekers_2019}.  
Research examining roles in OHCs has tended to focus on group \cite{maloney-krichmar_multilevel_2005,benne_functional_1948} or social roles \cite{yang_seekers_2019,sharma_engagement_2020} that are defined by behaviors.  
For example, Sharma et al.\ define a ``seeker'' role in a mental health forum as a person who makes a new thread~\cite{sharma_engagement_2020}.
In contrast, we examine structural roles that arise from a health event that creates a \textit{patient} and any number of non-professional auxiliary \textit{caregivers}; these structural roles are adopted by patients and caregivers and are defined by accompanying expectations and responsibilities~\cite{ebaugh_becoming_1988,brim_properties_1980}.
The expectations of each role are associated with (but not defined by) a set of behaviors that are ``characteristic of the person in a particular setting''~\cite{stewart_exploration_2005}.
Thus, patients and caregivers have different behaviors as they enact their role in an OHC.
Patient and caregiver responsibilities and behaviors may change frequently~\cite{levonian_bridging_2020}, but their role is relatively stable. This stability contrasts with the frequently-shifting behavior roles identified by Yang et al.\ in an online cancer forum~\cite{yang_seekers_2019}.
Note that structural roles are not explicitly afforded via the technical interface, in contrast to e.g. moderator roles on Wikipedia and other explicit roles that have been studied online~\cite{burke_mopping_2008,yan_network_2015}.
People with the same structural role may be more likely to interact with each other; Xu et al.\ found that online communication was more likely to occur between Twitter users who had the same health role, such as ``provider'' or ``engaged consumer''~\cite{xu_twitter_2015}.
We explore the interaction dynamics between patients and caregivers in detail on CaringBridge.

Patients and caregivers communicate differently online, although these differences have received little explicit focus.  
Lu et al.---a notable exception---identified differences in topic and sentiment in posts written by patients and caregivers in an online health forum~\cite{lu_understanding_2017}.  
In this work, however, we focus on supportive connections and the \textit{target} of online interactions by health role.
OHCs may provide a particular opportunity for caregivers seeking support online~\cite{colvin_caregivers_2004}, as patients are given ``interpretive precedence'' in dealing with a health condition, leaving caregivers without supportive relationships to understand their own role in a broader health journey~\cite{sanden_cancer_2019}.
When caregivers can communicate with other caregivers digitally, they develop more effective coping strategies for caregiving stress~\cite{namkoong_creating_2012}.
Offline, caregiver connections with other caregivers may be more passive than active; Gage suggests an important role of serendipity in the formation of new connections~\cite{gage_social_2013}.  
We find that serendipity plays a role in at least some of the connections on CaringBridge.
We aim to understand the differences in patient and caregiver communication on OHCs in response to a call for developing a deeper empirical understanding of OHC participation~\cite{rains_coping_2018}, a context in which caregivers are understudied~\cite{bloom_patients_2017}.


\subsection{OHCs as sources of interaction and support}


OHCs are associated with a variety of positive and negative health outcomes for their users, specifically due to the interaction that occurs on them~\cite{rains_coping_2018}.
Understanding this interaction in more detail---particularly with respect to author role---creates opportunities to improve the provisioning of support on OHCs.
For patients, use of OHCs is associated with greater perceived support \cite{oh_how_2014,rains_social_2016}, writing higher sentiment posts after interacting with others \cite{qiu_get_2011}, engagement measures such as duration of stay in a community \cite{wang_stay_2012,ma_write_2017,yang_commitment_2017}, perceptions of control over illness \cite{seckin_satisfaction_2013}, and decreased mortality~\cite{holt-lunstad_social_2010,uchino_social_2018}.
Caregivers also benefit from use of OHCs in terms of reduced stress, although evidence linking specific behaviors to outcomes for caregivers is more mixed than for patients \cite{hu_reducing_2015, kaltenbaugh_using_2015}.
Despite many benefits, making connections in OHCs can also have negative impacts on users, increasing their stress and leading them to leave the community~\cite{rains_coping_2018}. For patients, directly making comparisons with other patients can be distressing \cite{malik_computer-mediated_2008}, as can the sudden drop-out of key community members~\cite{attard_thematic_2012}.  
Furthermore, who an OHC user interacts with and the type of their interactions with others mediate both length of stay in the community and the benefits derived from using it~\cite{wang_stay_2012}.
These mixed and contextual outcomes make designing for the formation of new connections risky. We address this risk by examining the specific connections made between OHC users in order to identify user communication preferences.

We build on foundational OHC research from Bambina examining a health forum's network structure and its impact on the transmission of social support~\cite{bambina_online_2007}. Bambina analyzed a static snapshot of the posts in a cancer forum with 84 active participants, finding a core of highly supportive participants with a long tail of periphery members in the social network.
Our dataset enables research that addresses two key limitations of Bambina's work: (a) examining connections as a dynamic process rather than as fixed in a static network snapshot, and (b) including non-interacting ``lurkers'', which in our research is the population of CaringBridge authors who never interact with a fellow author.  With a complete and dynamic view of the interaction network, we are able to focus on the \textit{initiation} of new peer connections by authors.



Multiple factors are associated with new connection formation in OHCs.
New connection formation is often motivated by shared social identity \cite{sun_how_2019} and experience \cite{holbrey_qualitative_2013}.
Meng examined new connection formation in a weight management social networking site, finding substantial homophily effects related to health condition \cite{meng_your_2016}.  Centola and van de Rijt find similar strong homophily effects, noting that platform-specific traits---such as exercise preference in a fitness community---were less important in new health contact identification than demographic traits~\cite{centola_choosing_2015}.  In general, health-relevant traits contribute to the creation of specific connections~\cite{smith_social_2008}.
Outside of health, Seering et al. consider the immediate context (e.g. recent messages) that leads to first participation on Twitch.tv~\cite{seering_proximate_2020}; their examination of the factors that affect initiation with others broadly inspires our approach, although we focus less on immediate context and more on socio-cultural factors such as author role.

By studying initiations between users of an OHC without technical infrastructure for finding peers, we learn about the factors associated with the formation of supportive peer connections.










\section{Dataset \& CaringBridge Research Collaborative}  

This work was conducted during a research collaboration between CaringBridge (CB) and the University of Minnesota.
CB is a global, nonprofit social network dedicated to helping family and friends communicate with and support loved ones during a health journey.\footnote{Some of the text in this section is identical to that appearing in other works that result from this collaboration~\cite{ma_write_2017,levonian_bridging_2020,smith_i_2020}.}

CaringBridge.org offers individual \textit{sites} for users---free, personal, protected websites for patients and caregivers to share health updates and gather their community's support. 
Each site prominently features a \textit{journal}, which is a collection of timestamped, textual health \textit{updates} by or about a patient.  We use this terminology in concordance with previous CB research~\cite{ma_write_2017,levonian_bridging_2020}.
\textit{Authors} are CB users who have posted one or more updates on one or more sites.  
One author may publish updates on multiple sites, and multiple authors may publish updates on the same site.
Other registered users can comment on authors' sites but do not have sites of their own and are omitted from analysis, as we study peer interaction specifically and non-author users are likely not peers: they are primarily the friends, family, and acquaintances of site authors~\cite{smith_i_2020}.

To motivate our focus on CB specifically, we briefly discuss the affordances CB offers for connection with peer authors.
Rains argues that four primary affordances of communication technologies are most relevant for health-related social connection: visibility, availability, control, and reach \cite{rains_coping_2018}.  The design of CB offers reach---``potential to contact specific individuals, groups, or communities''---in only a limited way.
The search function on CB retrieves only sites with matching titles, which in most cases means a patient's full name is required to find a site. This barrier to social discovery means that achieving reach and the formation of a broader community requires additional effort.
Thus, we study peer connections in an environment where ``finding and interacting with peers facing the same health issue'' is not specifically supported \cite{rains_coping_2018}. CB authors' appropriative use of CB for peer connection provides an opportunity to understand the ways that peers connect with each other during their health journeys without the mediation of an explicit social discovery mechanism.

The CB dataset used in this work includes de-identified information from 535,481 authors and 588,210 sites created between June 1, 2005 and June 3, 2016, collectively containing 19M journal updates.  The data were acquired through collaboration with CB leadership in accordance with CB's Privacy Policy \& Terms of Use Agreement. This study was reviewed and deemed exempt from further IRB review by the University of Minnesota Institutional Review Board.
We opt not to publicly release the dataset used for analysis in this paper in order to protect participants' privacy~\cite{bruckman_cscw_2017}.  We welcome further inquiries about the dataset by contacting the authors.

\section{Methods: Dataset Preparation}  
\label{sec:background_methods}

\begin{figure}
\centering
\includegraphics{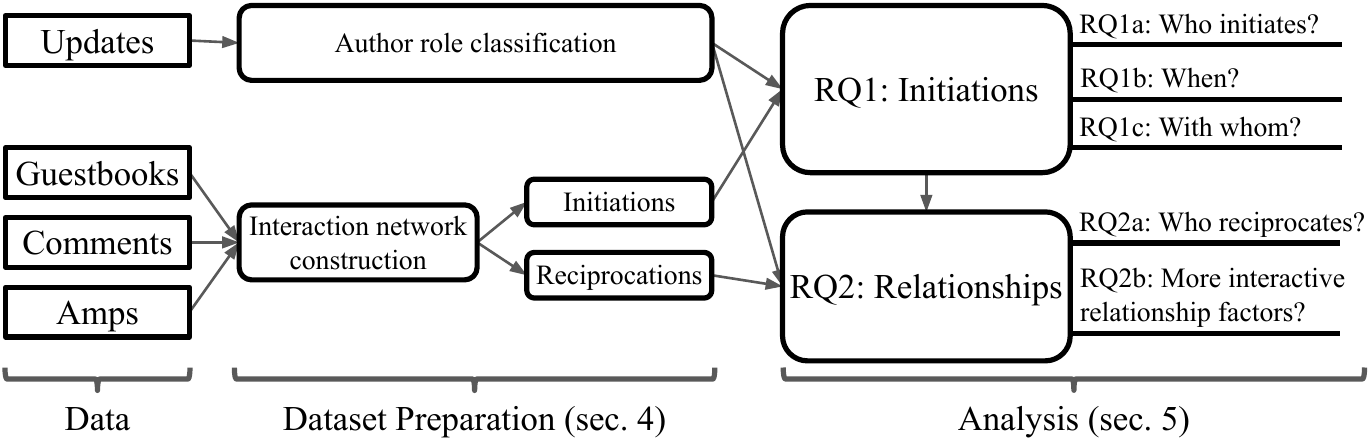}
\caption{Research dependency map. Background data preparation methods are necessary to address RQ1, which is built upon by RQ2.}
\label{fig:research_plan}
\end{figure}

To study peer connections and address our research questions, we first classified the role of individual CB authors using machine learning (sec. \ref{sec:background_authorship}) and constructed the network of interactions from the log data (sec. \ref{sec:background_network_structure}).  
Figure \ref{fig:research_plan} provides a high-level overview of the dependencies between these data preparation methods (sec. \ref{sec:background_methods}) and the analysis methods that address the research questions (sec. \ref{sec:analysis_methods}).
In constructing the network, we identified both initiations between authors and reciprocated dyads, which we used to study initiations (RQ1) and relationships (RQ2). 

We release our code on GitHub.\footnote{\url{https://github.com/levon003/cscw-caringbridge-interaction-network}}
Analysis code makes primary use of Python's scikit-learn \cite{scikit-learn}, NetworkX \cite{networkx}, StatsModels \cite{statsmodels}, NumPy \cite{van_der_walt_numpy_2011}, Pandas \cite{pandas}, and Matplotlib \cite{matplotlib} packages and R's mlogit \cite{mlogit_package} and stargazer \cite{stargazer_package} packages.

\subsection{Terminology}

For ease of reference, we list here the key terms we use in this paper:

\begin{itemize}
\item \textbf{Valid authors} are CB authors who meet basic requirements for being included in the study, such as publishing at least two journal updates over at least a 24-hour period.  Authors are the subset of registered CB users who have published at least one journal update on a CB site. (See section \ref{sec:background_authorship_valid})
\item \textbf{Author role} is the perspective from which an author account writes and publishes updates---either \textit{patient}, \textit{caregiver},  or \textit{mixed}. (See section \ref{sec:background_authorship_classification})
\item \textbf{Interactions} are one of three types of directed engagement (i.e. guestbooks, amps, and comments, introduced in Section \ref{sec:background_network_structure}) between an initiating author and a receiving author. Figure \ref{fig:caringbridge_interface} shows the interaction UI.
\item \textbf{Initiations} are the subset of interactions that compose the first interaction between an initiating author and a receiving author. A \textit{first initiation} is the first interaction an author makes on CB to any receiving author.
An \textit{initiating author} or initiator has made at least one initiation.
A \textit{non-initiating author} is an author who has made no initiations. 
A \textit{non-receiving author} is an author who has not been the target of any interaction.
\item \textbf{Reciprocations}, or reciprocal initiations, are the subset of initiations that reciprocate an earlier initiation from another author. In a dyad, the reciprocation is the initiation that comes second and closes the loop.
\item A \textbf{connection} exists between two different authors if at least one interaction has occurred between them.
\item \textbf{Relationships} are dyads with reciprocated initiations and the associated history of interactions between the two authors. The minimum number of interactions in a relationship is two: the original initiation and the reciprocal initiation. 
\item \textbf{Initiations period}---January 1st, 2014 to June 1st, 2016: the 2.5 year period of interactions used for RQ1. (See section \ref{sec:background_network_structure_types}.) RQ2 uses data from the full data collection period: January 1st, 2005 to June 1st, 2016.
\end{itemize}

\subsection{Authorship on CB}  
\label{sec:background_authorship}

In order to study inter-author interactions, we first select a sample of valid authors and then develop a machine learning classifier to identify author role (patient or caregiver) from their text updates.

\subsubsection{Valid author selection}\label{sec:background_authorship_valid}

Our dataset contains 535,481 total author users.  In this paper, we analyze 362,345 \textit{valid} authors 
with at least 2 updates published more than 24 hours apart.  
We exclude authors who have written any posts on sites determined to be spam by CB internal tooling ($-$10,776 authors), who we manually identified as either spammers ($-$17) or CB-internal test accounts ($-$9), or who only published updates within a 24-hour period ($-$162,334).  We selected the 24 hour threshold for author exclusion based on the distribution of author tenure (see Appendix \ref{app:sec:valid_author_identification}).


As one author may publish updates on multiple sites and one site may have updates published by multiple authors, identifying which authors interact with which other authors is challenging.
We use the term \textit{valid sites} to refer to the 340,414 sites (57.9\% of total) on which a valid author has written at least one update.
We identified 18,691 \textit{multi-site authors} (5.2\%) publishing on 2+ sites and 79,115 \textit{mixed-site authors} (21.8\%) publishing on at least one site on which other valid authors have (co-)authored updates. 
This estimate broadly aligns with estimates of multi-authorship from early studies of group blogs \cite{hearst_blogging_2009}.
As we discuss in the next section, the percentage of mixed-site authors is likely a conservative lower bound since author account sharing is common on CB.

\subsubsection{Author role classification}\label{sec:background_authorship_classification}


\begin{figure}
\centering
\begin{subfigure}[t]{0.5\textwidth}
  \includegraphics[width=\textwidth]{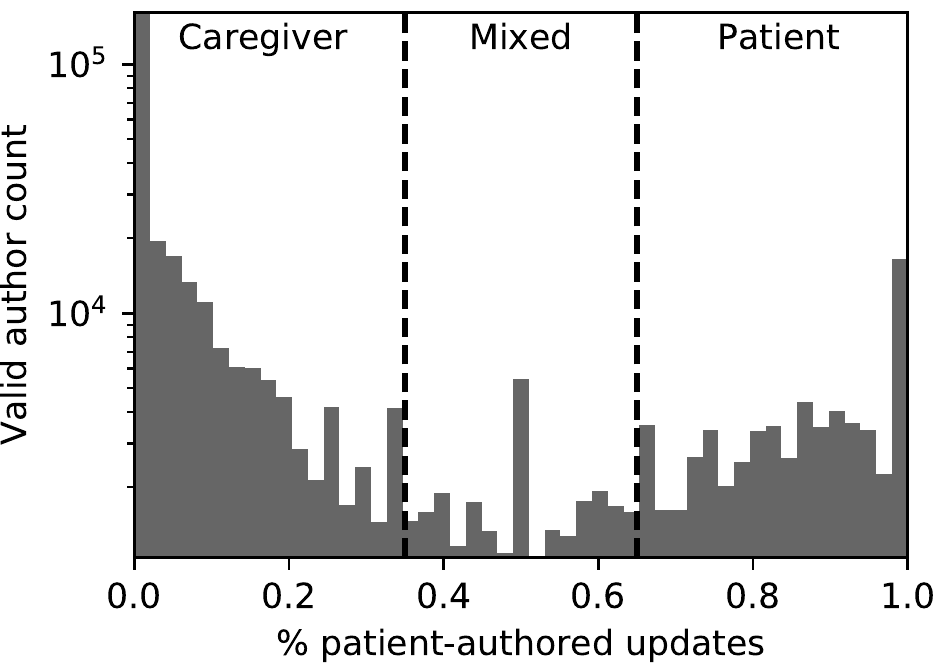}
\caption{Distribution of the proportion of each author's\\ updates that are classified as patient-authored.\\ Vertical lines indicate the thresholds used to assign\\ a role to each author.}
\label{fig:pct_patient_authored_distribution}
\end{subfigure}%
\begin{subfigure}[t]{0.5\textwidth}
\centering
  \includegraphics[height=2in]{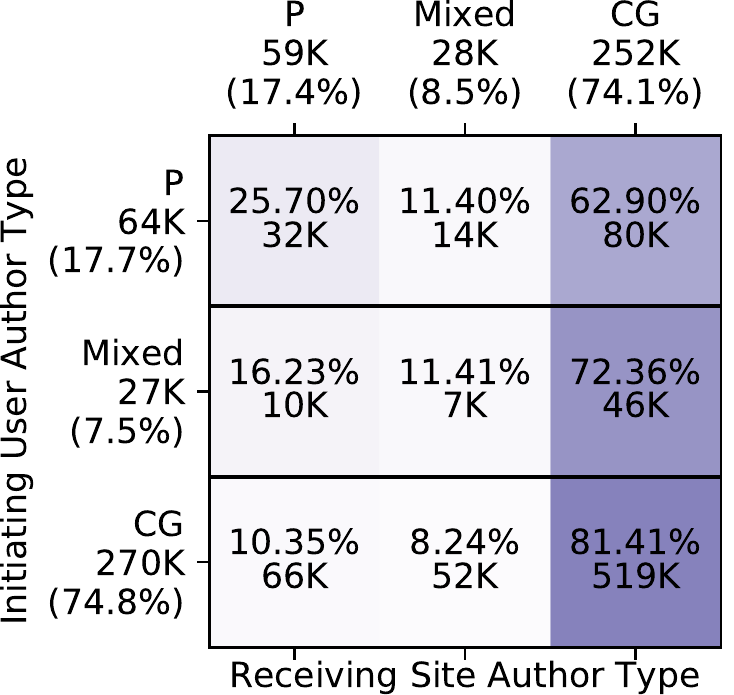}
\caption{Initiation counts broken down by the author role assigned to the initiating author and the receiving site.}
\label{fig:init_author_role_matrix}
\end{subfigure}
\caption{Classification of author role based on the text of an author's updates (a) enables a comparison of author initiations by assigned role (b).
}
\label{fig:author_role_assignment_summary}
\end{figure}

Authors on CB take the role of either \textit{patients} or \textit{caregivers}.\footnote{In figures and tables, we use the abbreviations P and CG respectively.}  However, authors can take on multiple roles or switch roles in three cases:
(1) multi-site authors may take a caregiver role on one site and a patient role on a second, 
(2) an author may use a single site for recounting two health journeys from the perspective of both a patient and a caregiver,
or (3) account sharing may result in updates published from the same author account but written by both a patient and one or more of that patient's caregivers.  
To tease apart these factors, we first classify authorship at the journal update level. 
We then classify an author's role as either Patient, Caregiver, or Mixed based on the classification of the updates published by that author.  \textit{Mixed} indicates one of the three observed cases above.

We trained a machine learning classifier to predict the author role of 15,850,052 updates that were authored on valid sites and contained at least 50 characters.
We combined human annotations of updates' author role from two previous CB studies \cite{levonian_bridging_2020,smith_i_2020} with additional annotations created while doing exploratory data analysis and using active sampling on earlier iterations of the classifier.  
Two of the authors independently annotated 429 updates, resulting in a Cohen's $\kappa$ of 0.829, which indicates sufficient reliability for this study~\cite{landis_measurement_1977}.  Combined with the annotations from previous studies, we had 6,932 human-annotated updates from 305 sites.\footnote{Note: 44 ground-truth updates (0.63\%) were assigned ambiguous or mixed labels that were reassigned to Caregiver for training, reflecting a predominant interest in patients: the binary classifier is trained to predict if updates are patient-authored or not.}

We used a linear SVM classifier on TFIDF-transformed unigram and bigram features, a common approach to binary text classification~\cite{wang_baselines_2012}.
As the training data are not identically distributed due to sampling differences across the three annotation efforts, we train with balanced classes by randomly downsampling the majority-class updates, an approach we found to outperform other training regimens (e.g. training with all annotated data).  We used hold-one-out cross validation to evaluate the performance of the model.\footnote{Since a shared site/author could leak information about the held-out data and give an overly optimistic view of classifier performance, we hold out at the site level rather than holding out individual updates.} 
Accuracy was 95.08\% and micro-averaged F1 score was 0.95.  
Patient-annotated updates ($n$ = 5,938, precision = 0.99, recall = 0.95, F1 = 0.97) were classified correctly at a greater rate than caregiver-annotated updates ($n$ = 994, precision = 0.77, recall = 0.93, F1 = 0.84).  

We categorized individual authors as either Patient, Caregiver, or Mixed.  
To assign a role to an individual author, we aggregated from the author role predictions of updates published by that author.  We used the same approach to categorize sites. Through manual investigation of 30 sites, we identified a variety of usage patterns, including a high frequency of sites with both patient- and caregiver-classified updates.  To assess the general patterns in author role and to allow for error introduced by the classifier, we use a consistent set of thresholds to define author role: \textit{Caregiver} sites/authors have less than one third of their updates classified as patient-authored, \textit{Mixed} sites/authors have between one third and two thirds of their updates classified as patient-authored, and \textit{Patient} sites/authors have more than two thirds of their updates classified as patient-authored.  
Figure \ref{fig:pct_patient_authored_distribution} show the distribution of the proportion of each author's updates that are classified as patient-authored, along with the thresholds.  The use of permissive thresholds to assign a role label captures the general perspective from which an author writes and keeps cascading classifier error to a minimum.
In the cross-validated ground truth data, 87.87\% of sites were classified at least two-thirds correctly. 
Thus, in the case of sites with all-Patient or all-Caregiver updates, the site-level error rate using these thresholds is at most 12.12\%, which we deem acceptable.

Applying the definitions above, we find 74.77\% of author accounts are classified as Caregiver, 17.74\% as Patient, and 7.49\% as Mixed. 
The distribution is similar for sites.
Mixed-author accounts may indicate either a single author embodying multiple roles or multiple people sharing the same account credentials.
96\% of Mixed-author accounts are shared by a patient and a caregiver (see Appendix \ref{app:sec:account_sharing}), which complicates analyses treating interactions between accounts as interactions between two people and suggests caution when interpreting Mixed-author results.
Overall, using the classifier predictions, we estimate that 22.06\% of updates are patient-authored. (See Appendix \ref{app:sec:p_update_proportion} for proportion estimation details.)

\subsection{Author interactions \& network structure}\label{sec:background_network_structure} 

\subsubsection{Inter-author interaction types \& analysis period}\label{sec:background_network_structure_types}

\begin{table}[]
\begin{tabular}{@{}lrrrr@{}}
\toprule
           & All interactions     & From valid authors & Non-self-interactions & Initiations\\ \midrule
Guestbooks & 82,980,359 & 5,864,304            & 5,212,720  & 654,192 \\
Amps       & 63,314,738 & 3,536,819            & 3,037,844  & 148,787 \\
Comments   & 31,052,715 & 1,094,435            & 881,781 & 111,623 \\ 
\textbf{Total} & \textbf{177,347,812} & \textbf{10,495,558} & \textbf{9,132,345} & \textbf{914,602} \\ \bottomrule
\end{tabular}%
\caption{Interaction counts broken down by type. This study considers only interactions from valid authors to valid sites. Self-interactions are interactions on sites where the interacting author has published an update and are excluded when building the network. Initiation counts are the number of non-self author/site pairs that were initiated by each type of interaction.}
\label{tab:int_type_counts}
\end{table}

\begin{figure}
\centering
\includegraphics[width=\textwidth]{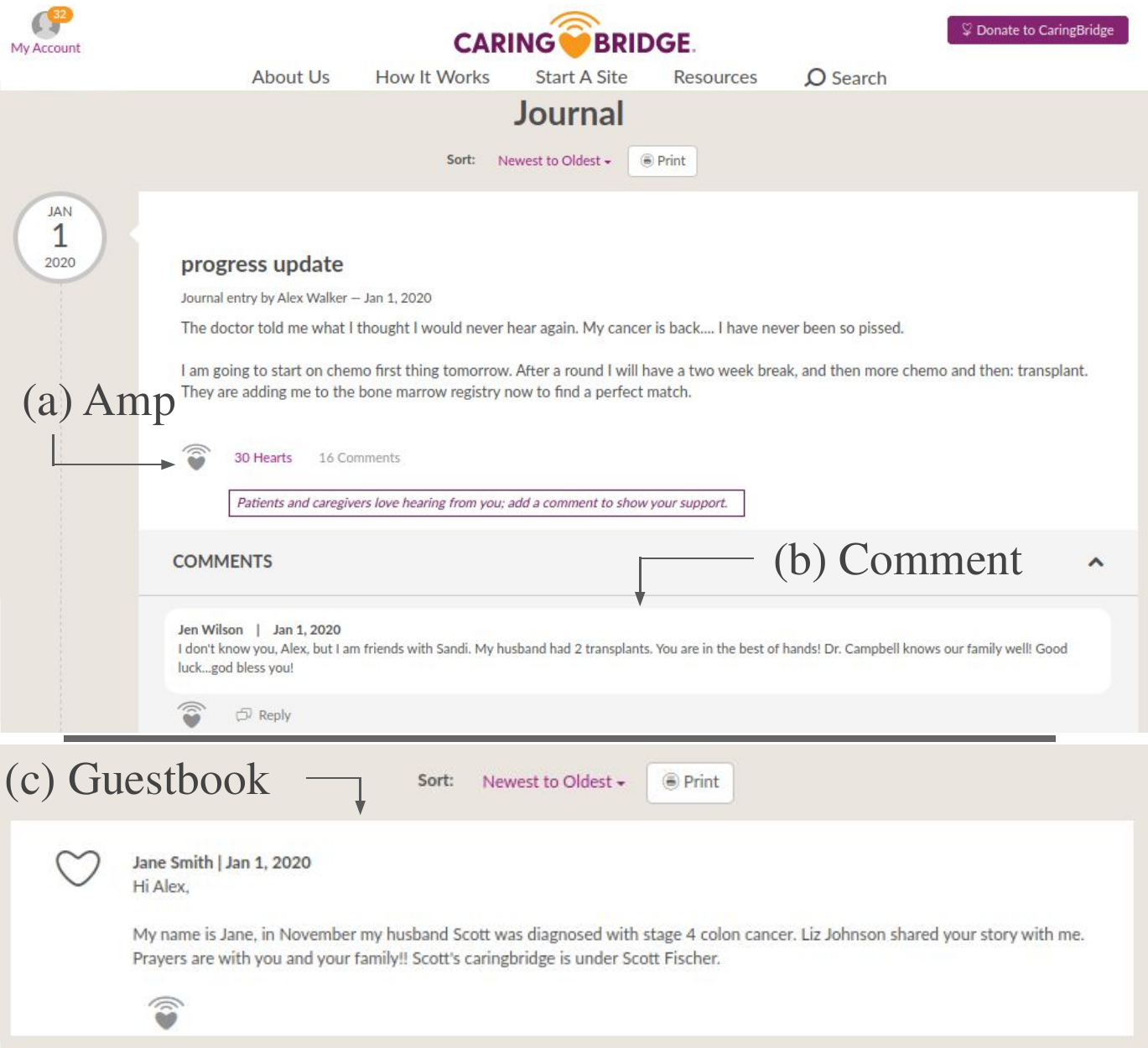}
\caption{CaringBridge interactions used in this study. Amps (a) and comments (b) are associated with a specific journal update, while guestbooks (c) are free-standing text posts left at the site level. Names and dates changed and texts paraphrased and anonymized~\cite{markham_fabrication_2012}.}
\label{fig:caringbridge_interface}
\end{figure}

To study the interactions between authors, we construct a network from the log data in the CB dataset.  
Direct messaging is not supported on CB; instead, all interactions are by an author on a site.
\textit{Guestbooks} are text posts left by a CB user on a site.\footnote{Guestbooks were renamed ``Well Wishes'' by CB, but we exclusively use the older name.}  \textit{Comments} are text posts left by a CB user on a specific journal update on a site.  \textit{Amps}\footnote{``Amps'' represent the idea of ``amplifying'' the love, hope, and compassion of the visitor.} are ``likes'' represented with a small heart icon and left by a CB user on a specific journal update on a site.  
The interface for these interactions is shown in Figure \ref{fig:caringbridge_interface}.
In this paper, we consider only interactions from valid authors to valid sites, with counts as shown in Table \ref{tab:int_type_counts}.  
Each interaction is associated with a unique identifier for the user and the site, as well as a timestamp.  Amps lack timestamp information, so we assume that amps occur at the publication time of the associated journal update. (We analyze this assumption in Appendix \ref{app:sec:amps_timestamp}.)
Other interactions are possible (see Appendix \ref{app:sec:other_interactions}), but we focus on guestbooks, comments, and amps as they are publicly visible, identifiable to a specific author/site pair, and result in a notification for the receiving author(s).

\begin{figure}
\centering
\includegraphics[width=\textwidth]{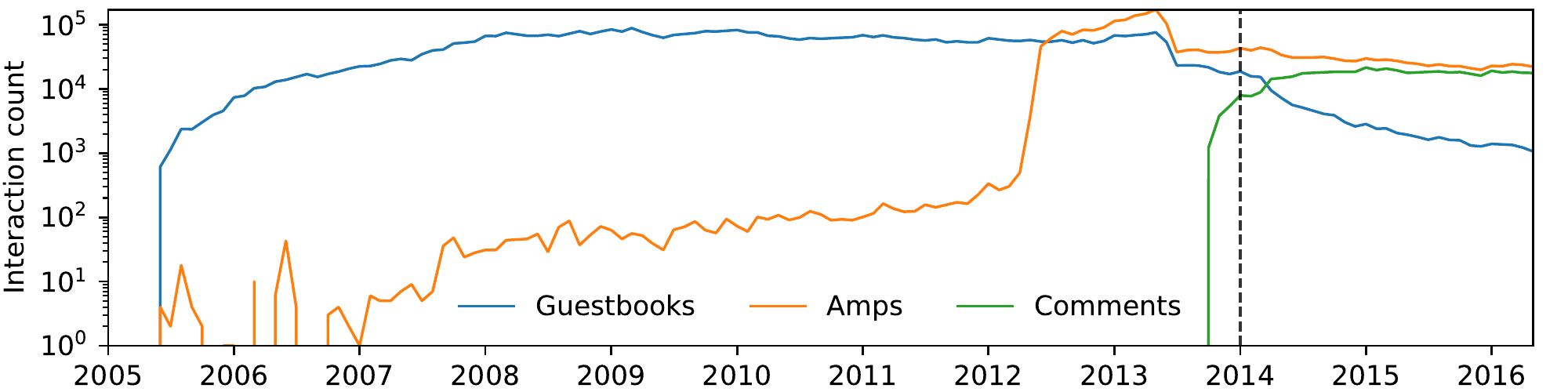}
\caption{Counts of each interaction type over time, on a log scale. The vertical dashed line indicates the beginning of the initiations period.}
\label{fig:interaction_counts_over_time}
\end{figure}

Figure \ref{fig:interaction_counts_over_time} shows the number of each interaction type on CB over time.  Note the introduction of amps and comments as features on CB.  
In order to avoid irregularities related to the introduction of new interaction types and to analyze a more established version of the network, all RQ1 analyses will focus on the state of the network from January 1, 2014 to June 1, 2016, which we refer to as the \textit{initiations period}.  
When models are fit to the initiations data, Jan 2014 -- Jan 2016 (80\% of the initiations period) is treated as training and inference data, and Jan 2016 -- Jun 2016 (the remaining 20\%) is treated as the test data and target for prediction. 
For RQ2 we use data from the full period (January 2005 to June 2016) because reciprocations and relationship interactions are susceptible to being right-censored i.e. interactions within the relationship occur after the end of the data collection period.

\subsubsection{Constructing the author interaction network}

We constructed a directed author interaction network in which initiations form edges between author nodes.
We analyzed this network, which includes reciprocal initiations, for RQ1.
We analyzed the interactions between reciprocated dyads for RQ2.

Interactions occur by an author on a \textit{site}. Therefore, to construct a directed one-mode network containing only author nodes requires assumptions about the intended target \textit{user} when an author interacts with a site. 
However, no assumptions are needed to construct a two-mode network \cite{latapy_basic_2008} containing both site and author nodes; simply make an edge between an author and a site if any interaction exists between that author and that site.  When constructing our network, we exclude interactions from authors to sites on which they have published any update as \textit{self-interactions} (12.99\% of all interactions). Table \ref{tab:int_type_counts} shows that the resulting network has 915K initiations that form edges.  
To convert this two-mode network into a one-mode network, we assume that each interaction links the interacting author to \textit{all authors} who have previously written an update on that site.  
In addition, we noted during data exploration that many guestbooks and comments are directed to both the caregiver author(s) of a site and the patient themselves, even if the patient had not yet published an update on the site for which they are the subject (or perhaps had not even yet created a CaringBridge account). 
Bloom et al. made a similar observation during their study of caregiver-authored CB sites. They found that support is not directed solely at the caregiver and instead is almost always directed to the patient in combination with the caregiver~\cite{bloom_patients_2017}.
Thus, we also draw an edge between the interacting author and any authors who publish a patient-classified update on the site at a later time. Such patient-specific edges are uncommon; only 4.9\% of all edges in our network are drawn to patient authors who had not yet published at the time of the interaction.\footnote{We also observe similar quantitative results when these patient-specific edges are not included for the three models (RQ1c, RQ2a, RQ2b) that are affected by this assumption.}

Our construction process resulted in 1,144,492 edges in the one-mode author interaction network.
Using the assumptions above, 9.1M author$\to$site interactions results in 14.8M author$\to$author interactions, of which 1.1M are initiations and thus form the edges within the interactions network.


\subsubsection{Interaction network structure}

\begin{table}[]
\begin{tabular}{@{}ll|ll@{}}
\toprule
Active Authors    & 66,440          & SCC Count        & 2,590          \\
Connected Authors & 55,655 (83.8\%) & WCC Count        & 2,335          \\
Isolates          & 10,785 (16.2\%) & Largest SCC Size & 16,946 (25.5\%)        \\
Max In-degree     & 612            & Largest WCC Size & 45,038 (67.8\%)        \\
Max Out-degree    & 409            & Largest SCC Diameter    & 38           \\ \bottomrule
\end{tabular}
\caption{Summary statistics for the interactions network at the end of the dataset (2016-06-01).}
\label{tab:background_network_summary}
\end{table}

We offer a brief description of the overall structure of the network formed by author interactions on CaringBridge in order to understand the context in which interactions are occurring.
Given that the network is directed, components of connected authors can be identified as \textit{strongly connected}---meaning a set of authors all reachable following the directed edges in the network---or \textit{weakly connected}---meaning a set of authors all reachable following edges in any direction in the network.  The simplest strongly-connected component (SCC) is two authors who have interacted with each other. The simplest weakly-connected component (WCC) is two authors where one author has interacted with the second, but that interaction has not been reciprocated.

Table \ref{tab:background_network_summary} provides summary statistics for the network at the end of the data collection period. These statistics summarize the subgraph consisting of \textit{active} authors only---the subset of valid authors that were active on CB within 6 months of the end of the data collection period---in order to capture only the recent connections.\footnote{Statistics are similar for the full network without inactive authors removed.}
At the end of the data collection period, 30.4\% of active authors are in one or more of 21,910 total reciprocated dyads.
The network is dominated by a single large WCC, in which a large SCC is embedded.  This pattern is consistent with the structure of other online health groups~\cite{rains_coping_2018,introne_sociotechnical_2016}.
We observe a lack of large isolated subnetworks, echoing findings in Urbanoski et al.~\cite{urbanoski_investigating_2017}; the second-largest SCC and WCC contain only 14 and 18 authors respectively.\footnote{The full distribution of the connected components is shown in Appendix Figure \ref{fig:background_cc_distribution_plot}.}
The vast majority of initiations (new edges) either grow or occur within the largest weakly-connected component (see further analysis in Appendix \ref{app:sec:initiation_type_classification}).
The number of active authors is generally decreasing during the initiations period (from 79.7K to 66.4K), as is the proportion of active authors in the largest connected component.  See Appendix Figure \ref{fig:background_network_summary_timeline} for a temporal view of connectivity within the network.
Author indegree and outdegree are positively correlated (r$=$0.468, p $<$0.001), consistent with prior work suggesting that online support-giving is highly reciprocal~\cite{pan_you_2017}.

\section{Methods: Analysis} \label{sec:analysis_methods}

Having classified authors by role and built the author interaction network, we now present the methods used to address our research questions.  
To address RQ1, we isolate for analysis the initiations that form the edges in the interactions network.
Because initiations between authors on CaringBridge are unexpected, we need to understand ``what initiations look like''. 
To determine if and how connected authors know each other, we conducted a content analysis of comment and guestbook initiations  (sec. \ref{sec:init_background_methods}). 
Next, to address the three subquestions of RQ1, we fit three different quantitative models:
\begin{enumerate}
\item To identify the factors associated with which authors do any initiation, we used logistic regression to predict whether an author initiates or not (sec. \ref{sec:init_who_methods}).
\item To identify the factors associated with when authors initiate relative to their first published update, we used linear regression to predict the amount of time between an author's first update and their first initiation (sec. \ref{sec:init_when_methods}).
\item To identify the factors associated with whom authors choose to initiate, we used conditional multinomial logistic regression to predict the target of each initiation (sec. \ref{sec:init_withwhom_methods}).
\end{enumerate}
The factors we consider during modeling are represented in four types of features: network, author role, activity, and health condition.  We motivate these features (sec. \ref{sec:init_features}) prior to introducing the models.

To address RQ2, we isolated for analysis the reciprocated dyads within the network. 
To address the two subquestions of RQ2, we fit three quantitative models:
\begin{enumerate}
\item To identify the factors associated with which authors reciprocate initiations, we used logistic regression to predict if an initiation will be reciprocated (sec. \ref{sec:reciprocation_methods}). 
\item To identify the factors that result in more interactive relationships, we used negative binomial regression to predict the total number of interactions in a relationship.  Additionally, we used logistic regression to predict if a relationship is balanced or not (sec. \ref{sec:relationship_methods}).
\end{enumerate}



\subsection{RQ1 Methods: Initiations within the network} \label{sec:init_background_methods}

We conducted a content analysis to characterize the relationship between the initiator and the receiver.
We randomly sampled 400 comment initiations and 400 guestbook initiations made by valid authors in the initiations period. Two annotators independently coded the 800 initiations.
Annotators used the text of the initiation---and the text of the associated journal update in the case of comment initiations---to identify two aspects of the initiation: (1) whether the initiator's tie with the receiver existed before the health event that resulted in the creation of the CB site, and (2) the relation between the initiator and the receiver.
Aspect (1) utilized a closed code set to identify the initiator-receiver tie as pre-health-event, post-health-event, or unknown. Aspect (2) was coded in an open manner, allowing for any relationship descriptor that could be identified from the context of the text  e.g. friend, fellow patient, one-time site visitor, etc.
The two annotators met to discuss and resolve disagreements.  
Reliability was established through these disagreement discussion meetings~\cite{mcdonald_reliability_2019}.
We direct readers to Appendix~\ref{app:sec:initiation_annotation} for additional details regarding the annotation process.
By annotating these two aspects of initiations, the content analysis contextualizes the RQ1 quantitative results: it characterizes the relationship between interacting authors on CB and surfaces genuine first-time interactions of the type that digital interventions may try to facilitate.

\subsection{RQ1 Methods: Features for modeling}\label{sec:init_features}\label{sec:health_condition_explanation}

In identifying factors that are associated with the initiation of new connections by authors, we consider four sets of features that prior work suggests are associated with the initiation of connections.
Here, we introduce the four feature sets as motivated by prior work and discuss operationalization in the CB context at a high level.  
The specific features used in the models are introduced later in the relevant model description.

\begin{itemize}
\item 
\textbf{Network features}---Interaction between users is affected by the network context of the initiator and the receiver \cite{rains_coping_2018,berkman_social_2014}.  Since initiations between authors on CB are unexpected,  we explore the impact of CB network context on predicting who initiates with whom.  We include network features that capture the current position of the initiator and the receiver. For example, ``triadic closure'' is a well-known network phenomena in which two previously unconnected people with a mutual contact are likely to connect \cite{overgoor_choosing_2019}; on CB, we can explore triadic closure using a binary feature that indicates if two not-yet-connected authors share a mutual connection. Simpler features describe if an author has ever been interacted with or if an author has ever initiated with others.
\item 
\textbf{Author role features}---Structural health role may affect user interaction online.  For example, Hartzler et al. found that role (as patient, survivor, or caregiver) was important in finding peer mentors~\cite{hartzler_leveraging_2016}.  We include role as a categorical variable (Patient, Caregiver, or Mixed, see sec. \ref{sec:background_authorship_classification}) to quantify the importance of health role to connection on CB. Where appropriate, we also include features for site author configuration e.g. binary indicators for mixed-site authors, as mixed-site authorship may suggest multiple intended recipients of an interaction.
\item 
\textbf{Activity features}---An author's level of activity is associated with their engagement with others.
In general, receiving replies online is associated with retention~\cite{sharma_engagement_2020}.  On CB specifically, posting frequency is correlated with receiving comments~\cite{ma_write_2017}.  Therefore, an author's update frequency is a key confounder for understanding initiations behavior; highly active authors may have different connection patterns than less active authors.
\item 
\textbf{Health condition features}---Health condition and health status are important predictors in the formation of new connections in OHCs~\cite{meng_your_2016,smith_social_2008}.  On CB, we operationalize health condition from author-reported data.
When CB authors create a site, they have the option to self-report a broad health condition category such as Cancer or Injury.  While only 59.1\% of valid sites self-report a health condition and such single-label categorization may be overly reductive \cite{li_condition_2018}, the sites that do report a condition can inform us about differences in inter-author communication behavior based on the health condition under question.  This data enables us to confirm expected patterns in increased interaction between authors who have the same health condition~\cite{thoits_stress_2010,rains_social_2016} and compare the importance of health condition to other factors.
We assign health conditions to authors based on the reported health condition of the sites they've authored or `None' when no condition is reported. 
For the 446 (3.1\%) authors with multiple sites that report different health conditions, we assign the first non-None/`Condition Unknown' health condition reported.
Ten total health condition categories were assigned, with counts shown in Appendix Table \ref{app:tab:health_condition_category_counts}.
When used as a feature, health condition is included as a ten-level categorical variable and abbreviated ``HC''.

\end{itemize}

As this work is exploratory, we had no specific hypotheses to test; instead, we fit the most parsimonious model that still enabled us to explore the relative importance of the four factors described above.

\subsection{RQ1a Methods: Who? Initiating authors}\label{sec:init_who_methods}

Which authors initiate during their time on CB, and which do not? To understand which factors predict initiation, we fit a logistic regression model predicting whether an author has made any initiation.  To avoid bias in the model from right-censoring---some authors will initiate but only after the end of our data collection period---we predict an outcome of initiating within 1 year of first authorship.  Similarly, we include a feature describing if an author was interacted with by another author in the first year.  Only 11.7\% of authors make their first initiation more than 1 year from their first update.  Only 2.4\% of authors are first interacted with more than 1 year from their first update.  Models with non-bounded outcomes and predictors produced similar results.

\subsubsection{Features}
Author role was included as a categorical variable. Binary indicators for the mixed-site and multi-site author designations were included as potential confounders.  We included update count---the total number of updates published by that author on CB---as well as update frequency---the ratio of that count to the author's tenure in months---as indicators of author activity level. 
We included health condition as a categorical variable.

\subsection{RQ1b Methods: When? Initiation timing}\label{sec:init_when_methods}

\begin{figure}
\centering
\begin{subfigure}[t]{0.5\textwidth}
  \includegraphics[width=\textwidth]{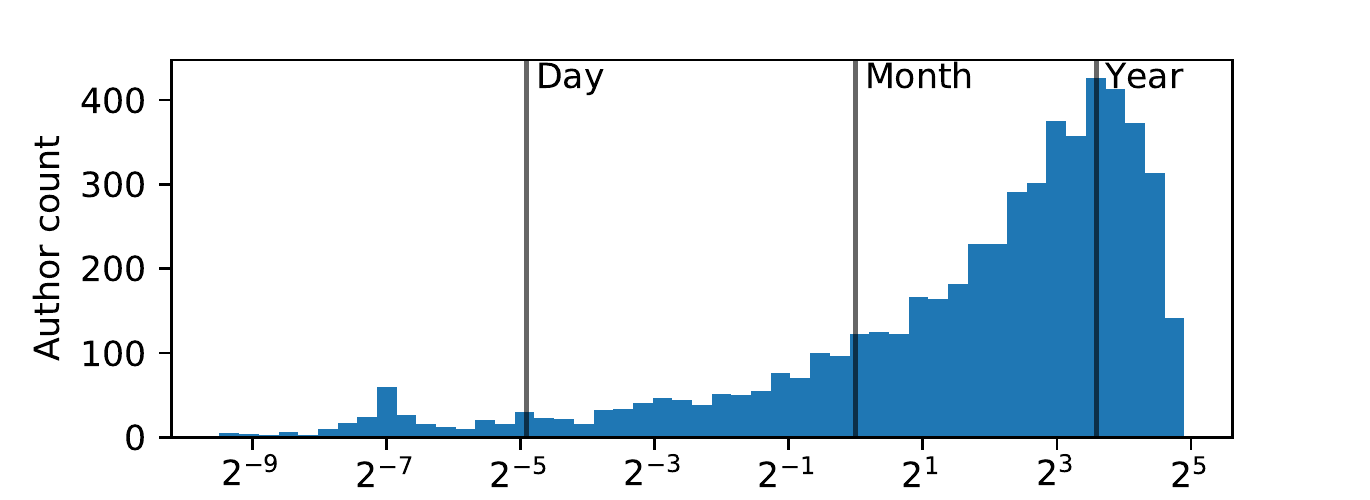}
\caption{Pre-authorship initiators ($n$=5,439): Time \\between first initiation and first authored update}
\label{fig:init_timing_pre_response}
\end{subfigure}%
\begin{subfigure}[t]{0.5\textwidth}
  \includegraphics[width=\textwidth]{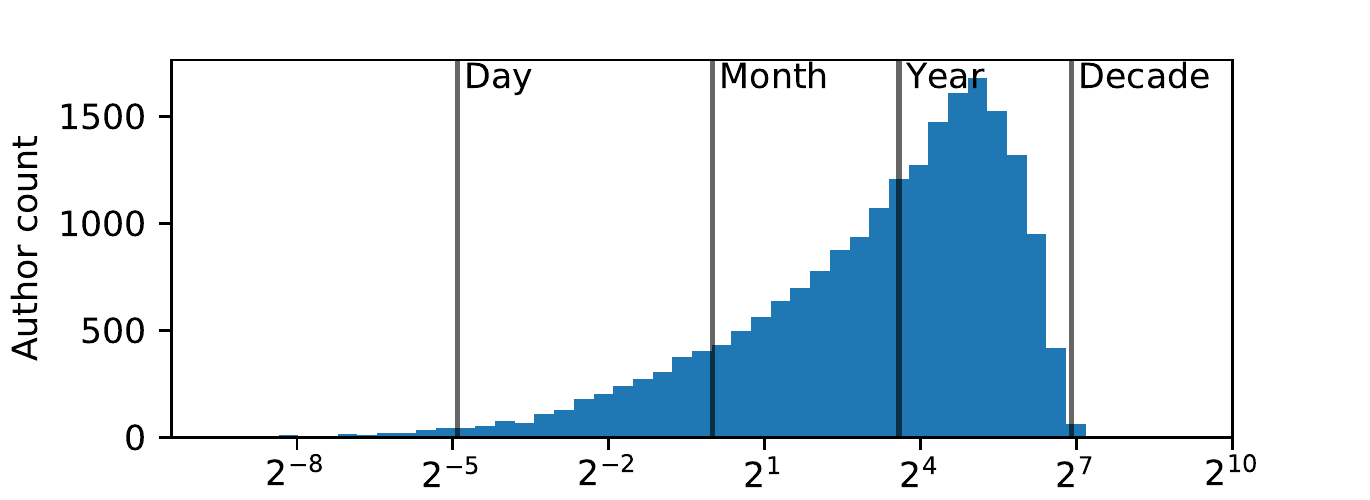}
\caption{Post-authorship initiators ($n$=20,687): Time \\between first authored update and first initiation}
\label{fig:init_timing_post_response}
\end{subfigure}
\caption{Distribution of time between first authorship and first initiation. 
For pre-authorship initiators, the median time to authorship is 5.8 months (mean 7.8mos).
For post-authorship initiators, the median time to initiation is 13.6 months (mean 22.3mos).
}
\label{fig:init_timing_response_plot}
\end{figure}

Given that an author is going to initiate, when is their first initiation likely to occur?  
We aim to understand the lifecycle of authors on CB by modeling when authors transition to peer-seeking behavior relative to their activities as authors.  
Thus, we differentiate between \textit{pre-authorship initiators} who first interact with another author before publishing their first journal update and \textit{post-authorship initiators} who first interact after publishing their first journal update.
20.82\% ($n$=5,439) of valid authors are pre-authorship initiators, with the remaining being post-authorship initiators.
We treat the pre-authorship initiator and post-authorship initiator cases separately, fitting linear regression models to predict the number of months between first authorship and initiation.  
Figure \ref{fig:init_timing_response_plot} shows the distribution of this interval. 

\subsubsection{Features}
To understand the relationship of the time between first initiation and first published update to the total time spent on CB, we compute ``total active time''---the number of months between an author's first published update or interaction and their last recorded update or interaction.  We include total active time in order to develop an understanding of initiator life-cycle and capture the relationship of initiation to the update-writing activities of authors.  
For post-authorship initiators, we add a binary feature ``Is interacted with?''--1 if that author was interacted with by any author pre-initiation and 0 otherwise.  Pre-authorship initiators cannot be interacted with by definition, as they lack sites to interact with until they become authors.

\subsection{RQ1c Methods: With whom? Initiation target}\label{sec:init_withwhom_methods}

We now turn to the question of whom an author initiates with given that they are initiating with someone.
We model initiations between authors as discrete choices to add a new directed edge to the graph, following Overgoor et al. \cite{overgoor_choosing_2019}.
In this paradigm, we fit a model to compute the conditional probability of a particular initiating author choosing the targeted receiving author---as opposed to all other authors who have sites on CB---given that the initiator is making a new initiation at this particular moment in the lifecycle of CB.  
Fitting a model to estimate this probability enables us to evaluate the relative importance of author traits such as health condition and role on the choice of a new connection target.

\subsubsection{Conditional multinomial logit models}

We used the Conditional Multinomial Logit Regression model, or conditional mlogit model, to estimate the probability of an author initiating with another author.
Specifically, the conditional mlogit model estimates the probability $P_{i,t}(j, C)$ of author $i$ initiating with author $j$ from among the set of candidates $C$ at time $t$.  
Given features $x_t$ for each author, we learn coefficients $\theta$ such that $P_{i,t}(j, C) = \exp( \theta^T x_{j,t}) / \big( \sum_{\ell \in C} \exp( \theta^T x_{\ell,t}) \big)$ \cite{overgoor_choosing_2019}.
As it is not computationally feasible to compare the initiation target against all other possible authors (i.e. $C=$ ``all authors''), one can employ negative sampling to select a subset of \textit{candidate} authors that were \textit{not} initiated with as a comparison group, without biasing the coefficient estimates \cite{overgoor_choosing_2019}.  We sample 24 candidate authors from the set of all valid authors with sites at the time of the initiation who have not previously been initiated with by the initiating author.\footnote{Negative candidates are sampled from the state of the network at the time of the initiation and so authors who have not yet posted an update cannot be negative candidates.} 
Thus, for each initiation, the model selects the true target from one of 25 candidate receiving authors. 
A model performing no better than random would achieve only 4\% accuracy at identifying the correct target. 

\subsubsection{Features}

For each initiation, features are computed for the pairs formed by the initiating author and each of the 25 candidate authors (which includes the target author).
We use features from all four sets:

\begin{itemize}
    \item \textbf{Network features}---Following \cite{overgoor_choosing_2019}, we include (1) a binary feature for any non-zero in-degree, as well as the log of the (2) outdegree and (3) indegree of each candidate, using a censored log that returns 0 when degree is zero.  We include additional binary features for (4) reciprocation, which is 1 if the candidate has previously interacted with the initiator, (5) weak connection (recommended in \cite{yan_network_2015}), which is 1 if the candidate is already in the same weakly-connected component as the initiator, and (6) friend-of-friend, which is 1 if the initiator is connected to a neighboring author that is already connected to the candidate (i.e. a feature for triadic closure \cite{granovetter_strength_1973}).  All network features are computed from the state of the network at the time of the initiation.
    \item \textbf{Author role features}---Includes (1) author role of the candidate, (2) a binary feature that is 1 if author role is the same between the initiator and the candidate, and binary features for if the candidate is (3) a multi-site author or (4) a mixed-site author.
    \item \textbf{Activity features}---Includes (1) count of updates made by candidate at the time of the initiation, (2) frequency of updates i.e. update count divided by author tenure in months, (3) number of days since the candidate's most recent published update prior to the initiation, and (4) number of days since the candidate's first published update.
    \item \textbf{Health condition features}---Includes only (1) a binary feature that is 1 if the initiator and the candidate author are assigned the same non-None health condition. (See section \ref{sec:health_condition_explanation}.)
\end{itemize}

We fit and present results for a full model containing all feature sets as well as models with each feature set independently. We checked inputs for colinearity, finding no two features were highly correlated.\footnote{The greatest correlation is $r=$0.44 between total update count and days since first published update.}
While the feature sets we use contain no demographic information, prior work suggests that demographic homophily plays only a small role in OHCs \cite{yan_network_2015}. However, we do fit a model using a proxy of author geography, discussed next.

\subsubsection{Geographic analysis}\label{sec:geo_analysis_methods}

One potentially important factor in predicting initiations on CB is the geographic proximity of CB authors.  Geographic proximity may indicate existing offline social relationships or post-diagnosis connections made in-person rather than on CaringBridge.  While a fine-grained investigation of geographic effects on the relationship between the CB interactions network and the social networks of P/CG is out-of-scope for this work given the available data, we generate a rough proxy for geographic proximity by assigning US states to authors based on the IP addresses of their journal updates and guestbooks.\footnote{Our data do not capture IP addresses for amps or comments.} To evaluate the impact of geography proximity on initiations, we fit a separate full mlogit model including this proxy as a feature.

We use the Maxmind GeoLite2 City database (from Aug 13, 2019) to do IP geolocation lookups.\footnote{\url{https://dev.maxmind.com/geoip/geoip2/geolite2/}} We refer to journal updates and guestbooks that are assigned identifiable geographic coordinates as \emph{geo-identifiable posts}. 
93\% of CB authors' geo-identifiable posts are entirely based in the United States.  Among these authors, we attempt to assign US states as our proxy for geographic proximity.  We avoid the direct use of latitude/longitude estimates to reduce the bias introduced through the use of IP geolocation \cite{poese_ip_2011}.

A US state is assigned to an author if that author has at least 10 geo-identifiable posts and among those posts the most-frequently-occurring state holds a plurality with at least a 20\% margin above the second most frequent state, with the intent of creating a high-precision, low-recall state assignment.  
We fit a conditional mlogit model that includes a dummy variable for same state assignment---1 when the initiating author and the candidate author have the same state assignment, and 0 otherwise---in order to assess the importance of geographic co-location.



\subsection{RQ2a Methods: Reciprocation}\label{sec:reciprocation_methods}

RQ2 addresses relationships---reciprocated dyads of valid authors.
Such relationships start with an initiation between an initiating author and a receiving author, followed by a reciprocal initiation from the receiving author to the initiating author after some period of time.
To understand which non-reciprocal initiations will result in a reciprocal initiation, we fit a logistic model to predict if the receiving author will reciprocate.
As with RQ1a (sec. \ref{sec:init_who_methods}), to account for potential right-censoring of reciprocations i.e. reciprocations that occurred after the end of the data collection period, we predict if a reciprocation will occur \textit{within one year} of the original initiation and train only on author pairs with an originating initiation occurring no later than one year before the end of the dataset.\footnote{74.8\% of reciprocations occur within one year.}
Thus, to broaden the scope of the reciprocations and relationships considered, we use initiations from the full eligible range (Jan 2015 to June 2015). 
Across all pairs of authors with at least one directed initiation between them in this time period, 12\% are reciprocated.

To understand which initiations will result in a reciprocal initiation,
we fit a logistic regression model predicting if an author pair with one initiation will be reciprocated within one year.  As features, we utilize initiator author role and receiver author role, including a full interaction term, in order to tease apart the impact of author role on reciprocation. In addition, we include the number of months (log transformed) between the receiver's first published update and the original initiation in order to understand when authors are most likely to reciprocate a connection in their time on CB. We include the same duration (log transformed) for the initiator, although as the initiator may not yet be a published author themselves we include a binary indicator variable that is 1 if the initiator published their first update before the initiation.

\subsection{RQ2b Methods: Relationships}\label{sec:relationship_methods}

A relationship is any reciprocated dyad between valid authors \textit{and} their associated history of interactions. 
Our analysis includes dyadic relationships from the full timeline in order to compensate for bias introduced by right censoring, as some interactions in a relationship will occur after the end of the data collection period (see Appendix \ref{app:sec:right_censoring} for additional analysis).
We fit two quantitative models in order to understand what factors---especially author role---are associated with more interactive and more balanced relationships.

\subsubsection{Number of interactions}

To explore the impact of author role on total number of interactions in a relationship, we fit a Negative Binomial regression model to estimate this count.
As the data are counts and significantly over-dispersed, a negative binomial model is appropriate; empirically, we observe a better fit with a negative binomial model than with Poisson or log-linear regression.
To improve the fit and reduce sensitivity to outliers, only relationships with fewer than 345 interactions (99th percentile) were included in the training data.\footnote{Results were similar excluding outliers at the 98th and 99.9th percentile.
The max number of interactions in one relationship was 18,340.
}
We included an interaction term between initiator author role and reciprocator author role to tease apart the impact of role for both authors.
We also included a feature for the duration of the relationship in months, which is measured from the first initiation to the last observed interaction within that relationship. 
Finally, to control for the level of balance in the relationship, we include a binary feature ``Is balanced?'' using the definition of balance introduced in the next section.

\subsubsection{Relationship balance}

Balanced relationships are potentially desirable because giving and receiving support are actions that reinforce each other but play distinct roles in realizing positive effects~\cite{kim_process_2012}. 
We wanted to contrast one-sided relationships to relationships where support is mutually exchanged.
We operationalize relationship balance as the percentage of interactions in a relationship made by the original initiator vs the reciprocator.
For all relationships, we classify a relationship as balanced if no author made more than 75\% of the interactions in the relationship. 
We fit a binary logistic regression model to predict if a relationship is balanced or not.
As in the interaction counts model, we include features for the relationship duration and a full interaction between the initiator and receiver author role.  We include the total number of interactions as an additional control.

\section{Results}\label{sec:results}

Results follow the same structure as the methods in Section \ref{sec:analysis_methods}.
We present the results for RQ1 in sections \ref{sec:init_background_results}-\ref{sec:init_with_whom_results}, followed by the results for RQ2 in sections \ref{sec:recip_results} and \ref{sec:relationship_results}.

\subsection{RQ1 Results: Initiations within the network}\label{sec:init_background_results}




We annotated 800 initiations to identify (1) whether the initiator's relationship with the receiver existed before the health event that led to the creation of the CB site and (2) the relation between the initiator and the receiver.  
We provide quotes from comments (marked `C') and guestbooks (marked `G') to illustrate relation categories, paraphrased and renamed to preserve anonymity and reduce traceability \cite{markham_fabrication_2012,bruckman_studying_2002}.
The majority of initiations are unidentifiable in both respects e.g. \textit{``I am praying for you! May the love and support of family and friends be of comfort to you.''} (G).
57.4\% ($n$=459) of initiations were coded with an unknown timing relative to the health event, while 63.9\% ($n$=511) were coded with an unknown relation.

The non-unknown initiations provide insight about initiators' goals and relationships to the receiver.  28.6\% ($n$=229) of initiations were coded as pre-health-event, while only 12.1\% ($n$=97) were coded as post-health-event.   
For each, we list the most common relations in order to demonstrate what pre-health-event and post-health-event initiations look like on CB.
The most common pre-health-event relations were:
\begin{itemize}
\item Friend ($n$=118) e.g. \textit{``My heart is heavy from hearing the news. .... Remember to search for answers and ask questions so that you understand everything.  Hugs to all the family.''} (C).
\item Family ($n$=20) e.g. \textit{``Hi Uncle Steve, A good day for scans, I will remember to sigh an extra prayer! Think of u often! Love, Anna''} (C).  
\end{itemize}
The most common post-health-event relations were:
\begin{itemize}
\item Third-party connections ($n$=35) e.g. \textit{``Hello, I came across your site because a mutual friend commented on facebook.  I am sorry you are going through this.  I am battling breast cancer (and also a mom of 6) and I wish you well.''} (C).
\item CG of similar patient ($n$=24) e.g. \textit{``"Hi Johnson family, we met at Parents of Preemies day. I loved reading the stories of strong little Timmy and especially the last update that he is home! Congratulations!''} (G).
\item One-time visitor ($n$=15) e.g. \textit{``My heart aches with your familiar story. I've never met you, but you and my husband now share a similar story. My husband is a STAGE IV PROSTATE CANCER SURVIVOR. You can do this. We are here for you guys and just like everyone else- we want to help. We are praying hard that you get some answers that give you HOPE. ''} (C). 
\end{itemize}
Some post-health-event initiations suggest that the initiation is a genuine ``first contact'' between these two authors, e.g. \textit{``I stumbled across a link to Terry's CaringBridge page and read through your loving entries.  I took care of my mother's CaringBridge page.  A labor of love and a nice way to keep loved ones informed.''} (G).  Others include explicit links to their CG sites: \textit{``Hi, my name is Kaylee and i came across your page! i look forward to following your story. My website is: (CB site link)''} (G). 
More detailed results, including the counts of each type of relation identified and additional quotes, are presented in Appendix \ref{app:sec:initiation_annotation}.

This content analysis suggests that \textbf{a small but important percentage of peer connections are formed between authors post-health-event, although a larger percentage are from existing connections re-established on CB as a result of the health event}. 
We now have an idea of what these initiations look like as we move into the quantitative modeling.
This content analysis also surfaces several interesting phenomena that are not addressed by our quantitative work. How were post-health-event one-time visitors getting links to the receiving CB site? What is the role of relation in the (re)forming of mutually supportive relationships?
These questions could be explored in future qualitative work.




\subsection{RQ1a Results: Who initiates?}\label{sec:init_who_results}

\begin{table}
\begin{center}
\begin{tabular}{lrrr}
      &  (1) Full  & (2) Int Received &    (3) No Int Received \\
\hline
Intercept & -0.323$^{*}$ &  0.242$^{*}$ & -0.845$^{*}$ \\
 & (0.017) & (0.025) & (0.026) \\
Role = Mixed & -0.104$^{*}$ &  0.048$^{***}$ & -0.311$^{*}$ \\
 & (0.038) & (0.052) & (0.064) \\
Role = P & -0.024$^{***}$ &  0.189$^{*}$ & -0.357$^{*}$ \\
 & (0.024) & (0.032) & (0.040) \\
Update count &  0.005$^{*}$ &  0.004$^{*}$ &  0.001$^{*}$ \\
 & (0.000) & (0.000) & (0.000) \\
Update frequency (updates/month) & -0.011$^{*}$ & -0.010$^{*}$ & -0.006$^{*}$ \\
 & (0.001) & (0.001) & (0.001) \\
Is a mixed-site author? &  0.074$^{*}$ & -0.210$^{*}$ &  0.014$^{***}$ \\
 & (0.019) & (0.025) & (0.033) \\
Is a multi-site author? &  0.012$^{***}$ &  0.275$^{*}$ &  0.012$^{***}$ \\
 & (0.042) & (0.062) & (0.063) \\
HC = Cancer &  0.104$^{*}$ &  0.098$^{*}$ &  0.108$^{*}$ \\
 & (0.021) & (0.029) & (0.036) \\
HC = Cardiovascular/Stroke &  0.048$^{***}$ &  0.071$^{***}$ &  0.040$^{***}$ \\
 & (0.048) & (0.065) & (0.077) \\
HC = Congenital/Immune Disorder &  0.087$^{***}$ &  0.135$^{***}$ & -0.067$^{***}$ \\
 & (0.112) & (0.146) & (0.200) \\
HC = Infant/Childbirth & -0.105$^{***}$ & -0.026$^{***}$ & -0.455$^{*}$ \\
 & (0.074) & (0.095) & (0.141) \\
HC = Injury & -0.116$^{*}$ & -0.221$^{*}$ & -0.056$^{***}$ \\
 & (0.056) & (0.071) & (0.097) \\
HC = Neurological Condition &  0.261$^{*}$ &  0.174$^{*}$ &  0.466$^{*}$ \\
 & (0.062) & (0.086) & (0.093) \\
HC = Other &  0.074$^{***}$ &  0.042$^{***}$ &  0.190$^{***}$ \\
 & (0.078) & (0.108) & (0.118) \\
HC = Surgery/Transplantation &  0.030$^{***}$ &  0.110$^{***}$ & -0.012$^{***}$ \\
 & (0.055) & (0.076) & (0.088) \\
 \hline
 Observations & 53,335 & 28,870 & 24,465 \\
 Log Likelihood & -35923.278 & -19434.315 & -14391.070 \\
 Test Accuracy & 58.99\% & 60.95\% & 73.23\% \\
\hline
\end{tabular}
\end{center}
\caption{Three logistic regression models for predicting if an author will initiate with other authors. Model (1) includes all authors. Model (2) includes only authors who receive at least one interaction from another author in their first year. Model (3) includes only authors who are not interacted with in their first year. Table \ref{tab:init_logreg_interaction_term} explores the interaction between author role and being interacted with. Note: $^{*}$p$<$0.05, $^{**}$p$<$0.01, $^{***}$p$<$0.001.
}
\label{tab:init_logreg_full_comparison}
\end{table}

\begin{table}
\begin{center}
\begin{tabular}{lrrrrrr}
Feature      &  Coef.  & Std.Err. &    t     & P$>$$|$t$|$ &  [0.025 &  0.975]  \\
\hline

Intercept                                        & -0.9049 &   0.0163 & -55.4107 &      0.0000 & -0.9369 & -0.8729  \\
Role = Mixed   & -0.2979 &   0.0648 &  -4.5992 &      0.0000 & -0.4248 & -0.1709  \\
Role = P                          & -0.2962 &   0.0400 &  -7.3971 &      0.0000 & -0.3747 & -0.2177  \\
Is interacted with?                             &  1.0857 &   0.0212 &  51.2408 &      0.0000 &  1.0441 &  1.1272  \\
Role = Mixed : Int'ed with? &  0.4356 &   0.0825 &   5.2807 &      0.0000 &  0.2739 &  0.5972  \\
Role = P : Int'ed with?     &  0.5932 &   0.0504 &  11.7622 &      0.0000 &  0.4943 &  0.6920  \\
\hline
\end{tabular}
\end{center}
\caption{Logistic regression model predicting if an author will initiate with other authors.  P authors are less likely than CG authors to initiate in the absence of interactions. Both P and CG become much more likely to initiate if interacted with (182\% increase in the odds of initiating), but this effect is stronger for P than for CG.  (Observations = 53,335, model d.f. = 5, log-likelihood = $-$34,147, test accuracy = 67.9\%)
}
\label{tab:init_logreg_interaction_term}
\end{table}

Which authors initiate peer connections?
We fit logistic regression models to identify the factors that differentiate authors who will never interact with a fellow author (42.7\% of authors, $n$=154,811) from \textit{initiating authors}---authors who have made at least one initiation (57.3\%). The models are trained on the 53,335 authors who published their first update in the initiations period.  
Table \ref{tab:init_logreg_full_comparison} shows three models predicting author initiation within their first year on CB. In exploratory modeling, we observed a strong impact of being interacted with on initiation behavior; being interacted with is associated with a 182\% increase in the odds of initiating. Thus, we fit two additional models, splitting the data by whether they had been interacted with, shown as models (2) and (3) in Table \ref{tab:init_logreg_full_comparison}.
When an author is not interacted with, being a patient rather than a caregiver is associated with a 30\% decrease in the odds of initiation.  When an author \textit{is} interacted with, being a patient is associated with a 21\% increase in the odds of initiation.  
This disparity suggested a statistical interaction effect between being interacted with and author role.  Table \ref{tab:init_logreg_interaction_term} shows this significant interaction effect, which demonstrates that patient authors are much more likely to be initiators after being interacted with compared to caregivers, although both patients and caregivers are more likely to initiate after being interacted with.
\textbf{Among non-receivers, caregivers are more likely to initiate than patients; among receivers, patients are more likely to initiate than caregivers.}

We observe differences in initiation probability by health condition. For example, compared to reporting no health condition, an author self-reporting Cancer is associated with an 11\% increase in the odds of initiating.  However, we caution against over-interpretation of the less-common health conditions categories such as Congenital/Immune Disorder.
Activity level and authorship configuration have small effects on probability of initiation.\footnote{In a separate model predicting if an author is interacted with, rather than if they initiate, these effects are more relevant, as could be expected. Each additional update published is associated with a 1\% increase in the odds of being interacted with, and being a mixed-site author is associated with a 123\% increase in the odds of being interacted with (versus a 15\% increase in the odds of initiating).}

\subsection{RQ1b Results: When?}\label{sec:init_when_results}

\begin{table}
\begin{center}
\begin{tabular}{lrrrr}
\toprule
& \multicolumn{2}{c}{(1) Pre-authorship} & \multicolumn{2}{c}{(2) Post-authorship} \\
Feature      &  Coef.  & SE &   Coef. & SE \\
\hline
Intercept &       -1.6215$^{***}$  &         0.078 &  -1.8437$^{***}$ &   0.032\\
Role = Mixed  &        	-0.6928$^{**}$  &        0.264 &  	-0.0897$^{}$ &   0.108 \\
Role = P    &  -1.2148$^{***}$  &        0.166 & -0.3996$^{***}$ &   0.071 \\
Total Active Time (log months)   & 1.1660$^{***}$  & 0.030 &  1.3251$^{***}$ &   0.010\\
Total Active Time : Role = Mixed 	 & 0.1657$^{}$ & 0.102 &  0.0187$^{}$ &   0.033 \\
Total Active Time : Role = P     & 0.2607$^{***}$ & 0.064  & 0.0616$^{**}$ &   0.023 \\
\hline
R$^{2}$  & \multicolumn{2}{c}{0.310} & \multicolumn{2}{c}{0.542}\\
F-stat  & \multicolumn{2}{c}{489.2$^{***}$} & \multicolumn{2}{c}{4,899$^{***}$}\\
Observations & \multicolumn{2}{c}{5,439} & \multicolumn{2}{c}{20,687}  \\
Log-likelihood & \multicolumn{2}{c}{-10,699} & \multicolumn{2}{c}{-33,771} \\
\bottomrule
\end{tabular}
\end{center}
\caption{Linear regression models (d.f.$=$5) predicting the time between an author's first published update and their first initiation (log months).  Model (1) includes only pre-authorship initiators, whereas model (2) includes only post-authorship initiators. Note: $^{*}$p$<$0.05, $^{**}$p$<$0.01, $^{***}$p$<$0.001.}
\label{tab:init_timing_author_type_interaction}
\end{table}

When do authors initiate peer connections?
To understand the lifecycle of CB users and the relationship between intended use (publishing updates) and appropriative use (peer connection), we use linear regression to model the time between first published update and first peer initiation. This interval is shown in Figure \ref{fig:init_timing_response_plot}.  
We conduct analyses of author timing for the 5,439 pre-authorship initiators and the 20,687 post-authorship initiators separately.
Only 3.72\% of initiating authors do so post-authorship but before receiving an interaction, with 75.46\% of initiating authors doing so after authoring their first update \textit{and} receiving an interaction.
Among pre-authorship initiators, patients publish their first update 1.28 months sooner after first initiation than caregivers (t=$-$5.31, p<0.001).
Among post-authorship initiators, patients initiate 4.7 months sooner after first authorship than caregivers (t=$-$11.91, p<0.001).
Controlling for total active time on CB by fitting a linear model to predict the number of months between first authorship and first initiation (log transformed), we observe the same pattern.
We find a significant interaction between total active time and author role for both pre-authorship initiators (ANOVA F=3.72, p<0.01) and post-authorship initiators (ANOVA F=8.91, p<0.001).
Both pre- and post-authorship model details are given in Table \ref{tab:init_timing_author_type_interaction}.


\begin{figure}
\centering
\begin{subfigure}[t]{0.5\textwidth}
  \includegraphics[width=\textwidth]{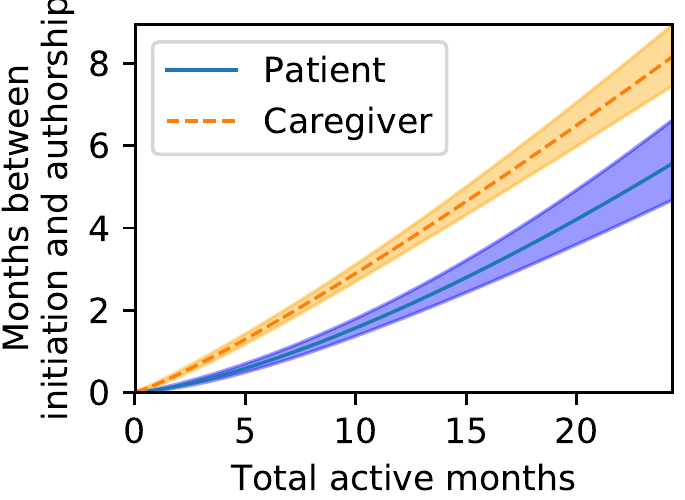}
\caption{Pre-authorship initiators ($n=5439$)}
\label{fig:init_timing_pre_effects}
\end{subfigure}%
\begin{subfigure}[t]{0.5\textwidth}
  \includegraphics[width=\textwidth]{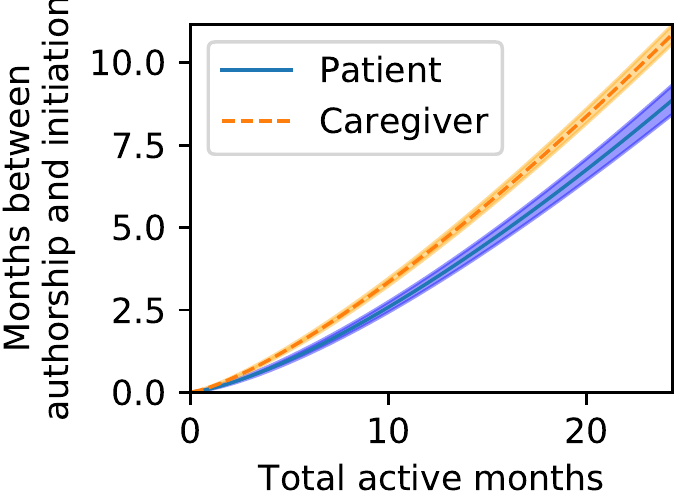}
\caption{Post-authorship initiators ($n=20687$)}
\label{fig:init_timing_post_effects}
\end{subfigure}
\caption{Effects plot of interaction between author role and total active time on CB pre- and post-initiation.  The sublinear trend indicates that a longer time active on CB is associated with a shorter time between first authorship and first initiation.
Shading indicates the 99\% confidence interval.
Mixed author role is not shown, as the difference from the CG condition is not significant (Table \ref{tab:init_timing_author_type_interaction}).
}
\label{fig:init_timing_effects_plot}
\end{figure}

An effects plot of the author role interaction among initiators is shown in Figure \ref{fig:init_timing_effects_plot}.  
\textbf{Compared to caregivers, patients initiate sooner after becoming an author.}
The effects plot shows a positive but sublinear trend, indicating that initiating earlier is associated with pre-initiation time forming a smaller percentage of an author's total time on CB.\footnote{If the trend were one-to-one, the percentage of total time that is between authorship and initiation would be the same on average for all authors, irrespective of their total time on CB. Instead, we see that total time is associated with a smaller percent of total time in the interval between authorship and initiation.}
Furthermore, the gap between patients and caregivers widens among users active on CB for a longer total period of time. 
For pre-authorship initiators who are only active for one month, patients are predicted to become authors 4.2 days sooner than caregivers. But among pre-authorship initiators who are active for two years, patients are predicted to become authors 77.2 days sooner than caregivers. 
For post-authorship initiators, the gap at one month is 1.6 days, widening to 59 days at two years.
This widening gap suggests a ``lifecycle'' model of CB use in which the authors active for a longer period of time are more likely to initiate earlier as a percentage of their total active time than those authors active for a shorter period of time, although we make no attempt to untangle the causal directionality of this effect.

We also fit linear regression models using a larger set of confounding features---but without total active time---in order to assess the predictability of time between first authorship and first initiation. 
For space reasons, model coefficients are presented in Appendix Table \ref{tab:init_timing_full_models}.  
These two models demonstrate that pre-authorship initiation is intrinsically high variance (R$^2 <0.01$). 
For post-authorship initiators, the author role, health condition, and ``interaction received'' features are significant predictors at the 99\% confidence level and so the proportion of variance explained is higher (R$^2 = 0.13$). 
For post-authorship initiators, receiving an interaction is associated with, on average, initiating 10.5 months sooner.  This large difference extends the results from RQ1a: \textbf{Not only are receiving authors more likely to initiate if interacted with, but also they will initiate much sooner than authors who are not receivers.}

\subsection{RQ1c Results: With whom?}\label{sec:init_with_whom_results}

\begin{table}[!htbp] 
\centering 
\begin{tabular}{@{\extracolsep{5pt}}lcccc} 
\toprule
\\[-1.8ex] & (1) All & (2) Network & (3) Role & (4) Activity \\ 
\hline \\[-1.8ex] 
Candidate out-degree (log) & $-$0.191$^{*}$ & $-$0.510$^{*}$ &  &  \\ 
  & (0.005) & (0.004) &  &  \\ 
Has in-degree? & 0.756$^{*}$ & 0.995$^{*}$ &  &  \\ 
  & (0.017) & (0.013) &  &  \\ 
Candidate in-degree (log) & 0.649$^{*}$ & 0.674$^{*}$ &  &  \\ 
  & (0.005) & (0.003) &  &  \\ 
Is reciprocal? & 20.016$^{*}$ & 13.068$^{*}$ &  &  \\ 
  & (0.460) & (0.174) &  &  \\ 
Is weakly connected? & 1.767$^{*}$ & 2.454$^{*}$ &  &  \\ 
  & (0.021) & (0.021) &  &  \\ 
Is friend-of-friend? & 5.220$^{*}$ & 4.881$^{*}$ &  &  \\ 
  & (0.097) & (0.050) &  &  \\ 
Candidate Role = Mixed & 0.020 &  & 0.095$^{*}$ &  \\ 
  & (0.018) &  & (0.012) &  \\ 
Candidate Role = P & $-$0.242$^{*}$ &  & 0.124$^{*}$ &  \\ 
  & (0.012) &  & (0.008) &  \\ 
Same author role? & 0.299$^{*}$ &  & 0.371$^{*}$ &  \\ 
  & (0.012) &  & (0.008) &  \\ 
Candidate multi-site author? & 0.315$^{*}$ &  & 0.249$^{*}$ &  \\ 
  & (0.015) &  & (0.010) &  \\ 
Candidate mixed-site author? & 0.474$^{*}$ &  & 1.365$^{*}$ &  \\ 
  & (0.008) &  & (0.005) &  \\ 
Candidate update count & $-$0.0003$^{*}$ &  &  & 0.001$^{*}$ \\ 
  & (0.00004) &  &  & (0.00002) \\ 
Candidate update frequency & 0.007$^{*}$ &  &  & 0.004$^{*}$ \\ 
  & (0.0002) &  &  & (0.0002) \\ 
Days since recent update & $-$0.011$^{*}$ &  &  & $-$0.013$^{*}$ \\ 
  & (0.00005) &  &  & (0.00005) \\ 
Days since first update & $-$0.001$^{*}$ &  &  & $-$0.001$^{*}$ \\ 
  & (0.00001) &  &  & (0.00001) \\ 
Same health condition? & 0.213$^{*}$ &  &  &  \\ 
  & (0.009) &  &  &  \\
\hline \\[-1.8ex] 
Observations & 155,141 & 155,141 & 155,141 & 155,141 \\ 
Log Likelihood & $-$133,746.600 & $-$353,610.400 & $-$465,555.600 & $-$206,743.700 \\ 
Test accuracy   &   77.3\% & 32.4\% & 9.8\% & 73.3\%  \\
\bottomrule
\end{tabular}
\caption{Conditional mlogit models predicting initiation probability for an initiating author and an arbitrary candidate author. The first model includes all features sets; models 2-4 include only one of the feature sets.
For space reasons, the model fit with only the health condition feature is not shown in the table above.  The model's single coefficient is 0.411 (s.e. 0.006), and it has log likelihood -496,923.3 and test accuracy 4.9\%.
Note: $^{*}$p$<$0.01.}
\label{tab:init_mlogit_full_results}
\end{table} 

With whom do authors initiate peer connections?
We fit conditional mlogit models to predict the probability of an author being the target of an initiation, as determined by both the traits of the initiator and the traits of the target.
Table \ref{tab:init_mlogit_full_results} shows the model coefficients and test accuracies. With the exception of the health condition model, all models predict the correct initiation target significantly above chance (i.e. 4\%), with the full model predicting the accurate target from among 25 candidates 77.3\% of the time.  
In isolation, the single most important feature is the number of days since the candidate's most recent update; a model that exclusively predicts as the target the candidate who has updated most recently actually achieves 81\% accuracy in the testing period, which explains the strong prediction performance of model \#4. However, we are interested in inference: the relative importance of the features.  Two features have a directionality difference between the focused models vs the full model: (a) the author being a patient rather than a caregiver, which overall makes an author more likely to be the target of an initiation but not when controlling for non-role factors; and (b) candidate update count, for which more updates is associated with a greater overall likelihood of being the initiation target, but not when controlling for non-activity factors. We thus focus on analysis of the full model with all features.

The most important binary feature is the reciprocation indicator---i.e. if choosing this candidate would result in a reciprocated connection---which is consistent with the importance of reciprocity in online interactions \cite{pan_you_2017}.  In general, network features are more predictive of initiation than having the same author role.  We see strong triadic closure effects i.e. ``Is friend-of-friend?'' has a large positive coefficient, and even being weakly connected with a candidate increases the likelihood of a connection.  Authors who have already received at least one initiation are more likely to be selected for subsequent initiations by other authors, even when controlling for author tenure and number of updates published by that author.
\textbf{An author's network position is an important factor in receiving new initiations,} with authors who have been interacted with by other authors being the most likely target for new initiations.

Author role is also important in the selection of a target author.
\textbf{Initiations are more likely to occur between two authors with the same role.} 
This effect is stronger for caregivers than for patients.  
We observe no significant difference between caregiver-role candidates and mixed-role candidates.
Authors of multiple sites and authors on mixed sites are also more likely to be the target of initiation, perhaps due to a larger set of interested readers that may include peer authors.

The activity features demonstrate that initiation is more likely when a candidate has more recently become an author, having written fewer updates but at a high frequency. As discussed above, authors are much more likely to be the target of initiations shortly after they publish an update, which may suggest update dissemination effects and textual content factors that are not captured in this analysis.

We also fit models using only the subset of the 66,616 first initiations, in order to evaluate whether the first initiation made by an author is somehow different. Feature directionality and relative magnitude remain the same, with the exception that first-time initiators are less likely to initiate with mixed-site authors.  

\subsubsection{Geographic analysis}\label{sec:geo_analysis_results}

32.7\% (118,534) of authors were assigned US states using the state-assignment procedure.  As a face validity check, the most frequent state assignments are Minnesota (6.3\%; CaringBridge was launched in Minnesota), California (2.8\%; the most populous US state), and Texas (2.8\%; the second most populous US state).  Authors assigned states are 
more active, more likely to have a plurality of their updates in a single US state, and more likely to initiate with other authors (2.41 vs 2.26 mean initiations, $p<0.001$) than the average CB author.  

In the initiations period, initiations between state-assigned authors account for only 4.5\% (7,007) of the total initiations.  49.5\% of these initiations were between two authors that have the same US state assignment, a percentage significantly above chance.  Fitting a full multinomial logit model that includes a dummy variable when the initiating author and the candidate author have the same state assignment confirms the importance of this feature: sharing a state assignment increases the odds of initiating with an author, holding other variables fixed.
(The model details and full comparison is shown in Appendix \ref{app:geo_analysis}, Table \ref{app:tab:geo_full_model_comparison}.) Fitting a model with only that feature results in a test accuracy (on 668 initiations in the test period) of 27.8\%.  This analysis suggests the importance of geographic co-location, although given the biased nature of the proxy used it is hard to reason about the magnitude of this effect relative to the other contextual factors.  At a minimum, \textbf{geographic co-location is an important predictor of initiation for some authors}.

\subsection{RQ2a Results: Reciprocations}\label{sec:recip_results}

\begin{table}
\begin{center}
\begin{tabular}{lr}
\begin{tabular}[b]{lrr}
Feature      &  Coef.  & Std.Err.  \\
\hline


Intercept                                                         & -2.320 & 0.008 \\
Initiator Role = Mixed                                       & -0.090 & 0.016 \\
Initiator Role = P                                           & -0.332 & 0.013 \\
Receiver Role = Mixed                                         & -0.178 & 0.017 \\
Receiver Role = P                                             & -0.110 & 0.015 \\
Init. Role = Mixed : Rcvr. Role = Mixed            & 0.597 & 0.043 \\
Init. Role = P : Rcvr. Role = Mixed                 & 0.687 & 0.034 \\
Init. Role = Mixed : Rcvr. Role = P            & 0.630 & 0.036 \\
Init. Role = P : Rcvr. Role = P                     & 1.204 & 0.025 \\
Was init. author?                                            & 1.115 & 0.010 \\
Was init. author? : Months after first init. update  & -0.310 & 0.002 \\
Months after first rcvr. update                          & 0.036 & 0.002 \\

\hline
\end{tabular}
& 
\begin{tabular}[b]{r}
Reciprocation Probs \\
\includegraphics[width=1.1in]{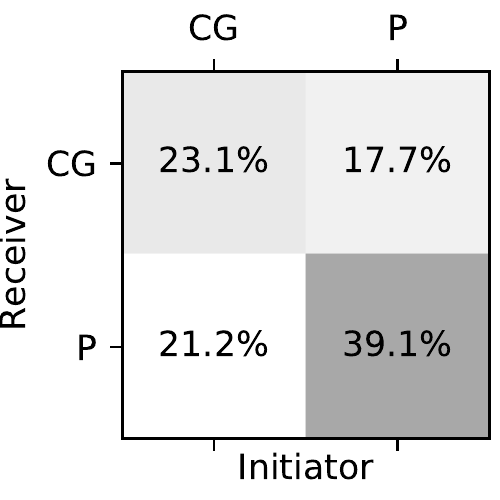} \\
\end{tabular}
\end{tabular}
\end{center}
\caption{Logistic regression model predicting if an author pair with one initiation will be reciprocated within one year. All coefficients are significant at p$<$0.001. Reciprocation is more likely between two authors that have the same role.
Predicted reciprocation probabilities are given for various initiating and receiving author roles fixing the other variables such that both the initiator and the receiver became authors one month before the original initiation.
(Observations = 737,747, model d.f. = 8, log-likelihood = $-$257,750).
}
\label{tab:recip_role_probs}
\end{table}

Which authors reciprocate and to which initiators?  
Among all pairs of authors with at least one directed initiation, only 12\% are reciprocated within one year.
We fit a model to predict if an author will reciprocate an initiation from another author within one year, as a function of both the initiator's and the receiver's author role.
Table \ref{tab:recip_role_probs} shows the coefficients for the logistic regression model.
\textbf{Authors are more likely to reciprocate an initiation when the initiator and the receiver have the same role.}
Patients are particularly likely to reciprocate in general, with patients who initiate with caregivers receiving the lowest reciprocation rates.  
Unsurprisingly, initiations from pre-authorship initiators are much less likely to be reciprocated; the initiator having already published their first update at the time of initiation is associated with a 205\% increase in the odds of reciprocation.  
When the initiator is already an author, reciprocation is less likely the longer that author has been on CB.  
In contrast, receivers are \textit{more} likely to initiate the longer they have been on CB.
To the right of Table \ref{tab:recip_role_probs}, we show estimated reciprocation probabilities based on author role, given that both the initiating and receiving author published their first journal update one month before the initiation.  
While only 12\% are initiated among all pairs, the probability of initiation among patient/patient pairs in which the initiator has already published their first update is more than twice that baseline.

\subsection{RQ2b Results: Relationships}\label{sec:relationship_results}

\begin{table}[]
\centering
\begin{tabular}{lrrrrr}
\toprule
Feature  &  IRR & Coef.  & Std.Err. & z & p \\ \hline
Intercept  & 17.479 &  2.8610 & 0.0061 & 471.7680 & 0.0000  \\
Initiator AR = Mixed    & 0.981 & -0.0194 & 0.0125 & -1.5531 & 0.1204  \\
Initiator AR = P & 0.928 & -0.0748 & 0.0099 & -7.5329 & 0.0000  \\
Reciprocator AR = Mixed & 0.967 & -0.0338 & 0.0128 & -2.6322 & 0.0085   \\
Reciprocator AR = P            & 0.946                            & -0.0553 & 0.0108 & -5.1382 & 0.0000 \\
Init. AR = Mixed : Recip. AR = Mixed & 1.044 &  0.0430 & 0.0328 & 1.3112 & 0.1898  \\
Init. AR = P : Recip. AR = Mixed & 1.187 & 0.1717 & 0.0254 & 6.7680 & 0.0000 \\
Init. AR = Mixed : Recip. AR = P    & 1.083 & 0.0798 & 0.0268 & 2.9724 & 0.0030 \\
Init. AR = P : Recip. AR = P        & 1.261 & 0.2321 & 0.0179 & 12.9835 & 0.0000 \\
Is balanced?        & 0.702 &  -0.3545 & 0.0056 & -63.0274 & 0.0000 \\
Duration (months) & 1.020 &  0.0198 & 0.0001 & 150.5271 & 0.0000 \\ \hline
Alpha & --- &  0.9337 & 0.0035 & 265.0840 & 0.0000 \\
\hline
\end{tabular}
\caption{Negative Binomial Regression model predicting dyadic relationship interaction counts. 
Incidence rate ratios (IRR) are given in the first column. Alpha is the estimated dispersion parameter, which is assumed non-zero.
(Observations = 125,629, model d.f. = 10, log-likelihood = $-$525,510)
}
\label{tab:dyad_int_count_results}
\end{table}

\begin{table}[]
\centering
\begin{tabular}{lrrrr}
\toprule
Feature  &  Coef.  & Std.Err. & z & p \\ \hline
Intercept                                                                  &  0.3867 &   0.0097 &  39.7880 &      0.0000  \\
Initiator AR = Mixed                                       & -0.0940 &   0.0253 &  -3.7129 &      0.0002  \\
Initiator AR = P                                            & -0.1148 &   0.0201 &  -5.7070 &      0.0000  \\
Reciprocator AR = Mixed & -0.1349 &   0.0260 &  -5.1841 &      0.0000 \\
Reciprocator AR = P & -0.1170 &   0.0218 &  -5.3628 &      0.0000  \\
Init. AR = Mixed : Recip. AR = Mixed &  0.2020 &   0.0666 &   3.0319 &      0.0024 \\
Init. AR = P : Recip. AR = Mixed     &  0.1186 &   0.0515 &   2.3035 &      0.0212   \\
Init. AR = Mixed : Recip. AR = P  &  0.1306 &   0.0544 &   2.4004 &      0.0164  \\
Init. AR = P : Recip. AR = P &  0.3156 &   0.0363 &   8.6853 &      0.0000 \\
Interaction count                                                                 & -0.0029 &   0.0001 & -26.0014 &      0.0000  \\
Duration (months)                                                           & -0.0060 &   0.0002 & -25.6988 &      0.0000 \\
\bottomrule
\end{tabular}
\caption{Logistic regression model predicting dyadic relationship balance. 
Dyadic relationships in which both authors have the same author role (AR) are more likely to be balanced.
(Observations = 124,377, model d.f. = 10, log-likelihood = $-$85,749)}
\label{tab:dyad_balance_results}
\end{table}

What factors lead to more interactive relationships? We present results for models predicting the total number of interactions in a relationship and the degree to which the interactions are balanced between both authors in a relationship. 

\subsubsection{Number of interactions}
We identified 125,629 relationships for analysis. The median relationship has 13 interactions, and 93.5\% of relationships have 100 or fewer interactions.
We evaluated the impact of having the same author role on the number of interactions in a relationship.
Table \ref{tab:dyad_int_count_results} shows the incidence rate ratios and coefficients for the negative binomial regression model predicting the total number of interactions in a relationship.
Features with incidence rate ratios greater than 1 are associated with an increased total number of interactions, while an incidence rate ratio less than 1 is associated with a decreased total.
\textbf{Relationships have more interactions when authors have the same role.}
Compared to caregiver/caregiver relationships, relationships where the initiator or the receiver is a patient is associated respectively with a 7.2\% and a 5.4\% decrease in the rate of interactions.
However, \textit{both} authors being patients is associated with a 10.7\% increase in interactions relative to caregiver/caregiver relationships.

\subsubsection{Relationship balance}
Balance refers to the difference in the number of interactions made by the original initiator vs the original reciprocator.
To control for the noise introduced by short relationships, we train the model using only relationships with at least 10 interactions ($n$=124,377).
The initiator of a relationship tends to interact more: 63\% of relationships involve a majority of the interactions coming from the initiator.   A majority (52.47\%) of relationships are balanced, with an additional 32.82\% dominated by the initiator and the final 14.71\% dominated by the reciprocator.  Appendix Figure \ref{fig:relationship_balance_distribution} shows the distribution of relationships by the percentage of interactions coming from the initiator.

We predict relationship balance as a function of author role while controlling for total interactions and duration.
Table \ref{tab:dyad_balance_results} show the coefficients for the logistic regression model predicting relationship balance.
\textbf{Reciprocated relationships are more balanced when authors have the same role.}
Having the same author role is associated with a 12\% increase in the odds of a relationship being balanced.
As with number of interactions, patient/patient relationships are more likely to be balanced compared with caregiver/caregiver relationships.





\section{Discussion}  

\begin{table}[]
\begin{tabular}{@{}lll@{}}
\toprule
  RQ                   & Result & Sec. \\ \midrule
RQ1a & \makecell[l]{Among non-receivers, caregivers are more likely to initiate than patients; \\among receivers, patients are more likely to initiate than caregivers.}        & \ref{sec:init_who_results}   \\[3mm]
RQ1b & \makecell[l]{Compared to caregivers, patients initiate sooner after becoming an author.}      & \ref{sec:init_when_results}   \\[3mm]
RQ1c & \makecell[l]{Initiations are more likely to occur between two authors who have the same role.} & \ref{sec:init_with_whom_results}   \\[3mm]
RQ2a & \makecell[l]{Authors are more likely to reciprocate an initiation when the initiator and the \\receiver have the same role.}              & \ref{sec:recip_results}   \\[3mm]
RQ2b                  & \makecell[l]{Reciprocated relationships have more interactions and are more balanced when \\authors have the same role.}                                 & \ref{sec:relationship_results}   \\[3mm]
                     \bottomrule
\end{tabular}
\caption{Summary of results.}
\label{tab:discussion_results_summary}
\end{table}

In the analyses presented in section \ref{sec:results}, we identified a variety of behavioral patterns and differences between patient and caregiver authors on CB.
Table \ref{tab:discussion_results_summary} highlights our key findings.
Why do we observe differences in connection behavior between patients and caregivers?
We suggest two primary interpretations for the observed differences:
First, the differences might indicate diverging \textit{preferences} of patients and caregivers for specific connection behaviors.
Second, the differences might indicate the existence of communicative or technical \textit{barriers} influencing the observed behaviors.
These interpretations are not mutually exclusive, with both aspects together contributing to the communication patterns observed in this study.

We emphasize that future qualitative work is needed to detangle these two interpretations.
For example, we found that while patients are more likely to initiate with peers than caregivers, caregivers are more likely to initiate after they have received an interaction.  Is this result due to patients' greater desire to actively seek out a support community (preference), to caregivers' lower knowledge of peer authors on CB (barrier), to caregivers' greater need for support alongside a stigma against asking for that support (preference+barrier), or to other factors?  Suggestively, Smith et al.\ found that---in the case of instrumental support---a greater proportion of caregivers do not ask for needed support compared to patients~\cite{smith_i_2020}. To identify the cause of the initiation gap, researchers could qualitatively study how these unmet caregiver support needs manifest in OHC peer connection behaviors. 

This section explores our results through the lens of preferences and barriers. 
First, we explore the RQ2 results with an eye toward fostering supportive online relationships, highlighting connections to prior work and unknowns that could be addressed in future work.  
Second, we discuss the implications of these findings for recommendation systems designed to facilitate new relationships.
Third, we discuss the implications of our results for future research that incorporates structural health roles, including plausible factors that affect caregivers' behavioral preferences for connection.



\subsection{Fostering online relationships}

Our results show that patient/patient and caregiver/caregiver author dyads are more likely to have highly interactive and balanced relationships than patient/caregiver dyads. Does this observed gap reveal a preference for interactions amongst authors who have the same role?  This interpretation is supported by Thoits' theory that experientially-similar others are important sources of support because they provide active coping assistance beyond the instrumental support provided by offline caregivers~\cite{thoits_mechanisms_2011}.  The prevalence of same-role dyads thus reflects a preference for authors who have had or are having similar experiences.  This preference could be supported through designs that aid authors in identifying experientially-similar others to engage for support.  For example, Ruthven proposes ``narrative retrieval'' as a novel IR task---one that could be applied in this context to identify similar-narrative authors who are sharing or who have shared experiences with the seeking author~\cite{ruthven_making_2019}. 
In addition to supporting same-role dyads, 
future qualitative work could investigate the specific types of support provided in \textit{different}-role dyads in order to identify the strategies used to make [currently rare] patient/caregiver relationships mutually beneficial and verify Thoits' theory about active coping assistance.

The observed gap could also indicate social barriers making it harder for patient/caregiver dyads to form relationships on CB.  Two salient barriers are that caregivers may not ``know what to say'' to patients \cite{jacobs_mypath:_2018} and that patients may perceive support offers as unhelpful \cite{iannarino_social_2014}. Tools such as MepsBot represent an opportunity to intervene during the comment-drafting process to increase the confidence of caregivers that they are writing comments that will be perceived as supportive to the receiving patient \cite{peng_exploring_2020}. Given the shifting needs of patients, sometimes non-response or a non-text response such as an amp may be the appropriate interaction for the relationship; designs for intervening in an on-going communcative process may benefit from incorporating and surfacing elements of the existing relationship's context in order to adapt to the complicated norms around response in sensitive health contexts~\cite{chang_respond_2018,andalibi_responding_2018}.
In summary, the gap between patient/caregiver and same-role author dyads indicates both preference for communication with experientially-similar others and socio-technical barriers to cross-role communication. 
OHCs should consider addressing this gap in pursuit of mutually supportive relationships.

\subsection{Designing peer recommendation systems}

Much research on OHCs is motivated by the goal of designing recommendation systems to form new relationships~\cite{cohen_social_2004,oleary_design_2017,eschler_im_2017,newman_its_2011,chen_make_2009,hartzler_leveraging_2016}.
Such systems have a goal of facilitating mutually supportive communication.
However, creating a recommendation system to facilitate supportive communication is hard; many online support interventions do not work as intended, producing minimal positive changes~\cite{lakey_social_1996,eysenbach_health_2004}.  
Peer recommendation systems that are faithful to the preferences of users have a better chance to succeed. 
Our work reveals the preferences of users ``in the wild'' and suggests three types of features that could be incorporated into peer recommendation systems in order to support the communication behaviors that OHC users are already doing.  

\begin{itemize}
\item \textit{Author role features.}
People benefit from and want relationships with experientially-similar others \cite{rains_coping_2018,thoits_mechanisms_2011}.
That people with the same author role tend to form these types of connections on CB provides evidence that the incorporation of author role information as additional recommendation system features could facilitate connections that have the shared experience qualities authors are seeking \cite{yang_seekers_2019,wright_measure_2010}. It is particularly notable that having the same author role is associated with more interactive relationships, a desirable outcome of recommendation to increase supportive social engagement.  
Author role should be considered alongside shared demographic or health condition traits as strong correlates with supportive connections~\cite{meng_your_2016}.
Our recommendation echoes calls to connect caregiver family members of cancer patients with others in a similar position~\cite{sanden_cancer_2019}.

\item \textit{Network features.}
Network features play an important role in who users initiate with; therefore, recommendation should also take network characteristics into account \cite{yan_network_2015}.  Since authors who have initiated with others are also more likely to be the target of future initiations, one goal of such a system may be to encourage a first interaction in order to facilitate the formation of future connections. Such an outcome could be encouraged through the recommendation of popular, central authors in the largest weakly-connected component with many existing connections.  Once connected with at least one other, friends-of-friends and others in the same component are natural choices given the importance of the network features in the RQ1c analysis.



\item \textit{Temporal features.}
Recommendation should also consider the impact of the timing of recommendations given; recommendations given in a particular period of a health journey may be more impactful in terms of positive health outcomes \cite{berkman_social_2014}.
We find that patients initiate sooner than caregivers do, which may suggest differential ``readiness'' for forming peer connections by author role.  Both patients and caregivers will initiate more quickly after being interacted with, so recommendation of previously-uninteracted-with authors may result in more communication overall.  Reciprocation is most likely when contacted by newer authors, so creating opportunities for newer authors to interact may be particularly beneficial for creating more reciprocated connections.  We also observe that initiating early is associated with a longer total time on CB. While longer-term use of CB is not necessarily beneficial for authors, it may indicate both need and opportunity for the cultivation of the longer-term mutually supportive relationships that form the foundation of a self-sustaining health community \cite{introne_sociotechnical_2016}.
\end{itemize}
Our results suggest that incorporating these features could facilitate more supportive connections. However, subsequent experimental work is necessary to verify the effectiveness of these features for the peer recommendation problem in OHCs.

\subsection{Incorporating structural roles in future research}

We highlight two implications of our results for research that incorporates structural roles.

\textit{Integrate structural and behavioral roles.}
Our results show that the structural roles \textit{patient} and \textit{caregiver} are associated with differences in OHC connection behavior.
While these roles have been used in previous literature \cite{gage_social_2013}, they have functioned mainly as descriptors without a clear definition.
Our results suggest an important research opportunity: understanding the relationship between behavioral roles and structural roles.
Behavioral roles describe patterns of user behavior. For example, Yang et al.\ define role as ``a set of interaction patterns regulated by explicit or implicit expectations and adopted by people in a social context to achieve specific social goals''~\cite{yang_seekers_2019}.  
Using this definition, they label patterns of behavior with names like ``story sharer''.
Structural roles function on a higher level, where adopting the role of \textit{patient} or \textit{caregiver} represents a personal transition: in terms of motivation, responsibilities, and relationships with others \cite{brim_properties_1980,meleis_transitions_2015}.
Identifying and tracking a person's behavioral roles during their transition into a structural role such as \textit{patient} could link motivations for use of OHCs with specific behaviors e.g. sharing stories. 
Such linking enables supporting a variety of structural roles by designing for the motivations that lead to the behaviors associated with those roles. 
Tensions between structural role and behavioral enactment of that role are relevant outside the health context as well. 
In contexts where structural roles have explicit technical support, such as content moderation~\cite{seering_moderator_2019}, separately examining moderator motivations and behaviors could motivate changes in the technical support provided for that role.

\textit{Focus on caregiver motivations.}
Our results about caregivers' use of CB has implications for understanding their motivations for connection. First, we find that---in the absence of receiving an interaction---caregivers are more likely than patients to initiate with other CB authors.
Caregivers' greater propensity to connect with others may reflect a lack of offline support for caregivers that creates stronger motivations for caregivers to initiate, particularly in the search for experientially-similar others \cite{schorch_designing_2016}.
Alternately, the observed appropriative use of CB for inter-caregiver communication may reflect a lack of appropriate channel for this communication in the readily-available communication technologies already in use \cite{sleeper_sharing_2016}.
However, further research is necessary to untangle the degree to which this gap between patients and caregivers is indicative of unmet support needs versus a simple homophily preference.  
Second, our results show that patients are particularly receptive to interactions from other authors and that caregivers are less affected by receiving an interaction.
Caregivers may be less likely to view themselves as the relevant recipient of the message \cite{sanden_cancer_2019,bloom_patients_2017} or may view reciprocation as an inappropriate articulation of personal concerns and challenges \cite{chen_caring_2013}.
In general, the appropriative use of CB for peer communication rather than health blogging presents opportunities to meet unmet needs.
Qualitative work is needed to understand the motivations that lead caregivers to initiate and to understand the particular importance these online peer relationships play in acquiring support for caregivers.

\section{Limitations \& Future Work}  

In this section, we discuss some limitations of our approach and sketch opportunities for future work to address these limitations.

The CB interaction network we study here is a partial view of the true social network, which includes both offline connections and online connections established or developed on other platforms. Furthermore, authors that do interact may differ from authors who do not interact on a variety of demographic and psychosocial factors \cite{han_lurking_2014,rains_social_2016}.
In studying the connection behavior of authors already on CB, we are engaging with a non-random sample of patients and caregivers, so application of these patterns to offline contexts should be done cautiously \cite{hargittai_potential_2018}.
Cross-platform and online/offline studies would contribute greatly to an understanding of the online health support ecosystem and the applicability of these findings to patients and caregivers more generally.

In studying peer connections, connections formed on CB between two strangers are the most similar to those created via hypothetical recommendation systems. However, identifying and isolating only these connections is challenging.  Detailed content analysis or other qualitative approaches to both identify and understand the formation of these connections would be valuable.  

In reasoning about the importance of author features for peer connections, we note the risk of unobserved confounds~\cite{jackson_social_2010}. While we attempted to address key confounders (e.g. geographic location) through additional analyses, the inclusion of additional likely confounding factors (e.g. existing offline relationships) would increase confidence in our findings.  Furthermore, while we found an increased likelihood of connection between authors with the same role or health condition, it is impossible to differentiate homophily effects from contagion effects in observational data~\cite{shalizi_homophily_2011}.  Future experimental work would enable exploration of these differing causes.
In examining the full network, we also did not account for tie strength \cite{burke_growing_2014}, which may be possible to estimate from the specific interactions between two authors.  Incorporating tie strength would enable the comparison of weak ties to strong ties formed on OHCs, which are known to have different supportive functions~\cite{albrecht_communicating_1987}.

In this study, we did not explore the impact of forming peer connections on specific outcomes such as engagement, perceived support, or stress. Further understanding of these outcomes is an important area for further research before the implementation of systems that facilitate the creation of these connections \cite{rains_coping_2018}.
While explorations of web-based social support have found correlations between social support and positive outcomes such as decreased stress, experimental interventions have not always found a decrease in stress even as received support increased for participants \cite{hill_influence_2006,rains_coping_2018,malik_computer-mediated_2008}.
Causal work is needed to understand the contexts in which peer connections are beneficial to participants.

Overall, author connections on OHCs provide a fruitful ground for further inter-disciplinary multi-method research.

\section{Conclusion}  

In this study, we explored the formation of peer connections in an OHC without explicit peer finding mechanisms.
By examining the peer connections that CaringBridge authors did form, we learned about their preferences.
We found significant differences in the initiation, reciprocation, and maintenance of these connections between two important structural roles: patients and caregivers. 
Our work indicates the importance of structural health roles to behavior in online health communities and suggests opportunities for the design of systems to actively facilitate or recommend these connections.
A focus on author roles opens up multiple opportunities for future research applying these results to the dynamics and design of mixed-role systems.  In particular, experimental work is needed to integrate author role into peer recommendation systems designed to facilitate interaction and foster mutually supportive relationships.

\begin{acks}
We would like to thank Daniel Kluver, C. Estelle Smith, Jacob Thebault-Spieker, and Sabirat Rubya for advice and feedback.  This work would not have been possible without our partners at CaringBridge and the Minnesota Supercomputing Institute (MSI) at the University of Minnesota. Finally, we thank the anonymous reviewers for exceptionally detailed and useful commentary.
\end{acks}

\bibliographystyle{ACM-Reference-Format}
\bibliography{umn-caringbridge,manual-entries}


\begin{thebibliography}{117}


\ifx \showCODEN    \undefined \def \showCODEN     #1{\unskip}     \fi
\ifx \showDOI      \undefined \def \showDOI       #1{#1}\fi
\ifx \showISBNx    \undefined \def \showISBNx     #1{\unskip}     \fi
\ifx \showISBNxiii \undefined \def \showISBNxiii  #1{\unskip}     \fi
\ifx \showISSN     \undefined \def \showISSN      #1{\unskip}     \fi
\ifx \showLCCN     \undefined \def \showLCCN      #1{\unskip}     \fi
\ifx \shownote     \undefined \def \shownote      #1{#1}          \fi
\ifx \showarticletitle \undefined \def \showarticletitle #1{#1}   \fi
\ifx \showURL      \undefined \def \showURL       {\relax}        \fi
\providecommand\bibfield[2]{#2}
\providecommand\bibinfo[2]{#2}
\providecommand\natexlab[1]{#1}
\providecommand\showeprint[2][]{arXiv:#2}

\bibitem[\protect\citeauthoryear{Albrecht and Adelman}{Albrecht and
  Adelman}{1987}]%
        {albrecht_communicating_1987}
\bibfield{author}{\bibinfo{person}{Terrance~L. Albrecht} {and}
  \bibinfo{person}{Mara~B. Adelman}.} \bibinfo{year}{1987}\natexlab{}.
\newblock \bibinfo{booktitle}{\emph{Communicating social support}}.
\newblock \bibinfo{publisher}{Sage Publications}.
\newblock
\showISBNx{978-0-8039-2679-0}
\newblock
\shownote{Google-Books-ID: nhtHAAAAMAAJ.}


\bibitem[\protect\citeauthoryear{Andalibi and Forte}{Andalibi and
  Forte}{2018}]%
        {andalibi_responding_2018}
\bibfield{author}{\bibinfo{person}{Nazanin Andalibi} {and}
  \bibinfo{person}{Andrea Forte}.} \bibinfo{year}{2018}\natexlab{}.
\newblock \showarticletitle{Responding to {Sensitive} {Disclosures} on {Social}
  {Media}: {A} {Decision}-{Making} {Framework}}.
\newblock \bibinfo{journal}{\emph{ACM Trans. Comput.-Hum. Interact.}}
  \bibinfo{volume}{25}, \bibinfo{number}{6} (\bibinfo{date}{Dec.}
  \bibinfo{year}{2018}), \bibinfo{pages}{31:1--31:29}.
\newblock
\showISSN{1073-0516}
\urldef\tempurl%
\url{https://doi.org/10.1145/3241044}
\showDOI{\tempurl}


\bibitem[\protect\citeauthoryear{Attard and Coulson}{Attard and
  Coulson}{2012}]%
        {attard_thematic_2012}
\bibfield{author}{\bibinfo{person}{Angelica Attard} {and}
  \bibinfo{person}{Neil~S. Coulson}.} \bibinfo{year}{2012}\natexlab{}.
\newblock \showarticletitle{A thematic analysis of patient communication in
  {Parkinson}’s disease online support group discussion forums}.
\newblock \bibinfo{journal}{\emph{Computers in Human Behavior}}
  \bibinfo{volume}{28}, \bibinfo{number}{2} (\bibinfo{date}{March}
  \bibinfo{year}{2012}), \bibinfo{pages}{500--506}.
\newblock
\showISSN{0747-5632}
\urldef\tempurl%
\url{https://doi.org/10.1016/j.chb.2011.10.022}
\showDOI{\tempurl}


\bibitem[\protect\citeauthoryear{Bambina}{Bambina}{2007}]%
        {bambina_online_2007}
\bibfield{author}{\bibinfo{person}{Antonina Bambina}.}
  \bibinfo{year}{2007}\natexlab{}.
\newblock \bibinfo{booktitle}{\emph{Online {Social} {Support}: {The}
  {Interplay} of {Social} {Networks} and {Computer}-{Mediated}
  {Communication}}}.
\newblock \bibinfo{publisher}{Cambria Press}.
\newblock
\showISBNx{978-1-62196-859-7}
\newblock
\shownote{Google-Books-ID: DIW0YKEBcPsC.}


\bibitem[\protect\citeauthoryear{Benne and Sheats}{Benne and Sheats}{1948}]%
        {benne_functional_1948}
\bibfield{author}{\bibinfo{person}{Kenneth~D. Benne} {and}
  \bibinfo{person}{Paul Sheats}.} \bibinfo{year}{1948}\natexlab{}.
\newblock \showarticletitle{Functional {Roles} of {Group} {Members}}.
\newblock \bibinfo{journal}{\emph{Journal of Social Issues}}
  \bibinfo{volume}{4}, \bibinfo{number}{2} (\bibinfo{year}{1948}),
  \bibinfo{pages}{41--49}.
\newblock
\showISSN{1540-4560}
\urldef\tempurl%
\url{https://doi.org/10.1111/j.1540-4560.1948.tb01783.x}
\showDOI{\tempurl}


\bibitem[\protect\citeauthoryear{Berkman, Kawachi, and Glymour}{Berkman
  et~al\mbox{.}}{2014}]%
        {berkman_social_2014}
\bibfield{author}{\bibinfo{person}{Lisa~F. Berkman}, \bibinfo{person}{Ichirō
  Kawachi}, {and} \bibinfo{person}{M.~Maria Glymour}.}
  \bibinfo{year}{2014}\natexlab{}.
\newblock \bibinfo{booktitle}{\emph{Social {Epidemiology}}}.
\newblock \bibinfo{publisher}{Oxford University Press}.
\newblock
\showISBNx{978-0-19-939533-0}
\newblock
\shownote{Google-Books-ID: qHpYCwAAQBAJ.}


\bibitem[\protect\citeauthoryear{Bloom}{Bloom}{2017}]%
        {bloom_patients_2017}
\bibfield{author}{\bibinfo{person}{Rosaleen~Duggan Bloom}.}
  \bibinfo{year}{2017}\natexlab{}.
\newblock \emph{\bibinfo{title}{Patients' and {Caregivers}' {Experience} of
  {Social} {Support} on {CaringBridge}}}.
\newblock Ph.{D}. \bibinfo{school}{The University of Utah},
  \bibinfo{address}{United States -- Utah}.
\newblock
\urldef\tempurl%
\url{http://search.proquest.com/docview/2187630344/abstract/7124738624E044AAPQ/1}
\showURL{%
\tempurl}
\newblock
\shownote{ISBN: 9780438889811.}


\bibitem[\protect\citeauthoryear{Bloom, Beck, Chou, Reblin, and
  Ellington}{Bloom et~al\mbox{.}}{2019}]%
        {bloom_their_2019}
\bibfield{author}{\bibinfo{person}{Rosaleen~D. Bloom}, \bibinfo{person}{Susan
  Beck}, \bibinfo{person}{Wen-Ying~Sylvia Chou}, \bibinfo{person}{Maija
  Reblin}, {and} \bibinfo{person}{Lee Ellington}.}
  \bibinfo{year}{2019}\natexlab{}.
\newblock \showarticletitle{In {Their} {Own} {Words}: {Experiences} of
  {Caregivers} of {Adults} {With} {Cancer} as {Expressed} on {Social} {Media}}.
\newblock \bibinfo{journal}{\emph{Oncology Nursing Forum}}
  \bibinfo{volume}{46}, \bibinfo{number}{5} (\bibinfo{date}{Sept.}
  \bibinfo{year}{2019}), \bibinfo{pages}{617--630}.
\newblock
\urldef\tempurl%
\url{https://doi.org/10.1188/19.ONF.617-630}
\showDOI{\tempurl}
\newblock
\shownote{Publisher: Oncology Nursing Society.}


\bibitem[\protect\citeauthoryear{Brim and Ryff}{Brim and Ryff}{1980}]%
        {brim_properties_1980}
\bibfield{author}{\bibinfo{person}{Orville~G Brim} {and}
  \bibinfo{person}{Carol~D Ryff}.} \bibinfo{year}{1980}\natexlab{}.
\newblock \showarticletitle{On the properties of life events}.
\newblock \bibinfo{journal}{\emph{Life-span development and behavior}}
  \bibinfo{volume}{3} (\bibinfo{year}{1980}).
\newblock


\bibitem[\protect\citeauthoryear{Bruckman}{Bruckman}{2002}]%
        {bruckman_studying_2002}
\bibfield{author}{\bibinfo{person}{Amy Bruckman}.}
  \bibinfo{year}{2002}\natexlab{}.
\newblock \showarticletitle{Studying the amateur artist: {A} perspective on
  disguising data collected in human subjects research on the {Internet}}.
\newblock \bibinfo{journal}{\emph{Ethics and Information Technology}}
  \bibinfo{volume}{4}, \bibinfo{number}{3} (\bibinfo{date}{Sept.}
  \bibinfo{year}{2002}), \bibinfo{pages}{217--231}.
\newblock
\showISSN{1572-8439}
\urldef\tempurl%
\url{https://doi.org/10.1023/A:1021316409277}
\showDOI{\tempurl}


\bibitem[\protect\citeauthoryear{Bruckman, Fiesler, Hancock, and
  Munteanu}{Bruckman et~al\mbox{.}}{2017}]%
        {bruckman_cscw_2017}
\bibfield{author}{\bibinfo{person}{Amy~S. Bruckman}, \bibinfo{person}{Casey
  Fiesler}, \bibinfo{person}{Jeff Hancock}, {and} \bibinfo{person}{Cosmin
  Munteanu}.} \bibinfo{year}{2017}\natexlab{}.
\newblock \showarticletitle{{CSCW} {Research} {Ethics} {Town} {Hall}: {Working}
  {Towards} {Community} {Norms}}. In \bibinfo{booktitle}{\emph{Companion of the
  2017 {ACM} {Conference} on {Computer} {Supported} {Cooperative} {Work} and
  {Social} {Computing}}} \emph{(\bibinfo{series}{{CSCW} '17 {Companion}})}.
  \bibinfo{publisher}{ACM}, \bibinfo{address}{New York, NY, USA},
  \bibinfo{pages}{113--115}.
\newblock
\showISBNx{978-1-4503-4688-7}
\urldef\tempurl%
\url{https://doi.org/10.1145/3022198.3022199}
\showDOI{\tempurl}


\bibitem[\protect\citeauthoryear{Burke and Kraut}{Burke and Kraut}{2008}]%
        {burke_mopping_2008}
\bibfield{author}{\bibinfo{person}{Moira Burke} {and} \bibinfo{person}{Robert
  Kraut}.} \bibinfo{year}{2008}\natexlab{}.
\newblock \showarticletitle{Mopping up: modeling wikipedia promotion
  decisions}. In \bibinfo{booktitle}{\emph{Proceedings of the {ACM} 2008
  conference on {Computer} supported cooperative work - {CSCW} '08}}.
  \bibinfo{publisher}{ACM Press}, \bibinfo{address}{San Diego, CA, USA},
  \bibinfo{pages}{27}.
\newblock
\showISBNx{978-1-60558-007-4}
\urldef\tempurl%
\url{https://doi.org/10.1145/1460563.1460571}
\showDOI{\tempurl}


\bibitem[\protect\citeauthoryear{Burke and Kraut}{Burke and Kraut}{2014}]%
        {burke_growing_2014}
\bibfield{author}{\bibinfo{person}{Moira Burke} {and}
  \bibinfo{person}{Robert~E. Kraut}.} \bibinfo{year}{2014}\natexlab{}.
\newblock \showarticletitle{Growing {Closer} on {Facebook}: {Changes} in {Tie}
  {Strength} {Through} {Social} {Network} {Site} {Use}}. In
  \bibinfo{booktitle}{\emph{Proceedings of the {SIGCHI} {Conference} on {Human}
  {Factors} in {Computing} {Systems}}} \emph{(\bibinfo{series}{{CHI} '14})}.
  \bibinfo{publisher}{ACM}, \bibinfo{address}{New York, NY, USA},
  \bibinfo{pages}{4187--4196}.
\newblock
\showISBNx{978-1-4503-2473-1}
\urldef\tempurl%
\url{https://doi.org/10.1145/2556288.2557094}
\showDOI{\tempurl}


\bibitem[\protect\citeauthoryear{Centola and van~de Rijt}{Centola and van~de
  Rijt}{2015}]%
        {centola_choosing_2015}
\bibfield{author}{\bibinfo{person}{Damon Centola} {and} \bibinfo{person}{Arnout
  van~de Rijt}.} \bibinfo{year}{2015}\natexlab{}.
\newblock \showarticletitle{Choosing your network: {Social} preferences in an
  online health community}.
\newblock \bibinfo{journal}{\emph{Social Science \& Medicine}}
  \bibinfo{volume}{125} (\bibinfo{date}{Jan.} \bibinfo{year}{2015}),
  \bibinfo{pages}{19--31}.
\newblock
\showISSN{0277-9536}
\urldef\tempurl%
\url{https://doi.org/10.1016/j.socscimed.2014.05.019}
\showDOI{\tempurl}


\bibitem[\protect\citeauthoryear{Chang, Whitlock, and Bazarova}{Chang
  et~al\mbox{.}}{2018}]%
        {chang_respond_2018}
\bibfield{author}{\bibinfo{person}{Pamara~F. Chang}, \bibinfo{person}{Janis
  Whitlock}, {and} \bibinfo{person}{Natalya~N. Bazarova}.}
  \bibinfo{year}{2018}\natexlab{}.
\newblock \showarticletitle{“{To} {Respond} or not to {Respond}, that is the
  {Question}”: {The} {Decision}-{Making} {Process} of {Providing} {Social}
  {Support} to {Distressed} {Posters} on {Facebook}:}.
\newblock \bibinfo{journal}{\emph{Social Media + Society}}
  (\bibinfo{date}{Feb.} \bibinfo{year}{2018}).
\newblock
\urldef\tempurl%
\url{https://doi.org/10.1177/2056305118759290}
\showDOI{\tempurl}
\newblock
\shownote{Publisher: SAGE PublicationsSage UK: London, England.}


\bibitem[\protect\citeauthoryear{Chen, Geyer, Dugan, Muller, and Guy}{Chen
  et~al\mbox{.}}{2009}]%
        {chen_make_2009}
\bibfield{author}{\bibinfo{person}{Jilin Chen}, \bibinfo{person}{Werner Geyer},
  \bibinfo{person}{Casey Dugan}, \bibinfo{person}{Michael Muller}, {and}
  \bibinfo{person}{Ido Guy}.} \bibinfo{year}{2009}\natexlab{}.
\newblock \showarticletitle{Make {New} {Friends}, but {Keep} the {Old}:
  {Recommending} {People} on {Social} {Networking} {Sites}}. In
  \bibinfo{booktitle}{\emph{Proceedings of the {SIGCHI} {Conference} on {Human}
  {Factors} in {Computing} {Systems}}} \emph{(\bibinfo{series}{{CHI} '09})}.
  \bibinfo{publisher}{ACM}, \bibinfo{address}{New York, NY, USA},
  \bibinfo{pages}{201--210}.
\newblock
\showISBNx{978-1-60558-246-7}
\urldef\tempurl%
\url{https://doi.org/10.1145/1518701.1518735}
\showDOI{\tempurl}
\newblock
\shownote{event-place: Boston, MA, USA.}


\bibitem[\protect\citeauthoryear{Chen, Ngo, and Park}{Chen
  et~al\mbox{.}}{2013}]%
        {chen_caring_2013}
\bibfield{author}{\bibinfo{person}{Yunan Chen}, \bibinfo{person}{Victor Ngo},
  {and} \bibinfo{person}{Sun~Young Park}.} \bibinfo{year}{2013}\natexlab{}.
\newblock \showarticletitle{Caring for {Caregivers}: {Designing} for
  {Integrality}}. In \bibinfo{booktitle}{\emph{Proceedings of the 2013
  {Conference} on {Computer} {Supported} {Cooperative} {Work}}}
  \emph{(\bibinfo{series}{{CSCW} '13})}. \bibinfo{publisher}{ACM},
  \bibinfo{address}{New York, NY, USA}, \bibinfo{pages}{91--102}.
\newblock
\showISBNx{978-1-4503-1331-5}
\urldef\tempurl%
\url{https://doi.org/10.1145/2441776.2441789}
\showDOI{\tempurl}
\newblock
\shownote{event-place: San Antonio, Texas, USA.}


\bibitem[\protect\citeauthoryear{Chou, Liu, Post, and Hesse}{Chou
  et~al\mbox{.}}{2011}]%
        {chou_health-related_2011}
\bibfield{author}{\bibinfo{person}{Wen-ying~Sylvia Chou},
  \bibinfo{person}{Benmei Liu}, \bibinfo{person}{Samantha Post}, {and}
  \bibinfo{person}{Bradford Hesse}.} \bibinfo{year}{2011}\natexlab{}.
\newblock \showarticletitle{Health-related {Internet} use among cancer
  survivors: data from the {Health} {Information} {National} {Trends} {Survey},
  2003–2008}.
\newblock \bibinfo{journal}{\emph{J Cancer Surviv}} \bibinfo{volume}{5},
  \bibinfo{number}{3} (\bibinfo{date}{Sept.} \bibinfo{year}{2011}),
  \bibinfo{pages}{263--270}.
\newblock
\showISSN{1932-2267}
\urldef\tempurl%
\url{https://doi.org/10.1007/s11764-011-0179-5}
\showDOI{\tempurl}


\bibitem[\protect\citeauthoryear{Cohen}{Cohen}{2004}]%
        {cohen_social_2004}
\bibfield{author}{\bibinfo{person}{Sheldon Cohen}.}
  \bibinfo{year}{2004}\natexlab{}.
\newblock \showarticletitle{Social relationships and health.}
\newblock \bibinfo{journal}{\emph{American psychologist}} \bibinfo{volume}{59},
  \bibinfo{number}{8} (\bibinfo{year}{2004}), \bibinfo{pages}{676}.
\newblock


\bibitem[\protect\citeauthoryear{Colvin, Chenoweth, Bold, and Harding}{Colvin
  et~al\mbox{.}}{2004}]%
        {colvin_caregivers_2004}
\bibfield{author}{\bibinfo{person}{Jan Colvin}, \bibinfo{person}{Lillian
  Chenoweth}, \bibinfo{person}{Mary Bold}, {and} \bibinfo{person}{Cheryl
  Harding}.} \bibinfo{year}{2004}\natexlab{}.
\newblock \showarticletitle{Caregivers of {Older} {Adults}: {Advantages} and
  {Disadvantages} of {Internet}-based {Social} {Support}}.
\newblock \bibinfo{journal}{\emph{Family Relations}} \bibinfo{volume}{53},
  \bibinfo{number}{1} (\bibinfo{year}{2004}), \bibinfo{pages}{49--57}.
\newblock
\showISSN{1741-3729}
\urldef\tempurl%
\url{https://doi.org/10.1111/j.1741-3729.2004.00008.x}
\showDOI{\tempurl}


\bibitem[\protect\citeauthoryear{Croissant}{Croissant}{2019}]%
        {mlogit_package}
\bibfield{author}{\bibinfo{person}{Yves Croissant}.}
  \bibinfo{year}{2019}\natexlab{}.
\newblock \bibinfo{booktitle}{\emph{mlogit: Multinomial Logit Models}}.
\newblock
\urldef\tempurl%
\url{https://CRAN.R-project.org/package=mlogit}
\showURL{%
\tempurl}
\newblock
\shownote{R package version 1.0-2.}


\bibitem[\protect\citeauthoryear{Cross, Parker, Prusak, and Borgatti}{Cross
  et~al\mbox{.}}{2004}]%
        {cross_knowing_2004}
\bibfield{author}{\bibinfo{person}{Rob Cross}, \bibinfo{person}{Andrew Parker},
  \bibinfo{person}{Laurence Prusak}, {and} \bibinfo{person}{Stephen~P.
  Borgatti}.} \bibinfo{year}{2004}\natexlab{}.
\newblock \bibinfo{booktitle}{\emph{Knowing {What} {We} {Know}: {Supporting}
  {Knowledge} {Creation} and {Sharing} in {Social} {Networks}}}.
\newblock \bibinfo{publisher}{Oxford University Press}.
\newblock
\showISBNx{978-0-19-983575-1}
\urldef\tempurl%
\url{http://www.oxfordscholarship.com/view/10.1093/0195165128.001.0001/acprof-9780195165128-chapter-5}
\showURL{%
\tempurl}


\bibitem[\protect\citeauthoryear{Dunkel-Schetter}{Dunkel-Schetter}{1984}]%
        {dunkel-schetter_social_1984}
\bibfield{author}{\bibinfo{person}{Christine Dunkel-Schetter}.}
  \bibinfo{year}{1984}\natexlab{}.
\newblock \showarticletitle{Social {Support} and {Cancer}: {Findings} {Based}
  on {Patient} {Interviews} and {Their} {Implications}}.
\newblock \bibinfo{journal}{\emph{Journal of Social Issues}}
  \bibinfo{volume}{40}, \bibinfo{number}{4} (\bibinfo{date}{Jan.}
  \bibinfo{year}{1984}), \bibinfo{pages}{77--98}.
\newblock
\showISSN{1540-4560}
\urldef\tempurl%
\url{https://doi.org/10.1111/j.1540-4560.1984.tb01108.x}
\showDOI{\tempurl}


\bibitem[\protect\citeauthoryear{Ebaugh}{Ebaugh}{1988}]%
        {ebaugh_becoming_1988}
\bibfield{author}{\bibinfo{person}{Helen Rose~Fuchs Ebaugh}.}
  \bibinfo{year}{1988}\natexlab{}.
\newblock \bibinfo{booktitle}{\emph{Becoming an {Ex}: {The} {Process} of {Role}
  {Exit}}}.
\newblock \bibinfo{publisher}{University of Chicago Press}.
\newblock
\showISBNx{978-0-226-18070-0}
\urldef\tempurl%
\url{https://www.press.uchicago.edu/ucp/books/book/chicago/B/bo5952245.html}
\showURL{%
\tempurl}


\bibitem[\protect\citeauthoryear{Eschler and Pratt}{Eschler and Pratt}{2017}]%
        {eschler_im_2017}
\bibfield{author}{\bibinfo{person}{Jordan Eschler} {and} \bibinfo{person}{Wanda
  Pratt}.} \bibinfo{year}{2017}\natexlab{}.
\newblock \showarticletitle{"{I}'m {So} {Glad} {I} {Met} {You}": {Designing}
  {Dynamic} {Collaborative} {Support} for {Young} {Adult} {Cancer}
  {Survivors}}. In \bibinfo{booktitle}{\emph{Proceedings of the 2017 {ACM}
  {Conference} on {Computer} {Supported} {Cooperative} {Work} and {Social}
  {Computing}}} \emph{(\bibinfo{series}{{CSCW} '17})}.
  \bibinfo{publisher}{ACM}, \bibinfo{address}{New York, NY, USA},
  \bibinfo{pages}{1763--1774}.
\newblock
\showISBNx{978-1-4503-4335-0}
\urldef\tempurl%
\url{https://doi.org/10.1145/2998181.2998326}
\showDOI{\tempurl}


\bibitem[\protect\citeauthoryear{Eysenbach, Powell, Englesakis, Rizo, and
  Stern}{Eysenbach et~al\mbox{.}}{2004}]%
        {eysenbach_health_2004}
\bibfield{author}{\bibinfo{person}{Gunther Eysenbach}, \bibinfo{person}{John
  Powell}, \bibinfo{person}{Marina Englesakis}, \bibinfo{person}{Carlos Rizo},
  {and} \bibinfo{person}{Anita Stern}.} \bibinfo{year}{2004}\natexlab{}.
\newblock \showarticletitle{Health related virtual communities and electronic
  support groups: systematic review of the effects of online peer to peer
  interactions}.
\newblock \bibinfo{journal}{\emph{BMJ}} \bibinfo{volume}{328},
  \bibinfo{number}{7449} (\bibinfo{date}{May} \bibinfo{year}{2004}),
  \bibinfo{pages}{1166}.
\newblock
\showISSN{0959-8138, 1468-5833}
\urldef\tempurl%
\url{https://doi.org/10.1136/bmj.328.7449.1166}
\showDOI{\tempurl}


\bibitem[\protect\citeauthoryear{Fox}{Fox}{2011}]%
        {fox_social_2011}
\bibfield{author}{\bibinfo{person}{Susannah Fox}.}
  \bibinfo{year}{2011}\natexlab{}.
\newblock \bibinfo{booktitle}{\emph{The social life of health information,
  2011}}.
\newblock \bibinfo{publisher}{Pew Internet \& American Life Project Washington,
  DC}.
\newblock


\bibitem[\protect\citeauthoryear{Frost and Massagli}{Frost and
  Massagli}{2008}]%
        {frost_social_2008}
\bibfield{author}{\bibinfo{person}{Jeana Frost} {and} \bibinfo{person}{Michael
  Massagli}.} \bibinfo{year}{2008}\natexlab{}.
\newblock \showarticletitle{Social {Uses} of {Personal} {Health} {Information}
  {Within} {PatientsLikeMe}, an {Online} {Patient} {Community}: {What} {Can}
  {Happen} {When} {Patients} {Have} {Access} to {One} {Another}’s {Data}}.
\newblock \bibinfo{journal}{\emph{Journal of Medical Internet Research}}
  \bibinfo{volume}{10}, \bibinfo{number}{3} (\bibinfo{year}{2008}),
  \bibinfo{pages}{e15}.
\newblock
\urldef\tempurl%
\url{https://doi.org/10.2196/jmir.1053}
\showDOI{\tempurl}


\bibitem[\protect\citeauthoryear{Gage}{Gage}{2013}]%
        {gage_social_2013}
\bibfield{author}{\bibinfo{person}{Elizabeth~A. Gage}.}
  \bibinfo{year}{2013}\natexlab{}.
\newblock \showarticletitle{Social networks of experientially similar others:
  {Formation}, activation, and consequences of network ties on the health care
  experience}.
\newblock \bibinfo{journal}{\emph{Social Science \& Medicine}}
  \bibinfo{volume}{95} (\bibinfo{date}{Oct.} \bibinfo{year}{2013}),
  \bibinfo{pages}{43--51}.
\newblock
\showISSN{0277-9536}
\urldef\tempurl%
\url{https://doi.org/10.1016/j.socscimed.2012.09.001}
\showDOI{\tempurl}


\bibitem[\protect\citeauthoryear{Gallagher, Stowell, Parker, and
  Welles}{Gallagher et~al\mbox{.}}{2019}]%
        {gallagher_reclaiming_2019}
\bibfield{author}{\bibinfo{person}{Ryan~J. Gallagher},
  \bibinfo{person}{Elizabeth Stowell}, \bibinfo{person}{Andrea~G. Parker},
  {and} \bibinfo{person}{Brooke~Foucault Welles}.}
  \bibinfo{year}{2019}\natexlab{}.
\newblock \showarticletitle{Reclaiming {Stigmatized} {Narratives}: {The}
  {Networked} {Disclosure} {Landscape} of \#{MeToo}}.
\newblock \bibinfo{journal}{\emph{Proceedings of the 2019 conference on
  Computer supported cooperative work - CSCW '19}} (\bibinfo{date}{May}
  \bibinfo{year}{2019}).
\newblock
\urldef\tempurl%
\url{https://doi.org/10.31235/osf.io/qsmce}
\showDOI{\tempurl}


\bibitem[\protect\citeauthoryear{Geiger, Yu, Yang, Dai, Qiu, Tang, and
  Huang}{Geiger et~al\mbox{.}}{2020}]%
        {geiger_garbage_2020}
\bibfield{author}{\bibinfo{person}{R.~Stuart Geiger}, \bibinfo{person}{Kevin
  Yu}, \bibinfo{person}{Yanlai Yang}, \bibinfo{person}{Mindy Dai},
  \bibinfo{person}{Jie Qiu}, \bibinfo{person}{Rebekah Tang}, {and}
  \bibinfo{person}{Jenny Huang}.} \bibinfo{year}{2020}\natexlab{}.
\newblock \showarticletitle{Garbage in, garbage out? do machine learning
  application papers in social computing report where human-labeled training
  data comes from?}. In \bibinfo{booktitle}{\emph{Proceedings of the 2020
  {Conference} on {Fairness}, {Accountability}, and {Transparency}}}
  \emph{(\bibinfo{series}{{FAT}* '20})}. \bibinfo{publisher}{Association for
  Computing Machinery}, \bibinfo{address}{Barcelona, Spain},
  \bibinfo{pages}{325--336}.
\newblock
\showISBNx{978-1-4503-6936-7}
\urldef\tempurl%
\url{https://doi.org/10.1145/3351095.3372862}
\showDOI{\tempurl}


\bibitem[\protect\citeauthoryear{Granovetter}{Granovetter}{1973}]%
        {granovetter_strength_1973}
\bibfield{author}{\bibinfo{person}{Mark~S. Granovetter}.}
  \bibinfo{year}{1973}\natexlab{}.
\newblock \showarticletitle{The {Strength} of {Weak} {Ties}}.
\newblock \bibinfo{journal}{\emph{Amer. J. Sociology}} \bibinfo{volume}{78},
  \bibinfo{number}{6} (\bibinfo{date}{May} \bibinfo{year}{1973}),
  \bibinfo{pages}{1360--1380}.
\newblock
\showISSN{0002-9602}
\urldef\tempurl%
\url{https://doi.org/10.1086/225469}
\showDOI{\tempurl}


\bibitem[\protect\citeauthoryear{Hagberg, Schult, and Swart}{Hagberg
  et~al\mbox{.}}{2008}]%
        {networkx}
\bibfield{author}{\bibinfo{person}{Aric~A. Hagberg}, \bibinfo{person}{Daniel~A.
  Schult}, {and} \bibinfo{person}{Pieter~J. Swart}.}
  \bibinfo{year}{2008}\natexlab{}.
\newblock \showarticletitle{Exploring Network Structure, Dynamics, and Function
  using NetworkX}. In \bibinfo{booktitle}{\emph{Proceedings of the 7th Python
  in Science Conference}}, \bibfield{editor}{\bibinfo{person}{Ga\"el
  Varoquaux}, \bibinfo{person}{Travis Vaught}, {and} \bibinfo{person}{Jarrod
  Millman}} (Eds.). \bibinfo{address}{Pasadena, CA USA}, \bibinfo{pages}{11 --
  15}.
\newblock


\bibitem[\protect\citeauthoryear{Halfaker, Keyes, Kluver, Thebault-Spieker,
  Nguyen, Shores, Uduwage, and Warncke-Wang}{Halfaker et~al\mbox{.}}{2015}]%
        {halfaker_user_2015}
\bibfield{author}{\bibinfo{person}{Aaron Halfaker}, \bibinfo{person}{Os Keyes},
  \bibinfo{person}{Daniel Kluver}, \bibinfo{person}{Jacob Thebault-Spieker},
  \bibinfo{person}{Tien Nguyen}, \bibinfo{person}{Kenneth Shores},
  \bibinfo{person}{Anuradha Uduwage}, {and} \bibinfo{person}{Morten
  Warncke-Wang}.} \bibinfo{year}{2015}\natexlab{}.
\newblock \showarticletitle{User {Session} {Identification} {Based} on {Strong}
  {Regularities} in {Inter}-activity {Time}}. In
  \bibinfo{booktitle}{\emph{Proceedings of the 24th {International}
  {Conference} on {World} {Wide} {Web}}} \emph{(\bibinfo{series}{{WWW} '15})}.
  \bibinfo{publisher}{International World Wide Web Conferences Steering
  Committee}, \bibinfo{address}{Republic and Canton of Geneva, Switzerland},
  \bibinfo{pages}{410--418}.
\newblock
\showISBNx{978-1-4503-3469-3}
\urldef\tempurl%
\url{https://doi.org/10.1145/2736277.2741117}
\showDOI{\tempurl}
\newblock
\shownote{event-place: Florence, Italy.}


\bibitem[\protect\citeauthoryear{Han, Hou, Kim, and Gustafson}{Han
  et~al\mbox{.}}{2014}]%
        {han_lurking_2014}
\bibfield{author}{\bibinfo{person}{Jeong~Yeob Han}, \bibinfo{person}{Jiran
  Hou}, \bibinfo{person}{Eunkyung Kim}, {and} \bibinfo{person}{David~H.
  Gustafson}.} \bibinfo{year}{2014}\natexlab{}.
\newblock \showarticletitle{Lurking as an {Active} {Participation} {Process}:
  {A} {Longitudinal} {Investigation} of {Engagement} with an {Online} {Cancer}
  {Support} {Group}}.
\newblock \bibinfo{journal}{\emph{Health Commun}} \bibinfo{volume}{29},
  \bibinfo{number}{9} (\bibinfo{year}{2014}), \bibinfo{pages}{911--923}.
\newblock
\showISSN{1041-0236}
\urldef\tempurl%
\url{https://doi.org/10.1080/10410236.2013.816911}
\showDOI{\tempurl}


\bibitem[\protect\citeauthoryear{Hargittai}{Hargittai}{2018}]%
        {hargittai_potential_2018}
\bibfield{author}{\bibinfo{person}{Eszter Hargittai}.}
  \bibinfo{year}{2018}\natexlab{}.
\newblock \showarticletitle{Potential {Biases} in {Big} {Data}: {Omitted}
  {Voices} on {Social} {Media}}.
\newblock \bibinfo{journal}{\emph{Social Science Computer Review}}
  (\bibinfo{date}{July} \bibinfo{year}{2018}),
  \bibinfo{pages}{0894439318788322}.
\newblock
\showISSN{0894-4393}
\urldef\tempurl%
\url{https://doi.org/10.1177/0894439318788322}
\showDOI{\tempurl}


\bibitem[\protect\citeauthoryear{Hartzler, Taylor, Park, Griffiths, Backonja,
  McDonald, Wahbeh, Brown, and Pratt}{Hartzler et~al\mbox{.}}{2016}]%
        {hartzler_leveraging_2016}
\bibfield{author}{\bibinfo{person}{Andrea~L Hartzler}, \bibinfo{person}{Megan~N
  Taylor}, \bibinfo{person}{Albert Park}, \bibinfo{person}{Troy Griffiths},
  \bibinfo{person}{Uba Backonja}, \bibinfo{person}{David~W McDonald},
  \bibinfo{person}{Sam Wahbeh}, \bibinfo{person}{Cory Brown}, {and}
  \bibinfo{person}{Wanda Pratt}.} \bibinfo{year}{2016}\natexlab{}.
\newblock \showarticletitle{Leveraging cues from person-generated health data
  for peer matching in online communities}.
\newblock \bibinfo{journal}{\emph{J Am Med Inform Assoc}} \bibinfo{volume}{23},
  \bibinfo{number}{3} (\bibinfo{date}{May} \bibinfo{year}{2016}),
  \bibinfo{pages}{496--507}.
\newblock
\showISSN{1067-5027}
\urldef\tempurl%
\url{https://doi.org/10.1093/jamia/ocv175}
\showDOI{\tempurl}


\bibitem[\protect\citeauthoryear{Hearst and Dumais}{Hearst and Dumais}{2009}]%
        {hearst_blogging_2009}
\bibfield{author}{\bibinfo{person}{Marti~A. Hearst} {and}
  \bibinfo{person}{Susan~T. Dumais}.} \bibinfo{year}{2009}\natexlab{}.
\newblock \showarticletitle{Blogging {Together}: {An} {Examination} of {Group}
  {Blogs}}. In \bibinfo{booktitle}{\emph{Third {International} {AAAI}
  {Conference} on {Weblogs} and {Social} {Media}}}.
\newblock
\urldef\tempurl%
\url{https://www.aaai.org/ocs/index.php/ICWSM/09/paper/view/182}
\showURL{%
\tempurl}


\bibitem[\protect\citeauthoryear{Hill, Weinert, and Cudney}{Hill
  et~al\mbox{.}}{2006}]%
        {hill_influence_2006}
\bibfield{author}{\bibinfo{person}{Wade Hill}, \bibinfo{person}{Clarann
  Weinert}, {and} \bibinfo{person}{Shirley Cudney}.}
  \bibinfo{year}{2006}\natexlab{}.
\newblock \showarticletitle{Influence of a {Computer} {Intervention} on the
  {Psychological} {Status} of {Chronically} {III} {Rural} {Women}}.
\newblock \bibinfo{journal}{\emph{Nurs Res}} \bibinfo{volume}{55},
  \bibinfo{number}{1} (\bibinfo{year}{2006}), \bibinfo{pages}{34--42}.
\newblock
\showISSN{0029-6562}
\urldef\tempurl%
\url{https://www.ncbi.nlm.nih.gov/pmc/articles/PMC1484522/}
\showURL{%
\tempurl}


\bibitem[\protect\citeauthoryear{Hlavac}{Hlavac}{2018}]%
        {stargazer_package}
\bibfield{author}{\bibinfo{person}{Marek Hlavac}.}
  \bibinfo{year}{2018}\natexlab{}.
\newblock \bibinfo{booktitle}{\emph{stargazer: Well-Formatted Regression and
  Summary Statistics Tables}}.
\newblock Central European Labour Studies Institute (CELSI), Bratislava,
  Slovakia.
\newblock
\urldef\tempurl%
\url{https://CRAN.R-project.org/package=stargazer}
\showURL{%
\tempurl}
\newblock
\shownote{R package version 5.2.2.}


\bibitem[\protect\citeauthoryear{Holbrey and Coulson}{Holbrey and
  Coulson}{2013}]%
        {holbrey_qualitative_2013}
\bibfield{author}{\bibinfo{person}{Sarah Holbrey} {and}
  \bibinfo{person}{Neil~S. Coulson}.} \bibinfo{year}{2013}\natexlab{}.
\newblock \showarticletitle{A qualitative investigation of the impact of peer
  to peer online support for women living with {Polycystic} {Ovary}
  {Syndrome}}.
\newblock \bibinfo{journal}{\emph{BMC Women's Health}} \bibinfo{volume}{13},
  \bibinfo{number}{1} (\bibinfo{date}{Dec.} \bibinfo{year}{2013}),
  \bibinfo{pages}{51}.
\newblock
\showISSN{1472-6874}
\urldef\tempurl%
\url{https://doi.org/10.1186/1472-6874-13-51}
\showDOI{\tempurl}


\bibitem[\protect\citeauthoryear{Holt-Lunstad, Smith, and Layton}{Holt-Lunstad
  et~al\mbox{.}}{2010}]%
        {holt-lunstad_social_2010}
\bibfield{author}{\bibinfo{person}{Julianne Holt-Lunstad},
  \bibinfo{person}{Timothy~B. Smith}, {and} \bibinfo{person}{J.~Bradley
  Layton}.} \bibinfo{year}{2010}\natexlab{}.
\newblock \showarticletitle{Social {Relationships} and {Mortality} {Risk}: {A}
  {Meta}-analytic {Review}}.
\newblock \bibinfo{journal}{\emph{PLOS Medicine}} \bibinfo{volume}{7},
  \bibinfo{number}{7} (\bibinfo{date}{July} \bibinfo{year}{2010}),
  \bibinfo{pages}{e1000316}.
\newblock
\showISSN{1549-1676}
\urldef\tempurl%
\url{https://doi.org/10.1371/journal.pmed.1000316}
\showDOI{\tempurl}


\bibitem[\protect\citeauthoryear{Hsiu-Chia and Feng-Yang}{Hsiu-Chia and
  Feng-Yang}{2009}]%
        {hsiu-chia_can_2009}
\bibfield{author}{\bibinfo{person}{Ko Hsiu-Chia} {and} \bibinfo{person}{Kuo
  Feng-Yang}.} \bibinfo{year}{2009}\natexlab{}.
\newblock \showarticletitle{Can {Blogging} {Enhance} {Subjective}
  {Well}-{Being} {Through} {Self}-{Disclosure}?}
\newblock \bibinfo{journal}{\emph{CyberPsychology \& Behavior}}
  \bibinfo{volume}{12}, \bibinfo{number}{1} (\bibinfo{date}{Feb.}
  \bibinfo{year}{2009}), \bibinfo{pages}{75--79}.
\newblock
\showISSN{1094-9313}
\urldef\tempurl%
\url{https://doi.org/10.1089/cpb.2008.016}
\showDOI{\tempurl}


\bibitem[\protect\citeauthoryear{Hu, Kung, Rummans, Clark, and Lapid}{Hu
  et~al\mbox{.}}{2015}]%
        {hu_reducing_2015}
\bibfield{author}{\bibinfo{person}{Chunling Hu}, \bibinfo{person}{Simon Kung},
  \bibinfo{person}{Teresa~A. Rummans}, \bibinfo{person}{Matthew~M. Clark},
  {and} \bibinfo{person}{Maria~I. Lapid}.} \bibinfo{year}{2015}\natexlab{}.
\newblock \showarticletitle{Reducing caregiver stress with internet-based
  interventions: a systematic review of open-label and randomized controlled
  trials}.
\newblock \bibinfo{journal}{\emph{J Am Med Inform Assoc}} \bibinfo{volume}{22},
  \bibinfo{number}{e1} (\bibinfo{date}{April} \bibinfo{year}{2015}),
  \bibinfo{pages}{e194--e209}.
\newblock
\showISSN{1067-5027}
\urldef\tempurl%
\url{https://doi.org/10.1136/amiajnl-2014-002817}
\showDOI{\tempurl}


\bibitem[\protect\citeauthoryear{Huh and Ackerman}{Huh and Ackerman}{2012}]%
        {huh_collaborative_2012}
\bibfield{author}{\bibinfo{person}{Jina Huh} {and} \bibinfo{person}{Mark~S.
  Ackerman}.} \bibinfo{year}{2012}\natexlab{}.
\newblock \showarticletitle{Collaborative {Help} in {Chronic} {Disease}
  {Management}: {Supporting} {Individualized} {Problems}}. In
  \bibinfo{booktitle}{\emph{Proceedings of the {ACM} 2012 {Conference} on
  {Computer} {Supported} {Cooperative} {Work}}} \emph{(\bibinfo{series}{{CSCW}
  '12})}. \bibinfo{publisher}{ACM}, \bibinfo{address}{New York, NY, USA},
  \bibinfo{pages}{853--862}.
\newblock
\showISBNx{978-1-4503-1086-4}
\urldef\tempurl%
\url{https://doi.org/10.1145/2145204.2145331}
\showDOI{\tempurl}


\bibitem[\protect\citeauthoryear{Hunter}{Hunter}{2007}]%
        {matplotlib}
\bibfield{author}{\bibinfo{person}{J.~D. Hunter}.}
  \bibinfo{year}{2007}\natexlab{}.
\newblock \showarticletitle{Matplotlib: A 2D graphics environment}.
\newblock \bibinfo{journal}{\emph{Computing in Science \& Engineering}}
  \bibinfo{volume}{9}, \bibinfo{number}{3} (\bibinfo{year}{2007}),
  \bibinfo{pages}{90--95}.
\newblock
\urldef\tempurl%
\url{https://doi.org/10.1109/MCSE.2007.55}
\showDOI{\tempurl}


\bibitem[\protect\citeauthoryear{Høybye, Johansen, and
  Tjørnhøj‐Thomsen}{Høybye et~al\mbox{.}}{2005}]%
        {hoybye_online_2005}
\bibfield{author}{\bibinfo{person}{Mette~Terp Høybye},
  \bibinfo{person}{Christoffer Johansen}, {and} \bibinfo{person}{Tine
  Tjørnhøj‐Thomsen}.} \bibinfo{year}{2005}\natexlab{}.
\newblock \showarticletitle{Online interaction. {Effects} of storytelling in an
  internet breast cancer support group}.
\newblock \bibinfo{journal}{\emph{Psycho-Oncology}} \bibinfo{volume}{14},
  \bibinfo{number}{3} (\bibinfo{year}{2005}), \bibinfo{pages}{211--220}.
\newblock
\showISSN{1099-1611}
\urldef\tempurl%
\url{https://doi.org/10.1002/pon.837}
\showDOI{\tempurl}


\bibitem[\protect\citeauthoryear{Iannarino}{Iannarino}{2014}]%
        {iannarino_social_2014}
\bibfield{author}{\bibinfo{person}{Nicholas Iannarino}.}
  \bibinfo{year}{2014}\natexlab{}.
\newblock \emph{\bibinfo{title}{Social {Support} in {Young} {Adult} {Cancer}
  {Survivors} and {Their} {Close} {Social} {Network} {Members}}}.
\newblock {PhD} {Thesis}. \bibinfo{school}{University of Kentucky}.
\newblock
\urldef\tempurl%
\url{https://uknowledge.uky.edu/comm_etds/27}
\showURL{%
\tempurl}


\bibitem[\protect\citeauthoryear{Introne, Semaan, and Goggins}{Introne
  et~al\mbox{.}}{2016}]%
        {introne_sociotechnical_2016}
\bibfield{author}{\bibinfo{person}{Joshua Introne}, \bibinfo{person}{Bryan
  Semaan}, {and} \bibinfo{person}{Sean Goggins}.}
  \bibinfo{year}{2016}\natexlab{}.
\newblock \showarticletitle{A {Sociotechnical} {Mechanism} for {Online}
  {Support} {Provision}}. In \bibinfo{booktitle}{\emph{Proceedings of the 2016
  {CHI} {Conference} on {Human} {Factors} in {Computing} {Systems}}}
  \emph{(\bibinfo{series}{{CHI} '16})}. \bibinfo{publisher}{ACM},
  \bibinfo{address}{New York, NY, USA}, \bibinfo{pages}{3559--3571}.
\newblock
\showISBNx{978-1-4503-3362-7}
\urldef\tempurl%
\url{https://doi.org/10.1145/2858036.2858582}
\showDOI{\tempurl}
\newblock
\shownote{event-place: San Jose, California, USA.}


\bibitem[\protect\citeauthoryear{Jackson}{Jackson}{2010}]%
        {jackson_social_2010}
\bibfield{author}{\bibinfo{person}{Matthew~O. Jackson}.}
  \bibinfo{year}{2010}\natexlab{}.
\newblock \bibinfo{booktitle}{\emph{Social and {Economic} {Networks}}}.
\newblock
\showISBNx{978-0-691-14820-5}
\urldef\tempurl%
\url{https://press.princeton.edu/books/paperback/9780691148205/social-and-economic-networks}
\showURL{%
\tempurl}


\bibitem[\protect\citeauthoryear{Jacobs, Cramer, and Barkhuus}{Jacobs
  et~al\mbox{.}}{2016}]%
        {jacobs_caring_2016}
\bibfield{author}{\bibinfo{person}{Maia Jacobs}, \bibinfo{person}{Henriette
  Cramer}, {and} \bibinfo{person}{Louise Barkhuus}.}
  \bibinfo{year}{2016}\natexlab{}.
\newblock \showarticletitle{Caring {About} {Sharing}: {Couples}' {Practices} in
  {Single} {User} {Device} {Access}}. In \bibinfo{booktitle}{\emph{Proceedings
  of the 19th {International} {Conference} on {Supporting} {Group} {Work}}}
  \emph{(\bibinfo{series}{{GROUP} '16})}. \bibinfo{publisher}{ACM},
  \bibinfo{address}{New York, NY, USA}, \bibinfo{pages}{235--243}.
\newblock
\showISBNx{978-1-4503-4276-6}
\urldef\tempurl%
\url{https://doi.org/10.1145/2957276.2957296}
\showDOI{\tempurl}
\newblock
\shownote{event-place: Sanibel Island, Florida, USA.}


\bibitem[\protect\citeauthoryear{Jacobs, Johnson, and Mynatt}{Jacobs
  et~al\mbox{.}}{2018}]%
        {jacobs_mypath:_2018}
\bibfield{author}{\bibinfo{person}{Maia Jacobs}, \bibinfo{person}{Jeremy
  Johnson}, {and} \bibinfo{person}{Elizabeth~D. Mynatt}.}
  \bibinfo{year}{2018}\natexlab{}.
\newblock \showarticletitle{{MyPath}: {Investigating} {Breast} {Cancer}
  {Patients}' {Use} of {Personalized} {Health} {Information}}.
\newblock \bibinfo{journal}{\emph{Proc. ACM Hum.-Comput. Interact.}}
  \bibinfo{volume}{2}, \bibinfo{number}{CSCW} (\bibinfo{date}{Nov.}
  \bibinfo{year}{2018}), \bibinfo{pages}{78:1--78:21}.
\newblock
\showISSN{2573-0142}
\urldef\tempurl%
\url{https://doi.org/10.1145/3274347}
\showDOI{\tempurl}


\bibitem[\protect\citeauthoryear{Jha and Elhadad}{Jha and Elhadad}{2010}]%
        {jha_cancer_2010}
\bibfield{author}{\bibinfo{person}{Mukund Jha} {and} \bibinfo{person}{Noemie
  Elhadad}.} \bibinfo{year}{2010}\natexlab{}.
\newblock \showarticletitle{Cancer {Stage} {Prediction} {Based} on {Patient}
  {Online} {Discourse}}. In \bibinfo{booktitle}{\emph{Proceedings of the 2010
  {Workshop} on {Biomedical} {Natural} {Language} {Processing}, {ACL} 2010}}.
  \bibinfo{address}{Sweden}, \bibinfo{pages}{64--71}.
\newblock


\bibitem[\protect\citeauthoryear{Kaltenbaugh, Klem, Hu, Turi, Haines, and
  Lingler}{Kaltenbaugh et~al\mbox{.}}{2015}]%
        {kaltenbaugh_using_2015}
\bibfield{author}{\bibinfo{person}{Donna~J. Kaltenbaugh},
  \bibinfo{person}{Mary~Lou Klem}, \bibinfo{person}{Lu Hu},
  \bibinfo{person}{Eleanor Turi}, \bibinfo{person}{Alice~J. Haines}, {and}
  \bibinfo{person}{Jennifer~Hagerty Lingler}.} \bibinfo{year}{2015}\natexlab{}.
\newblock \showarticletitle{Using {Web}-{Based} {Interventions} to {Support}
  {Caregivers} of {Patients} {With} {Cancer}: {A} {Systematic} {Review}}.
\newblock \bibinfo{journal}{\emph{Oncology Nursing Forum}}
  \bibinfo{volume}{42}, \bibinfo{number}{2} (\bibinfo{date}{Feb.}
  \bibinfo{year}{2015}), \bibinfo{pages}{156--164}.
\newblock
\urldef\tempurl%
\url{https://doi.org/10.1188/15.ONF.156-164}
\showDOI{\tempurl}


\bibitem[\protect\citeauthoryear{Kim, Han, Moon, Shaw, Shah, McTavish, and
  Gustafson}{Kim et~al\mbox{.}}{2012}]%
        {kim_process_2012}
\bibfield{author}{\bibinfo{person}{Eunkyung Kim}, \bibinfo{person}{Jeong~Yeob
  Han}, \bibinfo{person}{Tae~Joon Moon}, \bibinfo{person}{Bret Shaw},
  \bibinfo{person}{Dhavan~V. Shah}, \bibinfo{person}{Fiona~M. McTavish}, {and}
  \bibinfo{person}{David~H. Gustafson}.} \bibinfo{year}{2012}\natexlab{}.
\newblock \showarticletitle{The process and effect of supportive message
  expression and reception in online breast cancer support groups}.
\newblock \bibinfo{journal}{\emph{Psycho-Oncology}} \bibinfo{volume}{21},
  \bibinfo{number}{5} (\bibinfo{date}{May} \bibinfo{year}{2012}),
  \bibinfo{pages}{531--540}.
\newblock
\showISSN{1099-1611}
\urldef\tempurl%
\url{https://doi.org/10.1002/pon.1942}
\showDOI{\tempurl}


\bibitem[\protect\citeauthoryear{Lakey and Lutz}{Lakey and Lutz}{1996}]%
        {lakey_social_1996}
\bibfield{author}{\bibinfo{person}{Brian Lakey} {and}
  \bibinfo{person}{Catherine~J. Lutz}.} \bibinfo{year}{1996}\natexlab{}.
\newblock \showarticletitle{Social {Support} and {Preventive} and {Therapeutic}
  {Interventions}}.
\newblock In \bibinfo{booktitle}{\emph{Handbook of {Social} {Support} and the
  {Family}}}, \bibfield{editor}{\bibinfo{person}{Gregory~R. Pierce},
  \bibinfo{person}{Barbara~R. Sarason}, {and} \bibinfo{person}{Irwin~G.
  Sarason}} (Eds.). \bibinfo{publisher}{Springer US}, \bibinfo{address}{Boston,
  MA}, \bibinfo{pages}{435--465}.
\newblock
\showISBNx{978-1-4899-1388-3}
\urldef\tempurl%
\url{https://doi.org/10.1007/978-1-4899-1388-3_18}
\showDOI{\tempurl}


\bibitem[\protect\citeauthoryear{Lampinen}{Lampinen}{2014}]%
        {lampinen_account_2014}
\bibfield{author}{\bibinfo{person}{Airi M~I Lampinen}.}
  \bibinfo{year}{2014}\natexlab{}.
\newblock \showarticletitle{Account {Sharing} in the {Context} of {Networked}
  {Hospitality} {Exchange}}. In \bibinfo{booktitle}{\emph{Proceedings of the
  17th {ACM} {Conference} on {Computer} {Supported} {Cooperative} {Work} \&
  {Social} {Computing}}} \emph{(\bibinfo{series}{{CSCW} '14})}.
  \bibinfo{publisher}{ACM}, \bibinfo{address}{New York, NY, USA},
  \bibinfo{pages}{499--504}.
\newblock
\showISBNx{978-1-4503-2540-0}
\urldef\tempurl%
\url{https://doi.org/10.1145/2531602.2531665}
\showDOI{\tempurl}
\newblock
\shownote{event-place: Baltimore, Maryland, USA.}


\bibitem[\protect\citeauthoryear{Landis and Koch}{Landis and Koch}{1977}]%
        {landis_measurement_1977}
\bibfield{author}{\bibinfo{person}{J.~Richard Landis} {and}
  \bibinfo{person}{Gary~G. Koch}.} \bibinfo{year}{1977}\natexlab{}.
\newblock \showarticletitle{The {Measurement} of {Observer} {Agreement} for
  {Categorical} {Data}}.
\newblock \bibinfo{journal}{\emph{Biometrics}} \bibinfo{volume}{33},
  \bibinfo{number}{1} (\bibinfo{year}{1977}), \bibinfo{pages}{159--174}.
\newblock
\showISSN{0006-341X}
\urldef\tempurl%
\url{https://doi.org/10.2307/2529310}
\showDOI{\tempurl}


\bibitem[\protect\citeauthoryear{Latapy, Magnien, and Vecchio}{Latapy
  et~al\mbox{.}}{2008}]%
        {latapy_basic_2008}
\bibfield{author}{\bibinfo{person}{Matthieu Latapy}, \bibinfo{person}{Clémence
  Magnien}, {and} \bibinfo{person}{Nathalie~Del Vecchio}.}
  \bibinfo{year}{2008}\natexlab{}.
\newblock \showarticletitle{Basic notions for the analysis of large two-mode
  networks}.
\newblock \bibinfo{journal}{\emph{Social Networks}} \bibinfo{volume}{30},
  \bibinfo{number}{1} (\bibinfo{date}{Jan.} \bibinfo{year}{2008}),
  \bibinfo{pages}{31--48}.
\newblock
\showISSN{0378-8733}
\urldef\tempurl%
\url{https://doi.org/10.1016/j.socnet.2007.04.006}
\showDOI{\tempurl}


\bibitem[\protect\citeauthoryear{Levonian, Erikson, Luo, Narayanan, Rubya,
  Vachher, Terveen, and Yarosh}{Levonian et~al\mbox{.}}{2020}]%
        {levonian_bridging_2020}
\bibfield{author}{\bibinfo{person}{Zachary Levonian},
  \bibinfo{person}{Drew~Richard Erikson}, \bibinfo{person}{Wenqi Luo},
  \bibinfo{person}{Saumik Narayanan}, \bibinfo{person}{Sabirat Rubya},
  \bibinfo{person}{Prateek Vachher}, \bibinfo{person}{Loren Terveen}, {and}
  \bibinfo{person}{Svetlana Yarosh}.} \bibinfo{year}{2020}\natexlab{}.
\newblock \showarticletitle{Bridging {Qualitative} and {Quantitative} {Methods}
  for {User} {Modeling}: {Tracing} {Cancer} {Patient} {Behavior} in an {Online}
  {Health} {Community}}. In \bibinfo{booktitle}{\emph{Proceedings of the
  {Fourteenth} {International} {Conference} on {Web} and {Social} {Media}}}.
  \bibinfo{publisher}{AAAI}, \bibinfo{address}{Atlanta, GA},
  \bibinfo{pages}{12}.
\newblock


\bibitem[\protect\citeauthoryear{Li, Levonian, Ma, and Yarosh}{Li
  et~al\mbox{.}}{2018}]%
        {li_condition_2018}
\bibfield{author}{\bibinfo{person}{Changye Li}, \bibinfo{person}{Zachary
  Levonian}, \bibinfo{person}{Haiwei Ma}, {and} \bibinfo{person}{Svetlana
  Yarosh}.} \bibinfo{year}{2018}\natexlab{}.
\newblock \showarticletitle{Condition {Unknown}: {Predicting} {Patients}’
  {Health} {Conditions} in an {Online} {Health} {Community}}.
\newblock \bibinfo{journal}{\emph{CSCW}} (\bibinfo{date}{Nov.}
  \bibinfo{year}{2018}), \bibinfo{pages}{4}.
\newblock


\bibitem[\protect\citeauthoryear{Lipton, Wang, and Smola}{Lipton
  et~al\mbox{.}}{2018}]%
        {lipton_detecting_2018}
\bibfield{author}{\bibinfo{person}{Zachary~C. Lipton},
  \bibinfo{person}{Yu-Xiang Wang}, {and} \bibinfo{person}{Alex Smola}.}
  \bibinfo{year}{2018}\natexlab{}.
\newblock \showarticletitle{Detecting and {Correcting} for {Label} {Shift} with
  {Black} {Box} {Predictors}}.
\newblock \bibinfo{journal}{\emph{arXiv:1802.03916 [cs, stat]}}
  (\bibinfo{date}{Feb.} \bibinfo{year}{2018}).
\newblock
\urldef\tempurl%
\url{http://arxiv.org/abs/1802.03916}
\showURL{%
\tempurl}
\newblock
\shownote{arXiv: 1802.03916.}


\bibitem[\protect\citeauthoryear{Lu, Wu, Liu, Li, and Zhang}{Lu
  et~al\mbox{.}}{2017}]%
        {lu_understanding_2017}
\bibfield{author}{\bibinfo{person}{Yingjie Lu}, \bibinfo{person}{Yang Wu},
  \bibinfo{person}{Jingfang Liu}, \bibinfo{person}{Jia Li}, {and}
  \bibinfo{person}{Pengzhu Zhang}.} \bibinfo{year}{2017}\natexlab{}.
\newblock \showarticletitle{Understanding {Health} {Care} {Social} {Media}
  {Use} {From} {Different} {Stakeholder} {Perspectives}: {A} {Content}
  {Analysis} of an {Online} {Health} {Community}}.
\newblock \bibinfo{journal}{\emph{Journal of Medical Internet Research}}
  \bibinfo{volume}{19}, \bibinfo{number}{4} (\bibinfo{year}{2017}),
  \bibinfo{pages}{e109}.
\newblock
\urldef\tempurl%
\url{https://doi.org/10.2196/jmir.7087}
\showDOI{\tempurl}
\newblock
\shownote{Company: Journal of Medical Internet Research Distributor: Journal of
  Medical Internet Research Institution: Journal of Medical Internet Research
  Label: Journal of Medical Internet Research Publisher: JMIR Publications
  Inc., Toronto, Canada.}


\bibitem[\protect\citeauthoryear{Ma, Smith, He, Narayanan, Giaquinto, Evans,
  Hanson, and Yarosh}{Ma et~al\mbox{.}}{2017}]%
        {ma_write_2017}
\bibfield{author}{\bibinfo{person}{Haiwei Ma}, \bibinfo{person}{C.~Estelle
  Smith}, \bibinfo{person}{Lu He}, \bibinfo{person}{Saumik Narayanan},
  \bibinfo{person}{Robert~A. Giaquinto}, \bibinfo{person}{Roni Evans},
  \bibinfo{person}{Linda Hanson}, {and} \bibinfo{person}{Svetlana Yarosh}.}
  \bibinfo{year}{2017}\natexlab{}.
\newblock \showarticletitle{Write for {Life}: {Persisting} in {Online} {Health}
  {Communities} {Through} {Expressive} {Writing} and {Social} {Support}}.
\newblock \bibinfo{journal}{\emph{Proc. ACM Hum.-Comput. Interact.}}
  \bibinfo{volume}{1}, \bibinfo{number}{CSCW} (\bibinfo{date}{Dec.}
  \bibinfo{year}{2017}), \bibinfo{pages}{73:1--73:24}.
\newblock
\showISSN{2573-0142}
\urldef\tempurl%
\url{https://doi.org/10.1145/3134708}
\showDOI{\tempurl}


\bibitem[\protect\citeauthoryear{Malik and Coulson}{Malik and Coulson}{2008}]%
        {malik_computer-mediated_2008}
\bibfield{author}{\bibinfo{person}{Sumaira~H. Malik} {and}
  \bibinfo{person}{Neil~S. Coulson}.} \bibinfo{year}{2008}\natexlab{}.
\newblock \showarticletitle{Computer-mediated infertility support groups: {An}
  exploratory study of online experiences}.
\newblock \bibinfo{journal}{\emph{Patient Education and Counseling}}
  \bibinfo{volume}{73}, \bibinfo{number}{1} (\bibinfo{date}{Oct.}
  \bibinfo{year}{2008}), \bibinfo{pages}{105--113}.
\newblock
\showISSN{0738-3991}
\urldef\tempurl%
\url{https://doi.org/10.1016/j.pec.2008.05.024}
\showDOI{\tempurl}


\bibitem[\protect\citeauthoryear{Maloney-Krichmar and Preece}{Maloney-Krichmar
  and Preece}{2005}]%
        {maloney-krichmar_multilevel_2005}
\bibfield{author}{\bibinfo{person}{Diane Maloney-Krichmar} {and}
  \bibinfo{person}{Jenny Preece}.} \bibinfo{year}{2005}\natexlab{}.
\newblock \showarticletitle{A {Multilevel} {Analysis} of {Sociability},
  {Usability}, and {Community} {Dynamics} in an {Online} {Health} {Community}}.
\newblock \bibinfo{journal}{\emph{ACM Trans. Comput.-Hum. Interact.}}
  \bibinfo{volume}{12}, \bibinfo{number}{2} (\bibinfo{date}{June}
  \bibinfo{year}{2005}), \bibinfo{pages}{201--232}.
\newblock
\showISSN{1073-0516}
\urldef\tempurl%
\url{https://doi.org/10.1145/1067860.1067864}
\showDOI{\tempurl}


\bibitem[\protect\citeauthoryear{Markham}{Markham}{2012}]%
        {markham_fabrication_2012}
\bibfield{author}{\bibinfo{person}{Annette Markham}.}
  \bibinfo{year}{2012}\natexlab{}.
\newblock \showarticletitle{Fabrication as {Ethical} {Practice}}.
\newblock \bibinfo{journal}{\emph{Information, Communication \& Society}}
  \bibinfo{volume}{15}, \bibinfo{number}{3} (\bibinfo{date}{April}
  \bibinfo{year}{2012}), \bibinfo{pages}{334--353}.
\newblock
\showISSN{1369-118X}
\urldef\tempurl%
\url{https://doi.org/10.1080/1369118X.2011.641993}
\showDOI{\tempurl}
\newblock
\shownote{Publisher: Routledge \_eprint:
  https://doi.org/10.1080/1369118X.2011.641993.}


\bibitem[\protect\citeauthoryear{Massimi, Bender, Witteman, and Ahmed}{Massimi
  et~al\mbox{.}}{2014}]%
        {massimi_life_2014}
\bibfield{author}{\bibinfo{person}{Michael Massimi}, \bibinfo{person}{Jackie~L.
  Bender}, \bibinfo{person}{Holly~O. Witteman}, {and} \bibinfo{person}{Osman~H.
  Ahmed}.} \bibinfo{year}{2014}\natexlab{}.
\newblock \showarticletitle{Life {Transitions} and {Online} {Health}
  {Communities}: {Reflecting} on {Adoption}, {Use}, and {Disengagement}}. In
  \bibinfo{booktitle}{\emph{Proceedings of the 17th {ACM} {Conference} on
  {Computer} {Supported} {Cooperative} {Work} \& {Social} {Computing}}}
  \emph{(\bibinfo{series}{{CSCW} '14})}. \bibinfo{publisher}{ACM},
  \bibinfo{address}{New York, NY, USA}, \bibinfo{pages}{1491--1501}.
\newblock
\showISBNx{978-1-4503-2540-0}
\urldef\tempurl%
\url{https://doi.org/10.1145/2531602.2531622}
\showDOI{\tempurl}


\bibitem[\protect\citeauthoryear{McCosker and Darcy}{McCosker and
  Darcy}{2013}]%
        {mccosker_living_2013}
\bibfield{author}{\bibinfo{person}{Anthony McCosker} {and}
  \bibinfo{person}{Raya Darcy}.} \bibinfo{year}{2013}\natexlab{}.
\newblock \showarticletitle{Living with {Cancer}}.
\newblock \bibinfo{journal}{\emph{Information, Communication \& Society}}
  \bibinfo{volume}{16}, \bibinfo{number}{8} (\bibinfo{date}{Oct.}
  \bibinfo{year}{2013}), \bibinfo{pages}{1266--1285}.
\newblock
\showISSN{1369-118X}
\urldef\tempurl%
\url{https://doi.org/10.1080/1369118X.2012.758303}
\showDOI{\tempurl}


\bibitem[\protect\citeauthoryear{McDonald, Schoenebeck, and Forte}{McDonald
  et~al\mbox{.}}{2019}]%
        {mcdonald_reliability_2019}
\bibfield{author}{\bibinfo{person}{Nora McDonald}, \bibinfo{person}{Sarita
  Schoenebeck}, {and} \bibinfo{person}{Andrea Forte}.}
  \bibinfo{year}{2019}\natexlab{}.
\newblock \showarticletitle{Reliability and {Inter}-rater {Reliability} in
  {Qualitative} {Research}}.
\newblock \bibinfo{journal}{\emph{Proceedings of the ACM on Human-Computer
  Interaction}} \bibinfo{volume}{3}, \bibinfo{number}{CSCW}
  (\bibinfo{date}{Nov.} \bibinfo{year}{2019}).
\newblock
\urldef\tempurl%
\url{https://doi.org/10.1145/3359174}
\showDOI{\tempurl}


\bibitem[\protect\citeauthoryear{McKinney}{McKinney}{2010}]%
        {pandas}
\bibfield{author}{\bibinfo{person}{Wes McKinney}.}
  \bibinfo{year}{2010}\natexlab{}.
\newblock \showarticletitle{Data Structures for Statistical Computing in
  Python}. In \bibinfo{booktitle}{\emph{Proceedings of the 9th Python in
  Science Conference}}, \bibfield{editor}{\bibinfo{person}{St\'efan van~der
  Walt} {and} \bibinfo{person}{Jarrod Millman}} (Eds.). \bibinfo{pages}{51 --
  56}.
\newblock


\bibitem[\protect\citeauthoryear{Meleis}{Meleis}{2015}]%
        {meleis_transitions_2015}
\bibfield{author}{\bibinfo{person}{Afaf~lbrahim Meleis}.}
  \bibinfo{year}{2015}\natexlab{}.
\newblock \showarticletitle{Transitions {Theory}}.
\newblock In \bibinfo{booktitle}{\emph{Nursing theories \& nursing practice}
  (\bibinfo{edition}{4} ed.)}, \bibfield{editor}{\bibinfo{person}{Marlaine~C.
  Smith} {and} \bibinfo{person}{Marilyn~E. Parker}} (Eds.).
  \bibinfo{publisher}{F.A. Davis Company}, \bibinfo{address}{Philadelphia, PA},
  \bibinfo{pages}{361--380}.
\newblock
\showISBNx{978-0-8036-3312-4}


\bibitem[\protect\citeauthoryear{Meng}{Meng}{2016}]%
        {meng_your_2016}
\bibfield{author}{\bibinfo{person}{Jingbo Meng}.}
  \bibinfo{year}{2016}\natexlab{}.
\newblock \showarticletitle{Your {Health} {Buddies} {Matter}: {Preferential}
  {Selection} and {Social} {Influence} on {Weight} {Management} in an {Online}
  {Health} {Social} {Network}}.
\newblock \bibinfo{journal}{\emph{Health Communication}} \bibinfo{volume}{31},
  \bibinfo{number}{12} (\bibinfo{date}{Dec.} \bibinfo{year}{2016}),
  \bibinfo{pages}{1460--1471}.
\newblock
\showISSN{1041-0236}
\urldef\tempurl%
\url{https://doi.org/10.1080/10410236.2015.1079760}
\showDOI{\tempurl}


\bibitem[\protect\citeauthoryear{Namkoong, DuBenske, Shaw, Gustafson, Hawkins,
  Shah, McTavish, and Cleary}{Namkoong et~al\mbox{.}}{2012}]%
        {namkoong_creating_2012}
\bibfield{author}{\bibinfo{person}{Kang Namkoong}, \bibinfo{person}{Lori~L.
  DuBenske}, \bibinfo{person}{Bret~R. Shaw}, \bibinfo{person}{David~H.
  Gustafson}, \bibinfo{person}{Robert~P. Hawkins}, \bibinfo{person}{Dhavan~V.
  Shah}, \bibinfo{person}{Fiona~M. McTavish}, {and} \bibinfo{person}{James~F.
  Cleary}.} \bibinfo{year}{2012}\natexlab{}.
\newblock \showarticletitle{Creating a {Bond} between {Caregivers} {Online}:
  {Impact} on {Caregivers}' {Coping} {Strategies}}.
\newblock \bibinfo{journal}{\emph{J Health Commun}} \bibinfo{volume}{17},
  \bibinfo{number}{2} (\bibinfo{year}{2012}), \bibinfo{pages}{125--140}.
\newblock
\showISSN{1081-0730}
\urldef\tempurl%
\url{https://doi.org/10.1080/10810730.2011.585687}
\showDOI{\tempurl}


\bibitem[\protect\citeauthoryear{Newman, Lauterbach, Munson, Resnick, and
  Morris}{Newman et~al\mbox{.}}{2011}]%
        {newman_its_2011}
\bibfield{author}{\bibinfo{person}{Mark~W. Newman}, \bibinfo{person}{Debra
  Lauterbach}, \bibinfo{person}{Sean~A. Munson}, \bibinfo{person}{Paul
  Resnick}, {and} \bibinfo{person}{Margaret~E. Morris}.}
  \bibinfo{year}{2011}\natexlab{}.
\newblock \showarticletitle{It's {Not} {That} {I} {Don}'{T} {Have} {Problems},
  {I}'{M} {Just} {Not} {Putting} {Them} on {Facebook}: {Challenges} and
  {Opportunities} in {Using} {Online} {Social} {Networks} for {Health}}. In
  \bibinfo{booktitle}{\emph{Proceedings of the {ACM} 2011 {Conference} on
  {Computer} {Supported} {Cooperative} {Work}}} \emph{(\bibinfo{series}{{CSCW}
  '11})}. \bibinfo{publisher}{ACM}, \bibinfo{address}{New York, NY, USA},
  \bibinfo{pages}{341--350}.
\newblock
\showISBNx{978-1-4503-0556-3}
\urldef\tempurl%
\url{https://doi.org/10.1145/1958824.1958876}
\showDOI{\tempurl}


\bibitem[\protect\citeauthoryear{Oh, Ozkaya, and LaRose}{Oh
  et~al\mbox{.}}{2014}]%
        {oh_how_2014}
\bibfield{author}{\bibinfo{person}{Hyun~Jung Oh}, \bibinfo{person}{Elif
  Ozkaya}, {and} \bibinfo{person}{Robert LaRose}.}
  \bibinfo{year}{2014}\natexlab{}.
\newblock \showarticletitle{How does online social networking enhance life
  satisfaction? {The} relationships among online supportive interaction,
  affect, perceived social support, sense of community, and life satisfaction}.
\newblock \bibinfo{journal}{\emph{Computers in Human Behavior}}
  \bibinfo{volume}{30} (\bibinfo{date}{Jan.} \bibinfo{year}{2014}),
  \bibinfo{pages}{69--78}.
\newblock
\showISSN{0747-5632}
\urldef\tempurl%
\url{https://doi.org/10.1016/j.chb.2013.07.053}
\showDOI{\tempurl}


\bibitem[\protect\citeauthoryear{O'Leary, Bhattacharya, Munson, Wobbrock, and
  Pratt}{O'Leary et~al\mbox{.}}{2017}]%
        {oleary_design_2017}
\bibfield{author}{\bibinfo{person}{Kathleen O'Leary}, \bibinfo{person}{Arpita
  Bhattacharya}, \bibinfo{person}{Sean~A. Munson}, \bibinfo{person}{Jacob~O.
  Wobbrock}, {and} \bibinfo{person}{Wanda Pratt}.}
  \bibinfo{year}{2017}\natexlab{}.
\newblock \showarticletitle{Design {Opportunities} for {Mental} {Health} {Peer}
  {Support} {Technologies}}. In \bibinfo{booktitle}{\emph{Proceedings of the
  2017 {ACM} {Conference} on {Computer} {Supported} {Cooperative} {Work} and
  {Social} {Computing}}} \emph{(\bibinfo{series}{{CSCW} '17})}.
  \bibinfo{publisher}{ACM}, \bibinfo{address}{New York, NY, USA},
  \bibinfo{pages}{1470--1484}.
\newblock
\showISBNx{978-1-4503-4335-0}
\urldef\tempurl%
\url{https://doi.org/10.1145/2998181.2998349}
\showDOI{\tempurl}
\newblock
\shownote{event-place: Portland, Oregon, USA.}


\bibitem[\protect\citeauthoryear{Overgoor, Benson, and Ugander}{Overgoor
  et~al\mbox{.}}{2019}]%
        {overgoor_choosing_2019}
\bibfield{author}{\bibinfo{person}{Jan Overgoor}, \bibinfo{person}{Austin~R
  Benson}, {and} \bibinfo{person}{Johan Ugander}.}
  \bibinfo{year}{2019}\natexlab{}.
\newblock \showarticletitle{Choosing to {Grow} a {Graph}: {Modeling} {Network}
  {Formation} as {Discrete} {Choice}}. \bibinfo{pages}{12}.
\newblock
\urldef\tempurl%
\url{https://dl.acm.org/citation.cfm?doid=3308558.3313662}
\showURL{%
\tempurl}


\bibitem[\protect\citeauthoryear{Pan, Shen, and Feng}{Pan
  et~al\mbox{.}}{2017}]%
        {pan_you_2017}
\bibfield{author}{\bibinfo{person}{Wenjing Pan}, \bibinfo{person}{Cuihua Shen},
  {and} \bibinfo{person}{Bo Feng}.} \bibinfo{year}{2017}\natexlab{}.
\newblock \showarticletitle{You {Get} {What} {You} {Give}: {Understanding}
  {Reply} {Reciprocity} and {Social} {Capital} in {Online} {Health} {Support}
  {Forums}}.
\newblock \bibinfo{journal}{\emph{Journal of Health Communication}}
  \bibinfo{volume}{22}, \bibinfo{number}{1} (\bibinfo{date}{Jan.}
  \bibinfo{year}{2017}), \bibinfo{pages}{45--52}.
\newblock
\showISSN{1081-0730}
\urldef\tempurl%
\url{https://doi.org/10.1080/10810730.2016.1250845}
\showDOI{\tempurl}


\bibitem[\protect\citeauthoryear{Pedregosa, Varoquaux, Gramfort, Michel,
  Thirion, Grisel, Blondel, Prettenhofer, Weiss, Dubourg, Vanderplas, Passos,
  Cournapeau, Brucher, Perrot, and Duchesnay}{Pedregosa et~al\mbox{.}}{2011}]%
        {scikit-learn}
\bibfield{author}{\bibinfo{person}{F. Pedregosa}, \bibinfo{person}{G.
  Varoquaux}, \bibinfo{person}{A. Gramfort}, \bibinfo{person}{V. Michel},
  \bibinfo{person}{B. Thirion}, \bibinfo{person}{O. Grisel},
  \bibinfo{person}{M. Blondel}, \bibinfo{person}{P. Prettenhofer},
  \bibinfo{person}{R. Weiss}, \bibinfo{person}{V. Dubourg}, \bibinfo{person}{J.
  Vanderplas}, \bibinfo{person}{A. Passos}, \bibinfo{person}{D. Cournapeau},
  \bibinfo{person}{M. Brucher}, \bibinfo{person}{M. Perrot}, {and}
  \bibinfo{person}{E. Duchesnay}.} \bibinfo{year}{2011}\natexlab{}.
\newblock \showarticletitle{Scikit-learn: Machine Learning in {P}ython}.
\newblock \bibinfo{journal}{\emph{Journal of Machine Learning Research}}
  \bibinfo{volume}{12} (\bibinfo{year}{2011}), \bibinfo{pages}{2825--2830}.
\newblock


\bibitem[\protect\citeauthoryear{Peng, Guo, Tsang, and Ma}{Peng
  et~al\mbox{.}}{2020}]%
        {peng_exploring_2020}
\bibfield{author}{\bibinfo{person}{Zhenhui Peng}, \bibinfo{person}{Qingyu Guo},
  \bibinfo{person}{Ka~Wing Tsang}, {and} \bibinfo{person}{Xiaojuan Ma}.}
  \bibinfo{year}{2020}\natexlab{}.
\newblock \showarticletitle{Exploring the {Effects} of {Technological}
  {Writing} {Assistance} for {Support} {Providers} in {Online} {Mental}
  {Health} {Community}}. In \bibinfo{booktitle}{\emph{Proceedings of the 2020
  {CHI} {Conference} on {Human} {Factors} in {Computing} {Systems}}}
  \emph{(\bibinfo{series}{{CHI} '20})}. \bibinfo{publisher}{Association for
  Computing Machinery}, \bibinfo{address}{Honolulu, HI, USA},
  \bibinfo{pages}{1--15}.
\newblock
\showISBNx{978-1-4503-6708-0}
\urldef\tempurl%
\url{https://doi.org/10.1145/3313831.3376695}
\showDOI{\tempurl}


\bibitem[\protect\citeauthoryear{Poese, Uhlig, Kaafar, Donnet, and Gueye}{Poese
  et~al\mbox{.}}{2011}]%
        {poese_ip_2011}
\bibfield{author}{\bibinfo{person}{Ingmar Poese}, \bibinfo{person}{Steve
  Uhlig}, \bibinfo{person}{Mohamed~Ali Kaafar}, \bibinfo{person}{Benoit
  Donnet}, {and} \bibinfo{person}{Bamba Gueye}.}
  \bibinfo{year}{2011}\natexlab{}.
\newblock \showarticletitle{{IP} {Geolocation} {Databases}: {Unreliable}?}
\newblock \bibinfo{journal}{\emph{SIGCOMM Comput. Commun. Rev.}}
  \bibinfo{volume}{41}, \bibinfo{number}{2} (\bibinfo{date}{April}
  \bibinfo{year}{2011}), \bibinfo{pages}{53--56}.
\newblock
\showISSN{0146-4833}
\urldef\tempurl%
\url{https://doi.org/10.1145/1971162.1971171}
\showDOI{\tempurl}


\bibitem[\protect\citeauthoryear{Qiu, Zhao, Mitra, Wu, Caragea, Yen, Greer, and
  Portier}{Qiu et~al\mbox{.}}{2011}]%
        {qiu_get_2011}
\bibfield{author}{\bibinfo{person}{B. Qiu}, \bibinfo{person}{K. Zhao},
  \bibinfo{person}{P. Mitra}, \bibinfo{person}{D. Wu}, \bibinfo{person}{C.
  Caragea}, \bibinfo{person}{J. Yen}, \bibinfo{person}{G.~E. Greer}, {and}
  \bibinfo{person}{K. Portier}.} \bibinfo{year}{2011}\natexlab{}.
\newblock \showarticletitle{Get {Online} {Support}, {Feel} {Better} –
  {Sentiment} {Analysis} and {Dynamics} in an {Online} {Cancer} {Survivor}
  {Community}}. In \bibinfo{booktitle}{\emph{2011 {IEEE} {Third}
  {International} {Conference} on {Privacy}, {Security}, {Risk} and {Trust} and
  2011 {IEEE} {Third} {International} {Conference} on {Social} {Computing}}}.
  \bibinfo{pages}{274--281}.
\newblock
\urldef\tempurl%
\url{https://doi.org/10.1109/PASSAT/SocialCom.2011.127}
\showDOI{\tempurl}


\bibitem[\protect\citeauthoryear{Rains}{Rains}{2018}]%
        {rains_coping_2018}
\bibfield{author}{\bibinfo{person}{Stephen~A. Rains}.}
  \bibinfo{year}{2018}\natexlab{}.
\newblock \bibinfo{booktitle}{\emph{Coping with {Illness} {Digitally}}}.
\newblock \bibinfo{publisher}{MIT Press}.
\newblock
\showISBNx{978-0-262-34764-8}
\newblock
\shownote{Google-Books-ID: EbPkuQEACAAJ.}


\bibitem[\protect\citeauthoryear{Rains and Keating}{Rains and Keating}{2011}]%
        {rains_social_2011}
\bibfield{author}{\bibinfo{person}{Stephen~A. Rains} {and}
  \bibinfo{person}{David~M. Keating}.} \bibinfo{year}{2011}\natexlab{}.
\newblock \showarticletitle{The {Social} {Dimension} of {Blogging} about
  {Health}: {Health} {Blogging}, {Social} {Support}, and {Well}-being}.
\newblock \bibinfo{journal}{\emph{Communication Monographs}}
  \bibinfo{volume}{78}, \bibinfo{number}{4} (\bibinfo{date}{Dec.}
  \bibinfo{year}{2011}), \bibinfo{pages}{511--534}.
\newblock
\showISSN{0363-7751}
\urldef\tempurl%
\url{https://doi.org/10.1080/03637751.2011.618142}
\showDOI{\tempurl}


\bibitem[\protect\citeauthoryear{Rains and Wright}{Rains and Wright}{2016}]%
        {rains_social_2016}
\bibfield{author}{\bibinfo{person}{Stephen~A. Rains} {and}
  \bibinfo{person}{Kevin~B. Wright}.} \bibinfo{year}{2016}\natexlab{}.
\newblock \showarticletitle{Social {Support} and {Computer}-{Mediated}
  {Communication}: {A} {State}-of-the-{Art} {Review} and {Agenda} for {Future}
  {Research}}.
\newblock \bibinfo{journal}{\emph{Annals of the International Communication
  Association}} \bibinfo{volume}{40}, \bibinfo{number}{1} (\bibinfo{date}{Jan.}
  \bibinfo{year}{2016}), \bibinfo{pages}{175--211}.
\newblock
\showISSN{2380-8985}
\urldef\tempurl%
\url{https://doi.org/10.1080/23808985.2015.11735260}
\showDOI{\tempurl}


\bibitem[\protect\citeauthoryear{Riessman}{Riessman}{1965}]%
        {riessman_helper_1965}
\bibfield{author}{\bibinfo{person}{Frank Riessman}.}
  \bibinfo{year}{1965}\natexlab{}.
\newblock \showarticletitle{The “{Helper}” {Therapy} {Principle}}.
\newblock \bibinfo{journal}{\emph{Soc Work}} \bibinfo{volume}{10},
  \bibinfo{number}{2} (\bibinfo{date}{April} \bibinfo{year}{1965}),
  \bibinfo{pages}{27--32}.
\newblock
\showISSN{0037-8046}
\urldef\tempurl%
\url{https://doi.org/10.1093/sw/10.2.27}
\showDOI{\tempurl}


\bibitem[\protect\citeauthoryear{Ruthven}{Ruthven}{2019}]%
        {ruthven_making_2019}
\bibfield{author}{\bibinfo{person}{Ian Ruthven}.}
  \bibinfo{year}{2019}\natexlab{}.
\newblock \showarticletitle{Making {Meaning}: {A} {Focus} for {Information}
  {Interactions} {Research}}. In \bibinfo{booktitle}{\emph{Proceedings of the
  2019 {Conference} on {Human} {Information} {Interaction} and {Retrieval}}}
  \emph{(\bibinfo{series}{{CHIIR} '19})}. \bibinfo{publisher}{Association for
  Computing Machinery}, \bibinfo{address}{Glasgow, Scotland UK},
  \bibinfo{pages}{163--171}.
\newblock
\showISBNx{978-1-4503-6025-8}
\urldef\tempurl%
\url{https://doi.org/10.1145/3295750.3298938}
\showDOI{\tempurl}


\bibitem[\protect\citeauthoryear{Sandén, Nilsson, Thulesius, Hägglund, and
  Harrysson}{Sandén et~al\mbox{.}}{2019}]%
        {sanden_cancer_2019}
\bibfield{author}{\bibinfo{person}{Ulrika Sandén}, \bibinfo{person}{Fredrik
  Nilsson}, \bibinfo{person}{Hans Thulesius}, \bibinfo{person}{Maria
  Hägglund}, {and} \bibinfo{person}{Lars Harrysson}.}
  \bibinfo{year}{2019}\natexlab{}.
\newblock \showarticletitle{Cancer, a relational disease}.
\newblock \bibinfo{journal}{\emph{International Journal of Qualitative Studies
  on Health and Well-being}} \bibinfo{volume}{14}, \bibinfo{number}{1}
  (\bibinfo{date}{Jan.} \bibinfo{year}{2019}), \bibinfo{pages}{1622354}.
\newblock
\showISSN{null}
\urldef\tempurl%
\url{https://doi.org/10.1080/17482631.2019.1622354}
\showDOI{\tempurl}


\bibitem[\protect\citeauthoryear{Schorch, Wan, Randall, and Wulf}{Schorch
  et~al\mbox{.}}{2016}]%
        {schorch_designing_2016}
\bibfield{author}{\bibinfo{person}{Marén Schorch}, \bibinfo{person}{Lin Wan},
  \bibinfo{person}{David~William Randall}, {and} \bibinfo{person}{Volker
  Wulf}.} \bibinfo{year}{2016}\natexlab{}.
\newblock \showarticletitle{Designing for {Those} {Who} {Are} {Overlooked}:
  {Insider} {Perspectives} on {Care} {Practices} and {Cooperative} {Work} of
  {Elderly} {Informal} {Caregivers}}. In \bibinfo{booktitle}{\emph{Proceedings
  of the 19th {ACM} {Conference} on {Computer}-{Supported} {Cooperative} {Work}
  \& {Social} {Computing}}} \emph{(\bibinfo{series}{{CSCW} '16})}.
  \bibinfo{publisher}{ACM}, \bibinfo{address}{New York, NY, USA},
  \bibinfo{pages}{787--799}.
\newblock
\showISBNx{978-1-4503-3592-8}
\urldef\tempurl%
\url{https://doi.org/10.1145/2818048.2819999}
\showDOI{\tempurl}
\newblock
\shownote{event-place: San Francisco, California, USA.}


\bibitem[\protect\citeauthoryear{Seabold and Perktold}{Seabold and
  Perktold}{2010}]%
        {statsmodels}
\bibfield{author}{\bibinfo{person}{Skipper Seabold} {and}
  \bibinfo{person}{Josef Perktold}.} \bibinfo{year}{2010}\natexlab{}.
\newblock \showarticletitle{Statsmodels: Econometric and statistical modeling
  with python}. In \bibinfo{booktitle}{\emph{9th Python in Science
  Conference}}.
\newblock


\bibitem[\protect\citeauthoryear{Seering, Hammer, Kaufman, and Yang}{Seering
  et~al\mbox{.}}{2020}]%
        {seering_proximate_2020}
\bibfield{author}{\bibinfo{person}{Joseph Seering}, \bibinfo{person}{Jessica
  Hammer}, \bibinfo{person}{Geoff Kaufman}, {and} \bibinfo{person}{Diyi Yang}.}
  \bibinfo{year}{2020}\natexlab{}.
\newblock \showarticletitle{Proximate {Social} {Factors} in {First}-{Time}
  {Contribution} to {Online} {Communities}}.
\newblock \bibinfo{journal}{\emph{CHI}} (\bibinfo{year}{2020}),
  \bibinfo{pages}{14}.
\newblock
\urldef\tempurl%
\url{https://doi.org/10.1145/3313831.3376151}
\showDOI{\tempurl}


\bibitem[\protect\citeauthoryear{Seering, Wang, Yoon, and Kaufman}{Seering
  et~al\mbox{.}}{2019}]%
        {seering_moderator_2019}
\bibfield{author}{\bibinfo{person}{Joseph Seering}, \bibinfo{person}{Tony
  Wang}, \bibinfo{person}{Jina Yoon}, {and} \bibinfo{person}{Geoff Kaufman}.}
  \bibinfo{year}{2019}\natexlab{}.
\newblock \showarticletitle{Moderator engagement and community development in
  the age of algorithms}.
\newblock \bibinfo{journal}{\emph{New Media \& Society}} \bibinfo{volume}{21},
  \bibinfo{number}{7} (\bibinfo{date}{July} \bibinfo{year}{2019}),
  \bibinfo{pages}{1417--1443}.
\newblock
\showISSN{1461-4448}
\urldef\tempurl%
\url{https://doi.org/10.1177/1461444818821316}
\showDOI{\tempurl}


\bibitem[\protect\citeauthoryear{Seçkin}{Seçkin}{2013}]%
        {seckin_satisfaction_2013}
\bibfield{author}{\bibinfo{person}{Gül Seçkin}.}
  \bibinfo{year}{2013}\natexlab{}.
\newblock \showarticletitle{Satisfaction with health status among cyber
  patients: testing a mediation model of electronic coping support}.
\newblock \bibinfo{journal}{\emph{Behaviour \& Information Technology}}
  \bibinfo{volume}{32}, \bibinfo{number}{1} (\bibinfo{date}{Jan.}
  \bibinfo{year}{2013}), \bibinfo{pages}{91--101}.
\newblock
\showISSN{0144-929X}
\urldef\tempurl%
\url{https://doi.org/10.1080/0144929X.2011.603359}
\showDOI{\tempurl}


\bibitem[\protect\citeauthoryear{Shalizi and Thomas}{Shalizi and
  Thomas}{2011}]%
        {shalizi_homophily_2011}
\bibfield{author}{\bibinfo{person}{Cosma~Rohilla Shalizi} {and}
  \bibinfo{person}{Andrew~C. Thomas}.} \bibinfo{year}{2011}\natexlab{}.
\newblock \showarticletitle{Homophily and {Contagion} {Are} {Generically}
  {Confounded} in {Observational} {Social} {Network} {Studies}}.
\newblock \bibinfo{journal}{\emph{Sociological Methods \& Research}}
  \bibinfo{volume}{40}, \bibinfo{number}{2} (\bibinfo{date}{May}
  \bibinfo{year}{2011}), \bibinfo{pages}{211--239}.
\newblock
\showISSN{0049-1241}
\urldef\tempurl%
\url{https://doi.org/10.1177/0049124111404820}
\showDOI{\tempurl}


\bibitem[\protect\citeauthoryear{Sharma, Choudhury, Althoff, and Sharma}{Sharma
  et~al\mbox{.}}{2020}]%
        {sharma_engagement_2020}
\bibfield{author}{\bibinfo{person}{Ashish Sharma}, \bibinfo{person}{Monojit
  Choudhury}, \bibinfo{person}{Tim Althoff}, {and} \bibinfo{person}{Amit
  Sharma}.} \bibinfo{year}{2020}\natexlab{}.
\newblock \showarticletitle{Engagement {Patterns} of {Peer}-to-{Peer}
  {Interactions} on {Mental} {Health} {Platforms}}. In
  \bibinfo{booktitle}{\emph{{arXiv}:2004.04999 [cs]}}.
\newblock
\urldef\tempurl%
\url{http://arxiv.org/abs/2004.04999}
\showURL{%
\tempurl}
\newblock
\shownote{arXiv: 2004.04999.}


\bibitem[\protect\citeauthoryear{Simoni, Franks, Lehavot, and Yard}{Simoni
  et~al\mbox{.}}{2011}]%
        {simoni_peer_2011}
\bibfield{author}{\bibinfo{person}{Jane~M. Simoni}, \bibinfo{person}{Julie~C.
  Franks}, \bibinfo{person}{Keren Lehavot}, {and} \bibinfo{person}{Samantha~S.
  Yard}.} \bibinfo{year}{2011}\natexlab{}.
\newblock \showarticletitle{Peer interventions to promote health: {Conceptual}
  considerations}.
\newblock \bibinfo{journal}{\emph{American Journal of Orthopsychiatry}}
  \bibinfo{volume}{81}, \bibinfo{number}{3} (\bibinfo{year}{2011}),
  \bibinfo{pages}{351--359}.
\newblock
\showISSN{1939-0025(Electronic),0002-9432(Print)}
\urldef\tempurl%
\url{https://doi.org/10.1111/j.1939-0025.2011.01103.x}
\showDOI{\tempurl}


\bibitem[\protect\citeauthoryear{Sleeper, Melicher, Habib, Bauer, Cranor, and
  Mazurek}{Sleeper et~al\mbox{.}}{2016}]%
        {sleeper_sharing_2016}
\bibfield{author}{\bibinfo{person}{Manya Sleeper}, \bibinfo{person}{William
  Melicher}, \bibinfo{person}{Hana Habib}, \bibinfo{person}{Lujo Bauer},
  \bibinfo{person}{Lorrie~Faith Cranor}, {and} \bibinfo{person}{Michelle~L.
  Mazurek}.} \bibinfo{year}{2016}\natexlab{}.
\newblock \showarticletitle{Sharing {Personal} {Content} {Online}: {Exploring}
  {Channel} {Choice} and {Multi}-{Channel} {Behaviors}}. In
  \bibinfo{booktitle}{\emph{Proceedings of the 2016 {CHI} {Conference} on
  {Human} {Factors} in {Computing} {Systems} - {CHI} '16}}.
  \bibinfo{publisher}{ACM Press}, \bibinfo{address}{Santa Clara, California,
  USA}, \bibinfo{pages}{101--112}.
\newblock
\showISBNx{978-1-4503-3362-7}
\urldef\tempurl%
\url{https://doi.org/10.1145/2858036.2858170}
\showDOI{\tempurl}


\bibitem[\protect\citeauthoryear{Smith, Levonian, Giaquinto, Ma,
  Lein-Mcdonough, Li, O'Conner-Von, and Yarosh}{Smith et~al\mbox{.}}{2020}]%
        {smith_i_2020}
\bibfield{author}{\bibinfo{person}{C~Estelle Smith}, \bibinfo{person}{Zachary
  Levonian}, \bibinfo{person}{Robert Giaquinto}, \bibinfo{person}{Haiwei Ma},
  \bibinfo{person}{Gemma Lein-Mcdonough}, \bibinfo{person}{Zixuan Li},
  \bibinfo{person}{Susan O'Conner-Von}, {and} \bibinfo{person}{Svetlana
  Yarosh}.} \bibinfo{year}{2020}\natexlab{}.
\newblock \showarticletitle{``{I} {Cannot} {Do} {All} of this {Alone}'':
  {Sociotechnical} {Opportunities} for {Spiritual} and {Instrumental} {Support}
  on {Cancer} {Journeys}}.
\newblock \bibinfo{journal}{\emph{TOCHI}} (\bibinfo{year}{2020}).
\newblock
\urldef\tempurl%
\url{https://arxiv.org/abs/2005.11884}
\showURL{%
\tempurl}


\bibitem[\protect\citeauthoryear{Smith and Christakis}{Smith and
  Christakis}{2008}]%
        {smith_social_2008}
\bibfield{author}{\bibinfo{person}{Kirsten~P. Smith} {and}
  \bibinfo{person}{Nicholas~A. Christakis}.} \bibinfo{year}{2008}\natexlab{}.
\newblock \showarticletitle{Social {Networks} and {Health}}.
\newblock \bibinfo{journal}{\emph{Annual Review of Sociology}}
  \bibinfo{volume}{34}, \bibinfo{number}{1} (\bibinfo{year}{2008}),
  \bibinfo{pages}{405--429}.
\newblock
\urldef\tempurl%
\url{https://doi.org/10.1146/annurev.soc.34.040507.134601}
\showDOI{\tempurl}


\bibitem[\protect\citeauthoryear{Stewart, Fulmer, and Barrick}{Stewart
  et~al\mbox{.}}{2005}]%
        {stewart_exploration_2005}
\bibfield{author}{\bibinfo{person}{Greg~L. Stewart}, \bibinfo{person}{Ingrid~S.
  Fulmer}, {and} \bibinfo{person}{Murray~R. Barrick}.}
  \bibinfo{year}{2005}\natexlab{}.
\newblock \showarticletitle{An {Exploration} of {Member} {Roles} as a
  {Multilevel} {Linking} {Mechanism} for {Individual} {Traits} and {Team}
  {Outcomes}}.
\newblock \bibinfo{journal}{\emph{Personnel Psychology}} \bibinfo{volume}{58},
  \bibinfo{number}{2} (\bibinfo{year}{2005}), \bibinfo{pages}{343--365}.
\newblock
\showISSN{1744-6570}
\urldef\tempurl%
\url{https://doi.org/10.1111/j.1744-6570.2005.00480.x}
\showDOI{\tempurl}


\bibitem[\protect\citeauthoryear{Sun, Wang, and Rosson}{Sun
  et~al\mbox{.}}{2019}]%
        {sun_how_2019}
\bibfield{author}{\bibinfo{person}{Na Sun}, \bibinfo{person}{Xiying Wang},
  {and} \bibinfo{person}{Mary~Beth Rosson}.} \bibinfo{year}{2019}\natexlab{}.
\newblock \showarticletitle{How {Do} {Distance} {Learners} {Connect}? {Shared}
  {Identity}, {Focused} {Work} and {Future} {Possibilities}}.
  \bibinfo{pages}{13}.
\newblock


\bibitem[\protect\citeauthoryear{Tamersoy, De~Choudhury, and Chau}{Tamersoy
  et~al\mbox{.}}{2015}]%
        {tamersoy_characterizing_2015}
\bibfield{author}{\bibinfo{person}{Acar Tamersoy}, \bibinfo{person}{Munmun
  De~Choudhury}, {and} \bibinfo{person}{Duen~Horng Chau}.}
  \bibinfo{year}{2015}\natexlab{}.
\newblock \showarticletitle{Characterizing {Smoking} and {Drinking}
  {Abstinence} from {Social} {Media}}. In \bibinfo{booktitle}{\emph{Proceedings
  of the 26th {ACM} {Conference} on {Hypertext} \& {Social} {Media}}}
  \emph{(\bibinfo{series}{{HT} '15})}. \bibinfo{publisher}{ACM},
  \bibinfo{address}{New York, NY, USA}, \bibinfo{pages}{139--148}.
\newblock
\showISBNx{978-1-4503-3395-5}
\urldef\tempurl%
\url{https://doi.org/10.1145/2700171.2791247}
\showDOI{\tempurl}


\bibitem[\protect\citeauthoryear{Thoits}{Thoits}{2010}]%
        {thoits_stress_2010}
\bibfield{author}{\bibinfo{person}{Peggy~A. Thoits}.}
  \bibinfo{year}{2010}\natexlab{}.
\newblock \showarticletitle{Stress and {Health}: {Major} {Findings} and
  {Policy} {Implications}}.
\newblock \bibinfo{journal}{\emph{J Health Soc Behav}} \bibinfo{volume}{51},
  \bibinfo{number}{1\_suppl} (\bibinfo{date}{March} \bibinfo{year}{2010}),
  \bibinfo{pages}{S41--S53}.
\newblock
\showISSN{0022-1465}
\urldef\tempurl%
\url{https://doi.org/10.1177/0022146510383499}
\showDOI{\tempurl}


\bibitem[\protect\citeauthoryear{Thoits}{Thoits}{2011}]%
        {thoits_mechanisms_2011}
\bibfield{author}{\bibinfo{person}{Peggy~A. Thoits}.}
  \bibinfo{year}{2011}\natexlab{}.
\newblock \showarticletitle{Mechanisms {Linking} {Social} {Ties} and {Support}
  to {Physical} and {Mental} {Health}}.
\newblock \bibinfo{journal}{\emph{J Health Soc Behav}} \bibinfo{volume}{52},
  \bibinfo{number}{2} (\bibinfo{date}{June} \bibinfo{year}{2011}),
  \bibinfo{pages}{145--161}.
\newblock
\showISSN{0022-1465}
\urldef\tempurl%
\url{https://doi.org/10.1177/0022146510395592}
\showDOI{\tempurl}


\bibitem[\protect\citeauthoryear{Uchino, Bowen, Kent~de Grey, Mikel, and
  Fisher}{Uchino et~al\mbox{.}}{2018}]%
        {uchino_social_2018}
\bibfield{author}{\bibinfo{person}{Bert~N. Uchino}, \bibinfo{person}{Kimberly
  Bowen}, \bibinfo{person}{Robert Kent~de Grey}, \bibinfo{person}{Jude Mikel},
  {and} \bibinfo{person}{Edwin~B. Fisher}.} \bibinfo{year}{2018}\natexlab{}.
\newblock \showarticletitle{Social {Support} and {Physical} {Health}: {Models},
  {Mechanisms}, and {Opportunities}}.
\newblock In \bibinfo{booktitle}{\emph{Principles and {Concepts} of
  {Behavioral} {Medicine}: {A} {Global} {Handbook}}},
  \bibfield{editor}{\bibinfo{person}{Edwin~B. Fisher},
  \bibinfo{person}{Linda~D. Cameron}, \bibinfo{person}{Alan~J. Christensen},
  \bibinfo{person}{Ulrike Ehlert}, \bibinfo{person}{Yan Guo},
  \bibinfo{person}{Brian Oldenburg}, {and} \bibinfo{person}{Frank~J. Snoek}}
  (Eds.). \bibinfo{publisher}{Springer New York}, \bibinfo{address}{New York,
  NY}, \bibinfo{pages}{341--372}.
\newblock
\showISBNx{978-0-387-93826-4}
\urldef\tempurl%
\url{https://doi.org/10.1007/978-0-387-93826-4_12}
\showDOI{\tempurl}


\bibitem[\protect\citeauthoryear{Urbanoski, Mierlo, and Cunningham}{Urbanoski
  et~al\mbox{.}}{2017}]%
        {urbanoski_investigating_2017}
\bibfield{author}{\bibinfo{person}{Karen Urbanoski},
  \bibinfo{person}{Trevor~van Mierlo}, {and} \bibinfo{person}{John
  Cunningham}.} \bibinfo{year}{2017}\natexlab{}.
\newblock \showarticletitle{Investigating {Patterns} of {Participation} in an
  {Online} {Support} {Group} for {Problem} {Drinking}: a {Social} {Network}
  {Analysis}}.
\newblock \bibinfo{journal}{\emph{Int.J. Behav. Med.}} \bibinfo{volume}{24},
  \bibinfo{number}{5} (\bibinfo{date}{Oct.} \bibinfo{year}{2017}),
  \bibinfo{pages}{703--712}.
\newblock
\showISSN{1070-5503, 1532-7558}
\urldef\tempurl%
\url{https://doi.org/10.1007/s12529-016-9591-6}
\showDOI{\tempurl}


\bibitem[\protect\citeauthoryear{van~der Walt, Colbert, and Varoquaux}{van~der
  Walt et~al\mbox{.}}{2011}]%
        {van_der_walt_numpy_2011}
\bibfield{author}{\bibinfo{person}{Stefan van~der Walt},
  \bibinfo{person}{S.~Chris Colbert}, {and} \bibinfo{person}{Gael Varoquaux}.}
  \bibinfo{year}{2011}\natexlab{}.
\newblock \showarticletitle{The {NumPy} {Array}: {A} {Structure} for
  {Efficient} {Numerical} {Computation}}.
\newblock \bibinfo{journal}{\emph{Computing in Science Engineering}}
  \bibinfo{volume}{13}, \bibinfo{number}{2} (\bibinfo{date}{March}
  \bibinfo{year}{2011}), \bibinfo{pages}{22--30}.
\newblock
\showISSN{1558-366X}
\urldef\tempurl%
\url{https://doi.org/10.1109/MCSE.2011.37}
\showDOI{\tempurl}


\bibitem[\protect\citeauthoryear{Wang and Manning}{Wang and Manning}{2012}]%
        {wang_baselines_2012}
\bibfield{author}{\bibinfo{person}{Sida Wang} {and}
  \bibinfo{person}{Christopher Manning}.} \bibinfo{year}{2012}\natexlab{}.
\newblock \showarticletitle{Baselines and {Bigrams}: {Simple}, {Good}
  {Sentiment} and {Topic} {Classification}}. In
  \bibinfo{booktitle}{\emph{Proceedings of the 50th {Annual} {Meeting} of the
  {Association} for {Computational} {Linguistics} ({Volume} 2: {Short}
  {Papers})}}. \bibinfo{publisher}{Association for Computational Linguistics},
  \bibinfo{address}{Jeju Island, Korea}, \bibinfo{pages}{90--94}.
\newblock
\urldef\tempurl%
\url{https://www.aclweb.org/anthology/P12-2018}
\showURL{%
\tempurl}


\bibitem[\protect\citeauthoryear{Wang, Kraut, and Levine}{Wang
  et~al\mbox{.}}{2012}]%
        {wang_stay_2012}
\bibfield{author}{\bibinfo{person}{Yi-Chia Wang}, \bibinfo{person}{Robert
  Kraut}, {and} \bibinfo{person}{John~M. Levine}.}
  \bibinfo{year}{2012}\natexlab{}.
\newblock \showarticletitle{To {Stay} or {Leave}?: {The} {Relationship} of
  {Emotional} and {Informational} {Support} to {Commitment} in {Online}
  {Health} {Support} {Groups}}. In \bibinfo{booktitle}{\emph{Proceedings of the
  {ACM} 2012 {Conference} on {Computer} {Supported} {Cooperative} {Work}}}
  \emph{(\bibinfo{series}{{CSCW} '12})}. \bibinfo{publisher}{ACM},
  \bibinfo{address}{New York, NY, USA}, \bibinfo{pages}{833--842}.
\newblock
\showISBNx{978-1-4503-1086-4}
\urldef\tempurl%
\url{https://doi.org/10.1145/2145204.2145329}
\showDOI{\tempurl}


\bibitem[\protect\citeauthoryear{Wen and Rose}{Wen and Rose}{2012}]%
        {wen_understanding_2012}
\bibfield{author}{\bibinfo{person}{Miaomiao Wen} {and}
  \bibinfo{person}{Carolyn~Penstein Rose}.} \bibinfo{year}{2012}\natexlab{}.
\newblock \showarticletitle{Understanding {Participant} {Behavior}
  {Trajectories} in {Online} {Health} {Support} {Groups} {Using} {Automatic}
  {Extraction} {Methods}}. In \bibinfo{booktitle}{\emph{Proceedings of the 17th
  {ACM} {International} {Conference} on {Supporting} {Group} {Work}}}
  \emph{(\bibinfo{series}{{GROUP} '12})}. \bibinfo{publisher}{ACM},
  \bibinfo{address}{New York, NY, USA}, \bibinfo{pages}{179--188}.
\newblock
\showISBNx{978-1-4503-1486-2}
\urldef\tempurl%
\url{https://doi.org/10.1145/2389176.2389205}
\showDOI{\tempurl}


\bibitem[\protect\citeauthoryear{Wills}{Wills}{1981}]%
        {wills_downward_1981}
\bibfield{author}{\bibinfo{person}{Thomas~A. Wills}.}
  \bibinfo{year}{1981}\natexlab{}.
\newblock \showarticletitle{Downward comparison principles in social
  psychology}.
\newblock \bibinfo{journal}{\emph{Psychological Bulletin}}
  \bibinfo{volume}{90}, \bibinfo{number}{2} (\bibinfo{year}{1981}),
  \bibinfo{pages}{245--271}.
\newblock
\showISSN{1939-1455(Electronic),0033-2909(Print)}
\urldef\tempurl%
\url{https://doi.org/10.1037/0033-2909.90.2.245}
\showDOI{\tempurl}


\bibitem[\protect\citeauthoryear{Wright and Miller}{Wright and Miller}{2010}]%
        {wright_measure_2010}
\bibfield{author}{\bibinfo{person}{Kevin~B. Wright} {and}
  \bibinfo{person}{Claude~H. Miller}.} \bibinfo{year}{2010}\natexlab{}.
\newblock \showarticletitle{A {Measure} of {Weak}-{Tie}/{Strong}-{Tie}
  {Support} {Network} {Preference}}.
\newblock \bibinfo{journal}{\emph{Communication Monographs}}
  \bibinfo{volume}{77}, \bibinfo{number}{4} (\bibinfo{date}{Dec.}
  \bibinfo{year}{2010}), \bibinfo{pages}{500--517}.
\newblock
\showISSN{0363-7751}
\urldef\tempurl%
\url{https://doi.org/10.1080/03637751.2010.502538}
\showDOI{\tempurl}


\bibitem[\protect\citeauthoryear{Xu, Chiu, Chen, and Mukherjee}{Xu
  et~al\mbox{.}}{2015}]%
        {xu_twitter_2015}
\bibfield{author}{\bibinfo{person}{Weiai~Wayne Xu}, \bibinfo{person}{I-Hsuan
  Chiu}, \bibinfo{person}{Yixin Chen}, {and} \bibinfo{person}{Tanuka
  Mukherjee}.} \bibinfo{year}{2015}\natexlab{}.
\newblock \showarticletitle{Twitter hashtags for health: applying network and
  content analyses to understand the health knowledge sharing in a
  {Twitter}-based community of practice}.
\newblock \bibinfo{journal}{\emph{Qual Quant}} \bibinfo{volume}{49},
  \bibinfo{number}{4} (\bibinfo{date}{July} \bibinfo{year}{2015}),
  \bibinfo{pages}{1361--1380}.
\newblock
\showISSN{1573-7845}
\urldef\tempurl%
\url{https://doi.org/10.1007/s11135-014-0051-6}
\showDOI{\tempurl}


\bibitem[\protect\citeauthoryear{Yan, Peng, and Tan}{Yan et~al\mbox{.}}{2015}]%
        {yan_network_2015}
\bibfield{author}{\bibinfo{person}{Lu~(Lucy) Yan}, \bibinfo{person}{Jianping
  Peng}, {and} \bibinfo{person}{Yong Tan}.} \bibinfo{year}{2015}\natexlab{}.
\newblock \showarticletitle{Network {Dynamics}: {How} {Can} {We} {Find}
  {Patients} {Like} {Us}?}
\newblock \bibinfo{journal}{\emph{Information Systems Research}}
  \bibinfo{volume}{26}, \bibinfo{number}{3} (\bibinfo{date}{Aug.}
  \bibinfo{year}{2015}), \bibinfo{pages}{496--512}.
\newblock
\showISSN{1047-7047}
\urldef\tempurl%
\url{https://doi.org/10.1287/isre.2015.0585}
\showDOI{\tempurl}


\bibitem[\protect\citeauthoryear{Yang, Kraut, and Levine}{Yang
  et~al\mbox{.}}{2017}]%
        {yang_commitment_2017}
\bibfield{author}{\bibinfo{person}{Diyi Yang}, \bibinfo{person}{Robert Kraut},
  {and} \bibinfo{person}{John~M. Levine}.} \bibinfo{year}{2017}\natexlab{}.
\newblock \showarticletitle{Commitment of {Newcomers} and {Old}-timers to
  {Online} {Health} {Support} {Communities}}. In
  \bibinfo{booktitle}{\emph{Proceedings of the 2017 {CHI} {Conference} on
  {Human} {Factors} in {Computing} {Systems}}} \emph{(\bibinfo{series}{{CHI}
  '17})}. \bibinfo{publisher}{ACM}, \bibinfo{address}{New York, NY, USA},
  \bibinfo{pages}{6363--6375}.
\newblock
\showISBNx{978-1-4503-4655-9}
\urldef\tempurl%
\url{https://doi.org/10.1145/3025453.3026008}
\showDOI{\tempurl}


\bibitem[\protect\citeauthoryear{Yang, Kraut, Smith, Mayfield, and
  Jurafsky}{Yang et~al\mbox{.}}{2019}]%
        {yang_seekers_2019}
\bibfield{author}{\bibinfo{person}{Diyi Yang}, \bibinfo{person}{Robert Kraut},
  \bibinfo{person}{Tenbroeck Smith}, \bibinfo{person}{Elijah Mayfield}, {and}
  \bibinfo{person}{Dan Jurafsky}.} \bibinfo{year}{2019}\natexlab{}.
\newblock \showarticletitle{Seekers, {Providers}, {Welcomers}, and
  {Storytellers}: {Modeling} {Social} {Roles} in {Online} {Health}
  {Communities}}. In \bibinfo{booktitle}{\emph{Proceedings of the {CHI}
  {Conference} on {Human} {Factors} in {Computing} {Systems}}}
  \emph{(\bibinfo{series}{{CHI} '19})}. \bibinfo{pages}{12}.
\newblock
\urldef\tempurl%
\url{http://kraut.hciresearch.org/sites/kraut.hciresearch.org/files/articles/Yang19-SocialRolesInOnlineHealthCommunities.pdf}
\showURL{%
\tempurl}


\end{thebibliography}

\appendix
\newpage
\section{Additional analyses}

This appendix contains details that provide additional context and analysis of potential confounders.

\subsection{Valid author identification}
\label{app:sec:valid_author_identification}

\begin{figure}
\centering
\includegraphics[height=1.5in]{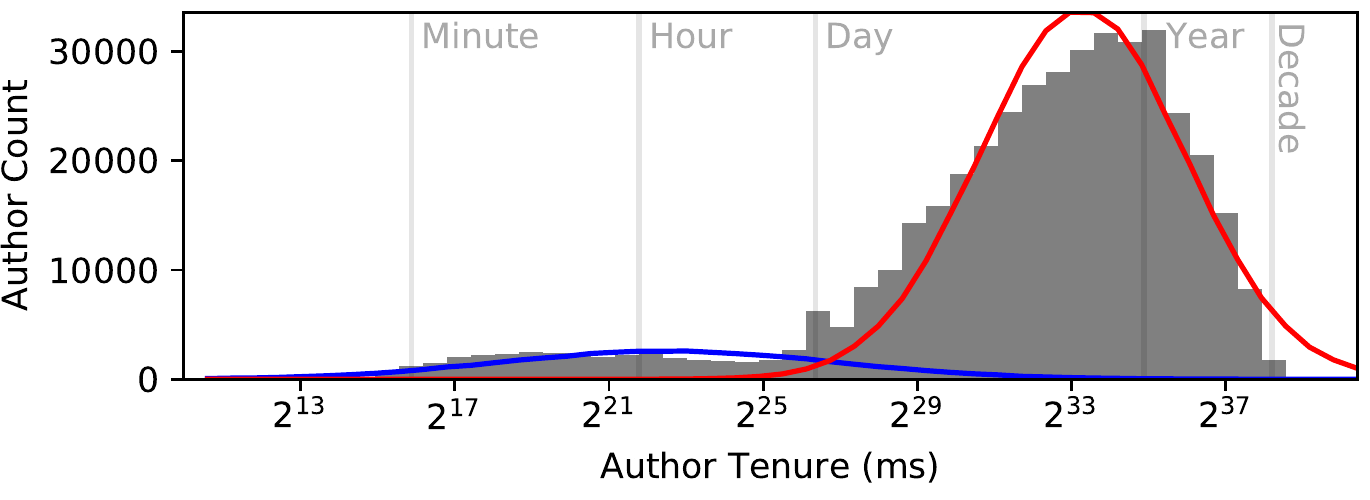}
\caption{Distribution of author tenure---the time between first and last journal update written by that author---for all 572,309 author users. The 161,799 (28.3\%) authors with only a single update (and thus 0 tenure) are not shown.
Median author tenure is 83 days (mean 311 days).
}
\label{fig:author_tenure_distribution}
\end{figure}

We omitted authors without at least two journal updates published more than 24 hours apart based on an analysis of author \textit{tenure}---the amount of time between an author's first and last published updates.
Figure \ref{fig:author_tenure_distribution} shows the distribution of tenure for all authors.  
We observe a bimodal lognormal distribution of author tenure and fit a two-component log-normal GMM in an approach adapted from Halfaker et al. \cite{halfaker_user_2015}. 
The lines overlayed on the histogram show the fit of the two Gaussian components to the tenure data.
We use the approximate visual intersection of the two GMM components, 24 hours, as a criterion for being a valid author.  
Self-reported health conditions by valid authors---which are used as features in the models fit to address RQ1---are shown in Table \ref{app:tab:health_condition_category_counts}.

\begin{table}[]
    \centering
    \begin{tabular}{lrr}
Health-condition category & Count & \% \\ \hline
None (not reported) & 152,818 & 42.17\% \\
Cancer & 109,339 & 30.18\% \\
Other & 37,556 & 10.36\% \\
Surgery/Transplantation & 15,415 & 4.25\% \\
Injury & 12,910 & 3.56\% \\
Cardiovascular/Stroke & 12,685 & 3.50\% \\
Neurological Condition & 9,376 & 2.59\% \\
Infant/Childbirth & 7,952 & 2.19\% \\
Condition Unknown & 2,252 & 0.62\% \\
Congenital/Immune Disorder & 2,042 & 0.56\% \\ \hline
    \end{tabular}
    \caption{Health condition assignments to valid authors based on site-level self reports.}
    \label{app:tab:health_condition_category_counts}
\end{table}

\subsection{Account sharing}\label{app:sec:account_sharing}

Author accounts classified as Mixed (7.49\% of authors, see Section \ref{sec:background_authorship_classification}) may indicate either a single author embodying multiple roles or multiple people sharing the same account credentials.  
Such account sharing generally occurs for convenience in the case of a trusted relationship between the patient and the caregiver \cite{jacobs_caring_2016}.  
Account sharing is potentially problematic for analyses treating interactions between accounts as interactions between two people, but particularly so if an account is shared by both a patient and a caregiver.
Thus, we classify an author account as Shared if, on any site, between one third and two thirds of that author's updates are classified as patient-authored. 
Using this conservative definition, we find only 7.46\% of accounts are Shared.%
\footnote{A more permissive definition labels author accounts as Shared if on any site that author has published both a Patient-classified and a Caregiver-classified update. While this definition---which labels 53.4\% of accounts as Shared---likely captures primarily classifier noise, it can be treated as an upper bound on author account sharing.}
Author account sharing is closely linked to the Mixed classification: 95.9\% of Mixed authors are Shared, suggesting that authors only rarely embody multiple roles e.g. writing as a patient author on one site and as a caregiver author on another.  Due to the high-proportion of Mixed-author accounts that are shared, in subsequent modeling we include as an author feature only the Mixed author role (as a dummy-coded categorical variable) and not a separate indicator variable of account sharing. We note that authors classified as Mixed are likely multiple people using the same account.

One implication of this analysis is that user accounts are nuanced and the assumption of one account being associated with one person or even one role is frequently mistaken.  
Further work on roles must grapple with the reality of user account sharing and the challenges it presents to both analysts and users \cite{lampinen_account_2014}.
For designers in particular, a ``one person, one account'' assumption may undermine the effectiveness of designed interventions, e.g. recommended articles to edit on Wikipedia or personalized social media feeds.

\subsection{Computing patient-authored update proportion}\label{app:sec:p_update_proportion}

We computed the proportion of patient-authored updates on valid sites using a random sample of 5,000 unlabeled journal updates.  We used the 305 models trained during hold-one-out cross validation in order to compute standard error as an estimate of the variability of this proportion.  The model predicted that 24.84\% (s.e. 0.03\%) of unlabeled journal updates were patient-authored.  As the label distribution in the training data is different from the label distribution over the target updates, we need to correct for this distribution shift as it will bias the estimate towards the balanced training distribution.  We use Black Box Shift Estimation \cite{lipton_detecting_2018} to quantify the shift in distribution and produce a revised estimate, finding that 22.06\% (s.e. 0.11\%) of unlabeled journal updates were patient-authored.

\subsection{Assumption analysis: Amp timestamps}\label{app:sec:amps_timestamp}

Amps (``likes'' on CB) lack timestamp information, so we assume that amps occur at the publication time of the associated journal update.
To assess whether this assumption is reasonable, we examine the moment when the amps feature was introduced, reasoning that amps on journal updates published before the amps feature launched indicate a lag time between the update publication and the amp interaction. 
Only 0.32\% of amps occur on updates published before the launch of the amps feature, and comparing updates published the week before the launch date to the updates published in the week after, only 23.1\% of amps are recorded pre-launch.
This analysis suggests that the majority of amps are given in the week the journal update was published.

\subsection{Interaction network details}

\begin{figure}
\centering
\begin{subfigure}[t]{0.5\textwidth}
  \includegraphics{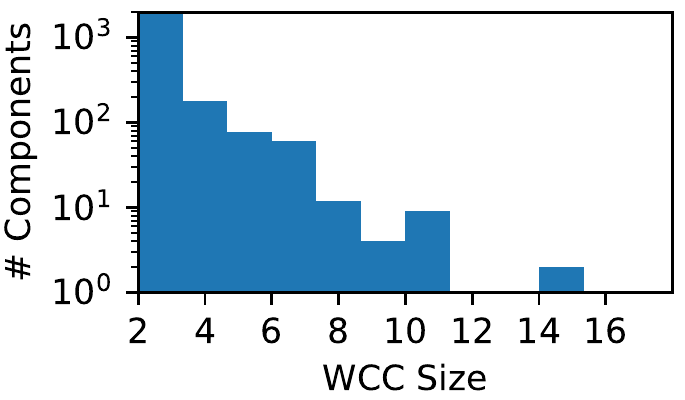}
\end{subfigure}%
\begin{subfigure}[t]{0.5\textwidth}
  \includegraphics{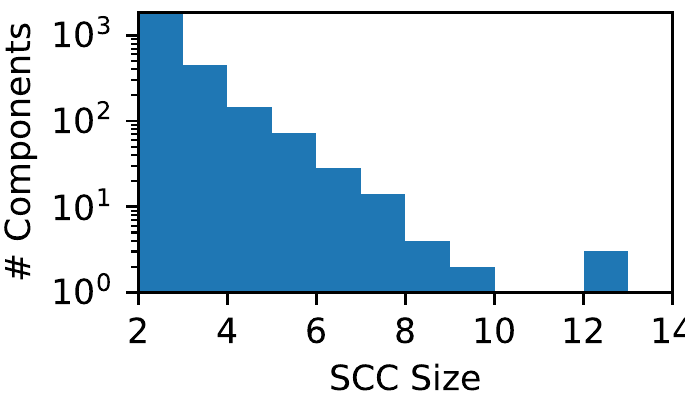}
\end{subfigure}
\caption{Distribution of the sizes of the 2335 weakly connected components (WCCs) and 2590 strongly connected components (SCCs) composed of two or more active authors at the end of the analysis period.  The largest WCC (size=45038) and largest SCC (size=16946) are not shown. Active authors are valid authors who were active on CB within six months of the end of the analysis period.
}
\label{fig:background_cc_distribution_plot}
\end{figure}
\begin{figure}
\centering
\includegraphics[width=\textwidth]{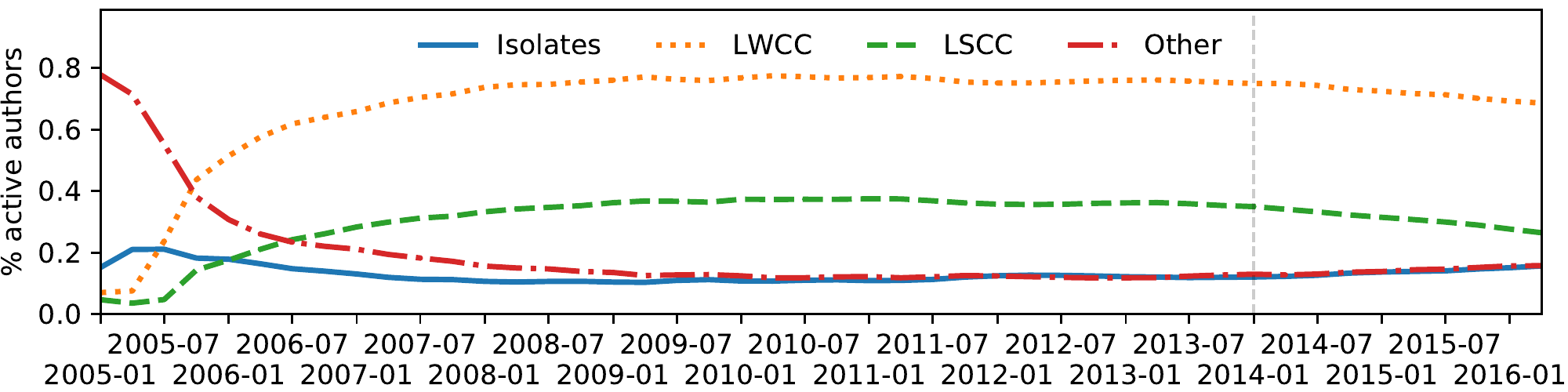}
\caption{Proportion of active authors---authors with public activity on CB within 6 months of the sampled date---based on their position within the network. 
``Isolates'' is the proportion of authors unconnected to any other author. 
``LWCC'' is the proportion of authors in the largest weakly connected component.
``Other'' is the proportion of authors weakly connected to at least one other user but not in the LWCC.
Together, these three account for all active authors. For comparison, we also show the proportion of authors in the largest strongly connected component (``LSCC''). 
The vertical dashed line indicates the beginning of the initiations period.}
\label{fig:background_network_summary_timeline}
\end{figure}

At the end of the initiations period, Figure \ref{fig:background_cc_distribution_plot} shows the distribution of the connected components, excluding the largest.
Figure \ref{fig:background_network_summary_timeline} shows the proportions of active authors in various network positions throughout the data range.

\subsection{Initiation type classification \& network growth}\label{app:sec:initiation_type_classification}

\begin{figure}
\centering
\includegraphics[width=\textwidth]{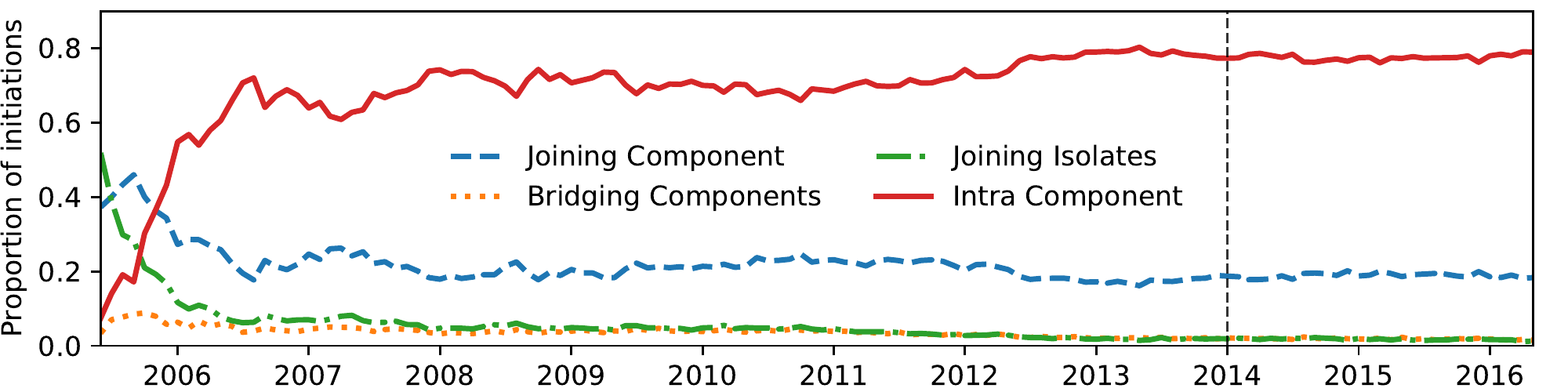}
\caption{Proportion of each initiation type over time. Initiation types are as defined by Gallagher et al.~\cite{gallagher_reclaiming_2019}. The vertical dashed line indicates the beginning of the initiations period.}
\label{fig:initiation_types_timeline}
\end{figure}

In order to understand how initiations relate to the growth of the network over time, we classified initiations according to the initiator's and the receivers' position within the network.
We identify four different initiation types, adapting definitions from Gallagher et al. to our context~\cite{gallagher_reclaiming_2019}:
(1) \textit{Joining Component} is an initiation connecting an unconnected author to an existing weakly-connected component.
(2) \textit{Bridging Component} is an initiation connecting two weakly-connected components, merging them.
(3) \textit{Joining Isolates} is an initiation that connects two previously unconnected authors (i.e. two ``isolates'').
(4) \textit{Intra Component} is an initiation between two authors who were already in the same weakly-connected component. 
By definition, all reciprocations are Intra Component initiations.
We classified all initiations as one of these four types.

Figure \ref{fig:initiation_types_timeline} shows the proportion of each type of initiation over time. 
Before the start of the initiations period in 2014, the network formed through Joining Isolates and Joining Component initiations before the majority of initiations became Intra Component.
Within the initiations period, 201,188 initiations were made, and over this period the proportion of each type remained quite consistent.
Bridging Component and Joining Isolates combined make up only 3.62\% of the initiations in the initiations period.  
The vast majority of initiations are between authors in the largest weakly-connected component (WCC) and other members of the largest WCC or previously unconnected authors.  
10.79\% of initiations in the initiations period are reciprocal, which is 13.91\% of the Intra Component initiations. The vast majority of Intra Component initiations are within the largest WCC specifically; only 3.5\% of Intra Component initiations involved components other than the largest.
Of the Joining Component initiations, 24.08\% of initiations-period initiations are initiated by the unconnected author and not someone in the component. 
These proportions suggest that the majority of initiations grow the largest WCC or occur within the largest WCC.
Overall, the network factor most associated with new initiations is the largest WCC. 

\subsection{Other interactions}\label{app:sec:other_interactions}

In the author interaction network, we include only guestbooks, amps, and comments, as described in Section \ref{sec:background_network_structure}.  However, we omit three types of interaction that may behave differently than the included interaction types: (1) author visits to other sites, (2) comments on other comments, and (3) explicit or text-based links to other sites in journal updates. 
First, we omit visits because our visit data is incomplete and can be only tenuously linked to specific authors at specific times and is invisible to the receiving site's authors.  Second, we omit comments left in response to guestbooks and update comments as the feature is relatively recent and minimally used.  Third, inter-site HTML links can be published in the text of journal updates.
To assess the impact of excluding such text-based inter-site links on the validity of the resulting network, we conducted a high-precision analysis of existing inter-site links.
Direct hyperlinks were extracted from the 19M journal updates. 101,923 valid links to CB sites were identified, of which 32.2\% (32,767) were determined to be self-links i.e. they linked to the site on which the update was authored.
Of the remaining 69,156 inter-site link interactions, 100\% were found to be redundant with existing interactions recorded via the other interaction types between the update author and the linked site.
Thus, we conclude that the three interaction types used provide a sufficiently detailed view of the inter-author interaction network.


\subsection{Initiation annotation details}\label{app:sec:initiation_annotation}

Initiation annotation (see section \ref{sec:init_background_methods}) proceeded in four rounds. Both annotators are authors of this paper and familiar with the CaringBridge dataset. In the first round, 30 guestbooks and 30 comments were coded and discussed to establish a codebook (or ``coding scheme'' \cite{geiger_garbage_2020}); these 60 initiations were discarded from further consideration.  Each subsequent round consisted of sampling an equal number of guestbooks and comments, two coders independently annotating them, and meeting to discuss disagreements and update the codebook.  200 initiations were sampled in the second and third rounds and 400 in the final round for a total of 800 initiations. 

As the codebook was not intended to be generalizable beyond this specific context and the relation codes were an open set, we did not compute a statistical measure of IRR and instead resolved all disagreements via discussion \cite{mcdonald_reliability_2019}.  Due to the inherent subjectivity in the annotation task, disagreements were relatively common; for the annotation of tie formation timing relative to the health event, raw agreement at the end of the second, third, and fourth round of annotation was 77.5\%, 70.0\%, and 76.8\% respectively. 
All disagreements were resolved quickly and centered on when sufficient evidence is present in the initiation to assign a non-Unknown label.  
Due to the lack of context, guestbooks are notably harder to annotate than comments.  The codebook is available in the GitHub repository.\footnote{\url{https://github.com/levon003/cscw-caringbridge-interaction-network}}

Result details beyond those in section \ref{sec:init_background_results} are presented here.
Table \ref{app:tab:initiation_annotation_counts} shows the annotated initiation counts, broken down by high-level relationship category. Categories were selected after annotation to summarize the data at a higher level than the raw codes e.g. a ``listserv contact'' becomes ''Other'' and multiple subtypes become ``Friend''.
Table \ref{app:tab:initiation_annotation_examples} shows representative initiations along with their annotated values.  Quotes are paraphrased to preserve poster anonymity and reduce traceability, which we deem ethically appropriate given the sensitive context \cite{markham_fabrication_2012,bruckman_studying_2002}.

\begin{table}[]
\begin{tabular}{@{}lrrr|r@{}}
Relation Category  & Post-health-event & Pre-health-event & Unknown & Total \\ \hline
Unknown                & 7             & 67             & 437     & 511   \\
Friend                 & 1             & 118            & 10      & 129   \\
Third-party connection & 35            & 3              & 9       & 47    \\
CG of similar patient  & 24            & 0              & 1       & 25    \\
Family                 & 1             & 20             & 1       & 22    \\
Other                  & 7             & 10             & 1       & 18    \\
One-time visitor       & 15            & 1              & 0       & 16    \\
Coworker or Schoolmate & 0             & 10             & 0       & 10    \\
Fellow patient         & 7             & 0              & 0       & 7     \\
(No text)              & -             & -              & -       & 15    \\ \hline
Total                  & 97 (12.1\%) & 229 (28.6\%) & 459 (57.4\%) & 800  
\end{tabular}
\caption{Annotated initiation counts, broken down by the two annotation types: ``Relation Category'', meaning the relation between the initiator and the receiver, and whether this tie existed before the health event that is the focus of the CB site. Fifteen initiations containing only whitespace characters were not annotated but are included in the total.}
\label{app:tab:initiation_annotation_counts} 
\end{table}

\begin{table}[]
\begin{tabular}{@{}llp{8.1cm}@{}}
\toprule
Relation Category  & Pre/post? & Initiation Text \\ \hline
CG of similar patient  & Post & Hi John, Our son Tommy, 16, is in the room next door. He is day+14. We are sending you lots of love, prayers and positive thoughts through the walls. We hope everyday you get a little stronger and feel better. \\
Third-party connection  & Post & Hi Don.  I am a friend of Ben's and through him I've been following your journey since last June.  I just want you to know that countless prayers have been said for you, your family and the doctors treating you.  I am thankful that Danny started this site so that we can all encourage you every step of the way. \\
One-time visitor  & Post & Diya, I know we have never met but tonight my heart and prayers go out to you. I traveled the road you are on seven years ago. Ella and the breast cancer support group, friends and family were my strength. Be strong. My story is under (CB site name) in Caring Bridge. If you ever need anyone to talk to, please call me. Anytime. \\
Fellow patient & Post & Congratulations on Day +2! I am the friend of Jenna's who also has multiple myeloma. Today is Day +84 for me. The next 30 days will be the toughest for you but try to walk and eat as much as you can to encourage all those little stem cells to grow! Sara Jones ((CB site name) on CaringBridge) \\
Unknown & Post & Patel family, I just read about your son in an article written by (local journalist). I had no idea. I am sending you prayers and positive thoughts. God bless you all. \\
Friend  & Pre & Sarah, Dan and I are so grateful to have this connection to you through Caring Bridge.  Our prayers have been winging your way since we heard the news on your hospitalization. We are traveling home tomorrow and will be back in church this Sunday.  We are holding you in our hearts.  We love you, Molly and Dan \\
Unknown  & Unknown & You are all in our thoughts on this wonderful day. Abigail \\
\bottomrule
\end{tabular}
\caption{Annotated initiation exemplars. A sample of paraphrased initiations and the annotations: (a) relation between the initiator and the receiver and (b) if the initiator was deemed to be a pre-health-event or a post-health-event connection. All names are aliases.}
\label{app:tab:initiation_annotation_examples} 
\end{table}

\subsection{RQ1b model details}

Table \ref{tab:init_timing_full_models} presents two linear regression models predicting the time between an author's first update and first initiation. Model \#1, for pre-authorship initiators, demonstrates that reasoning about the high-variance relationship between initiations and going on to become an author is extremely high variance, perhaps because the health event that precipitates the creation of a site has not yet occurred.  The post-authorship initiation model demonstrates the importance of receiving an interaction and also that multi-site authors initiate much later than single-site authors.

\subsection{Geographic model}\label{app:geo_analysis}

Table \ref{app:tab:geo_full_model_comparison} show the full model details comparing the conditional mlogit model for initiations (see section \ref{sec:init_withwhom_methods}) with a model fit using only the subset of initiations between users that are assigned US states (see section \ref{sec:geo_analysis_methods}).

\begin{table}[!htbp] \centering
\begin{tabular}{lcc}
\\[-1.8ex] & (1) Pre-authorship & (2) Post-authorship \\
\hline \\[-1.8ex]
 Intercept & 8.049$^{**}$ & 25.696$^{**}$ \\
  & (0.139) & (0.355) \\
 Role = Mixed & -0.568$^{}$ & 1.572$^{*}$ \\
  & (0.415) & (0.654) \\
 Role = P & -1.33$^{**}$ & -3.926$^{**}$ \\
  & (0.248) & (0.412) \\

HC = Cancer & 0.241$^{}$ & 3.449$^{**}$ \\
  & (0.234) & (0.374) \\
HC = Cardiovascular/Stroke & 0.597$^{}$ & 3.172$^{**}$ \\
  & (0.526) & (0.859) \\
HC = Condition Unknown & -3.106$^{}$ & 20.102$^{**}$ \\
  & (7.176) & (2.574) \\
HC = Congenital/Immune Disorder & -1.197$^{}$ & -3.44$^{}$ \\
  & (1.275) & (1.997) \\
HC = Infant/Childbirth & -1.356$^{}$ & 4.551$^{**}$ \\
  & (0.886) & (1.089) \\
HC = Injury & 0.38$^{}$ & 5.433$^{**}$ \\
  & (0.611) & (0.936) \\
HC = Neurological Condition & 0.405$^{}$ & 6.293$^{**}$ \\
  & (0.657) & (1.038) \\
HC = Other & -1.403$^{}$ & 19.134$^{**}$ \\
  & (0.813) & (0.695) \\
HC = Surgery/Transplantation & 0.426$^{}$ & 8.003$^{**}$ \\
  & (0.609) & (0.86) \\
 Will become multi-site author? & -0.845$^{}$ & \\
  & (0.438) & \\
 Is multi-site author? & & 12.671$^{**}$ \\
  & & (0.979) \\
 Int received? & & -10.475$^{**}$ \\
  & & (0.35) \\
 Int received? : Time to first received int & & 0.732$^{**}$ \\
  & & (0.024) \\
\hline \\[-1.8ex]
 Observations & 5,438 & 20,687 \\
 R$^{2}$ & 0.008 & 0.128 \\
 Residual SE & 7.175(df = 5425) & 23.191(df = 20672)  \\
 F Statistic & 3.531$^{**}$ (df = 12; 5425) & 216.265$^{**}$ (df = 14; 20672) \\
\hline
\end{tabular}
\caption{Linear regression models predicting the time between an author's first update and first initiation.  
Model (1) includes only pre-authorship initiators, whereas model (2) includes only post-authorship initiators.
``Time to first received int'' is the number of months between an author's first update and first received interaction.
Note: $^{*}$p$<$0.05; $^{**}$p$<$0.01.
}
\label{tab:init_timing_full_models}
\end{table}

\subsection{Right-censoring and survival analysis}\label{app:sec:right_censoring}

In addressing RQ2b, we predict the number of interactions in a relationship, rather than relationship duration (section \ref{sec:relationship_methods}).
We avoid doing a survival analysis due to long gaps between author interactions on CB making it hard to predict right-censoring. Simulating an end of the dataset 6 months earlier than the true end and assuming that any author with a published update or interaction within 6 months of the simulated dataset end is right-censored, we miss more than half of the authors that are actually censored (recall = 0.495); this level of inaccuracy occurs despite using a censorship threshold that is twice that used in prior work \cite{ma_write_2017}.

This difficulty leads us to avoid fitting survival models to predict relationship duration. Empirically, relationships initiated by caregivers are longer than those initiated by patients (27.8 months vs 26.7 months respectively, t=5.64, p<0.001), although as discussed above this effect may be due to caregivers staying on CB longer or some other effect introduced by the right-censored nature of the data.

\begin{figure}
\centering
\includegraphics[width=\textwidth]{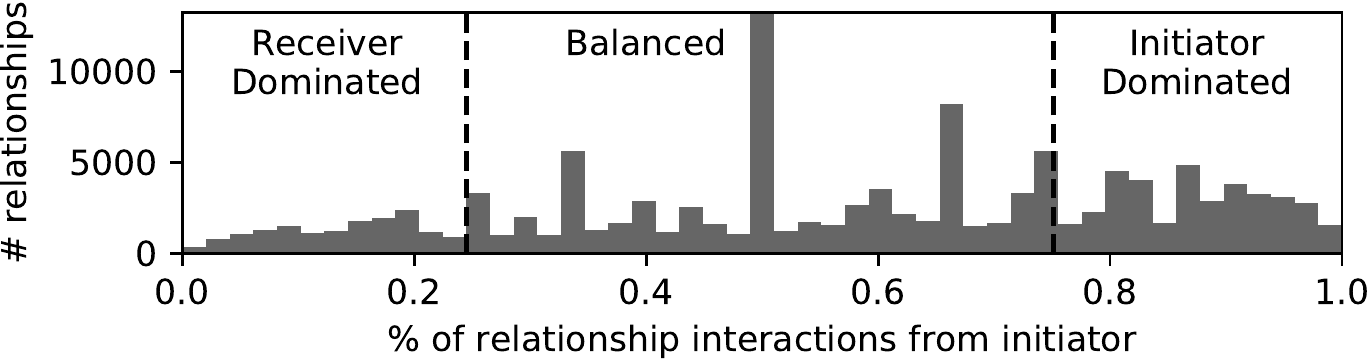}
\caption{Distribution of the proportion of a relationship's interactions made by the initiator of that relationship and the thresholds used to identify balanced relationships. 52.47\% of relationships are balanced using the indicated thresholds.}
\label{fig:relationship_balance_distribution}
\end{figure}

\begin{table}[!htbp] \centering 
\begin{tabular}{@{\extracolsep{5pt}}lccc} 
\\[-1.8ex] & (1) & (2) & (3)\\ 
\hline \\[-1.8ex] 
Candidate out-degree (log) & $-$0.191$^{***}$ & $-$0.225$^{***}$ & $-$0.231$^{***}$ \\ 
  & (0.005) & (0.025) & (0.029) \\ 
Has in-degree? & 0.756$^{***}$ & 1.907$^{***}$ & 1.795$^{***}$ \\ 
  & (0.017) & (0.170) & (0.182) \\ 
Candidate in-degree (log) & 0.649$^{***}$ & 0.960$^{***}$ & 0.974$^{***}$ \\ 
  & (0.005) & (0.027) & (0.030) \\ 
Is reciprocal? & 20.016$^{***}$ & 8.458$^{***}$ & 7.999$^{***}$ \\ 
  & (0.460) & (0.574) & (0.568) \\ 
Is weakly connected? & 1.767$^{***}$ & 4.647$^{***}$ & 4.281$^{***}$ \\ 
  & (0.021) & (0.430) & (0.435) \\ 
Is friend-of-friend? & 5.220$^{***}$ & 3.196$^{***}$ & 2.801$^{***}$ \\ 
  & (0.097) & (0.201) & (0.207) \\ 
Candidate Role = Mixed & 0.020 & 0.110 & 0.111 \\ 
  & (0.018) & (0.085) & (0.098) \\ 
Candidate Role = P & $-$0.242$^{***}$ & $-$0.187$^{***}$ & $-$0.247$^{***}$ \\ 
  & (0.012) & (0.063) & (0.072) \\ 
Same author role? & 0.299$^{***}$ & 0.331$^{***}$ & 0.295$^{***}$ \\ 
  & (0.012) & (0.059) & (0.068) \\ 
Same health condition? & 0.213$^{***}$ & 0.366$^{***}$ & 0.400$^{***}$ \\ 
  & (0.009) & (0.049) & (0.056) \\ 
Candidate multi-site author? & 0.315$^{***}$ & 0.752$^{***}$ & 0.772$^{***}$ \\ 
  & (0.015) & (0.058) & (0.066) \\ 
Candidate mixed-site author? & 0.474$^{***}$ & 0.367$^{***}$ & 0.352$^{***}$ \\ 
  & (0.008) & (0.059) & (0.067) \\ 
Candidate update count & $-$0.0003$^{***}$ & $-$0.001$^{***}$ & $-$0.0004$^{***}$ \\ 
  & (0.00004) & (0.0001) & (0.0001) \\ 
Candidate update frequency & 0.007$^{***}$ & 0.013$^{***}$ & 0.013$^{***}$ \\ 
  & (0.0002) & (0.002) & (0.002) \\ 
Days since recent update & $-$0.011$^{***}$ & $-$0.006$^{***}$ & $-$0.005$^{***}$ \\ 
  & (0.00005) & (0.0001) & (0.0001) \\ 
Days since first update & $-$0.001$^{***}$ & $-$0.001$^{***}$ & $-$0.0005$^{***}$ \\ 
  & (0.00001) & (0.00003) & (0.00004) \\ 
Same U.S. state assignment? &  &  & 2.723$^{***}$ \\ 
  &  &  & (0.069) \\ 
\hline \\[-1.8ex] 
Observations & 155,141 & 7,007 & 7,007 \\ 
Log Likelihood & $-$133,746.600 & $-$4,830.810 & $-$3,743.011 \\ 
Test Accuracy & 77.2\% & 84.3\% & 87.1\% \\
\hline \\[-1.8ex] 
\end{tabular}
\caption{Conditional mlogit models for initiation with the subset of authors given US state assignments via IP geolookup. Model (1) is the full model on all the initiations in the initiations period. Model (2) includes only the subset of authors with state assignments. Model (3) is that same subset with an additional dummy variable indicating matching state assignment between the initiator and the candidate. Comparing (1) and (2) demonstrates that this author subset is broadly similar in initiation factors compared to the full author sample, while (3) demonstrates the importance of matching state assignments.  Note that a matching state assignment is less important than the network-based features. \textit{Note:} $^{***}$ indicates p$<$0.01}
\label{app:tab:geo_full_model_comparison} 
\end{table}

\end{document}